\documentclass[9pt,twocolumn,twoside]{pnas-blank}
\include{pnasresearcharticle.sty}
\usepackage{comment}
\templatetype{pnasresearcharticle} 

\title{Motor-driven microtubule diffusion in a photobleached dynamical coordinate system}

\author[a,$\wedge$]{Soichi Hirokawa}
\author[a]{Heun Jin Lee}
\author[b]{Rachel A Banks}
\author[c]{Ana I Duarte}
\author[c]{Bibi Najma}
\author[c]{Matt Thomson}
\author[b,c,1]{Rob Phillips}

\affil[a]{Department of Applied Physics, California Institute of Technology, Pasadena, CA, USA}
\affil[b]{Division of Biology and Biological Engineering, California Institute of Technology, Pasadena, CA, USA}
\affil[c]{Department of Physics, California Institute of Technology, Pasadena, CA, USA}
\affil[$\wedge$]{Present address: Institut de Biologie du D\'{e}veloppement de Marseille, Aix-Marseille Universit\'{e}, Marseille, France}

\leadauthor{Hirokawa}

\significancestatement{Cytoskeletal reorganization can come at the
cost of local diffusive-like disordering effects from the same active elements,
but distinguishing these processes can be challenging. By
photobleaching an actively contracting microtubule network, we show that the bulk redistribution of filaments exhibit a diffusion-like
reorganization that can be tuned by the effective motor speed. By tuning these
parameters, we show a conserved relationship between active contraction rates and effective diffusion constants, suggesting that
while advection of the cytoskeletal network dominates over scales of 
tens to hundreds of microns, motors additionally induce a diffusive-like effect that begins
to compete with advection at micron scales.}

\authorcontributions{S.H., H.J.L., R.P. designed research. S.H., H.J.L., B.N. performed research. S.H., H.J.L., A.D., R.B. provided
reagents and analytical tools. S.H. and A.D. analyzed data. S.H., H.J.L., R.B., M.T., R.P. wrote the manuscript.}
\authordeclaration{The authors declare no conflict of interest.}
\correspondingauthor{\textsuperscript{1}To whom correspondence should be addressed. E-mail: phillips@pboc.caltech.edu}


\begin{abstract}
Motor-driven cytoskeletal remodeling in cellular systems can often
be accompanied by a diffusive-like effect at local scales,
but distinguishing the contributions of the ordering process, such as
active contraction of a network,
from this active diffusion is difficult to achieve. 
Using light-dimerizable kinesin motors to spatially control the formation 
and contraction of a microtubule network, we deliberately 
photobleach a grid pattern onto the filament 
network serving as a transient and dynamic coordinate system to observe the deformation and translation of the
remaining fluorescent squares of microtubules. We find that the network contracts at a rate set by motor speed 
but is accompanied by a diffusive-like spread throughout the bulk of the contracting network
with effective diffusion constant two orders of magnitude lower than that for a freely-diffusing microtubule. 
We further find that on micron scales,
the diffusive timescale is only a factor of $\approx3$ slower than
that of advection regardless of conditions, showing that the global contraction and long-time relaxation from this diffusive
behavior are both motor-driven but exhibit local competition within the network bulk.

\end{abstract}
\dates{This manuscript was compiled on \today}
\doi{\url{www.pnas.org/cgi/doi/10.1073/pnas.XXXXXXXXXX}}

\begin{document}

\verticaladjustment{-2pt}
\maketitle
\thispagestyle{firststyle}
\ifthenelse{\boolean{shortarticle}}{\ifthenelse{\boolean{singlecolumn}}{\abscontentformatted}{\abscontent}}{}

Whether for schools of fish evading a sea lion or in the ordered array of microtubules comprising the spindle of dividing cells, coordinated movement and emergent patterning is a hallmark of biological dynamics across all biological scales.
Curiosity surrounding the underlying principles dictating such a ubiquitous feature in biology have led to an explosion of theoretical \cite{vicsek1995,tonertu1995,ramaswamy2010,
denk2020} and experimental efforts \cite{nedelec1997,kudrolli2008,kumar2014,copenhagen2021,sanchez2012} to understand them. 
{\it In vitro} active matter systems offer a powerful means to study how cytoskeletal elements self-organize to generate a diverse array of networked structures. By mixing multimerized motors with filaments, a broad range of ordered patterns have been demonstrated, occurring in solutions which are spatially-homogeneous \cite{nedelec1997,surrey2001,sanchez2012} or are locally defined through patterned light \cite{schuppler2016,ross2019, zhang2021,lemma2023}. A common observation from these assays is that the constitutive filaments rearrange in time under dynamics that appear to be primarily advective in nature.  Recent efforts have led to several quantitative models that macroscopically describe the flow-like redistribution of microtubules under a range of conditions related to properties of the motors and filaments  \cite{lee2001,nedelec2001,sankararaman2004,belmonte2017,foster2015}. 
In addition to advective behavior, previous theoretical studies of contractile active gels have also shown that local fluctuations within a globally contracting network can give rise to a motor-driven diffusive-like effect among filaments \cite{mackintosh2008,furthauer2020}, a phenomenon that has been observed experimentally \cite{kulic2008}. 
This seeming competition between active diffusion and 
advection is
poorly understood and invites a rigorous approach to distinguish
these two effects.
\begin{figure}[t]
\centering{
\includegraphics[trim={0cm 0cm 0cm 0cm}, scale=0.75]{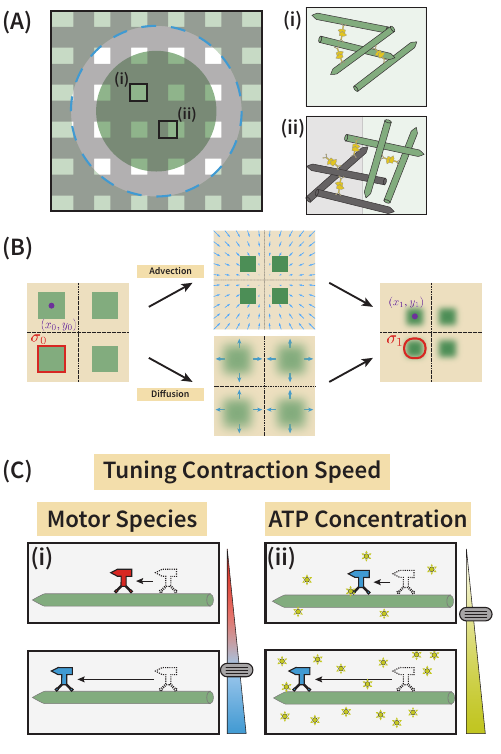}}
\caption{\textbf{FRAP-based approaches to studying advective and diffusive-like redistribution of cytoskeletal elements.} (A) Photobleaching a grid-like pattern leaves (i) squares of fluorescent microtubules (green) surrounded by (ii) non-fluorescent filaments (black) and allows us to examine the role of diffusive-like microtubule spread in the bulk of a global radially contracting network. Dashed blue circle outlines the edge of the dimerizing light inside of which 
the filaments couple and create a net contraction. (B) Tracking of centroids $\left[ (x_0, y_0) \,\mathrm{ to }\, (x_1, y_1) \right]$ and areas ($\sigma_0$ to $\sigma_1$) of the fluorescent squares allows us to quantify the advective and diffusive contributions in the contracting system. (C) The rates of these dynamics can be tuned by changing the effective motor speed through either (i) changes in the
motor species or (ii) changes to the ATP concentration in the system. We tune these parameters to examine rates of contraction
and bulk reorganization of microtubules in the contracting cytoskeletal network.}
\label{fig:cell_life}
\end{figure}

In the work presented here, we incorporate fluorescence recovery after photobleaching (FRAP) into a light-controllable kinesin motor dimerization system \cite{ross2019, qu2021, banks2023} to characterize the interplay of motor-driven advective and diffusive dynamics. FRAP studies have typically been accompanied by 
various theory-based extensions of the diffusion equation to account for convective flow, reaction of molecules, or transport \cite{axelrod1976,nedelec2001,braga2007,kumar2010,ciocanel2017} and have been effectively applied to active {\it in vitro} systems to determine how filaments are redistributing into or elastically contracting relative to the photobleached region \cite{furthauer2020,tayar2021,lemma2021,colin2023}. For our study, we photobleach a grid pattern onto a contracting microtubule network, which creates square fluorescent regions (Fig. \ref{fig:cell_life}(A)).
By tracking the area and centroids of these regions, we are able to account for the advective contraction of the network.  By measuring how the darkened photobleached lines blur, we can account for how much the microtubules in the network undergo diffusive behavior (Fig. \ref{fig:cell_life}(B)).
\begin{figure*}[t!]
\centering{
\includegraphics[trim={0cm 0cm 0cm 0cm}, scale=0.94]{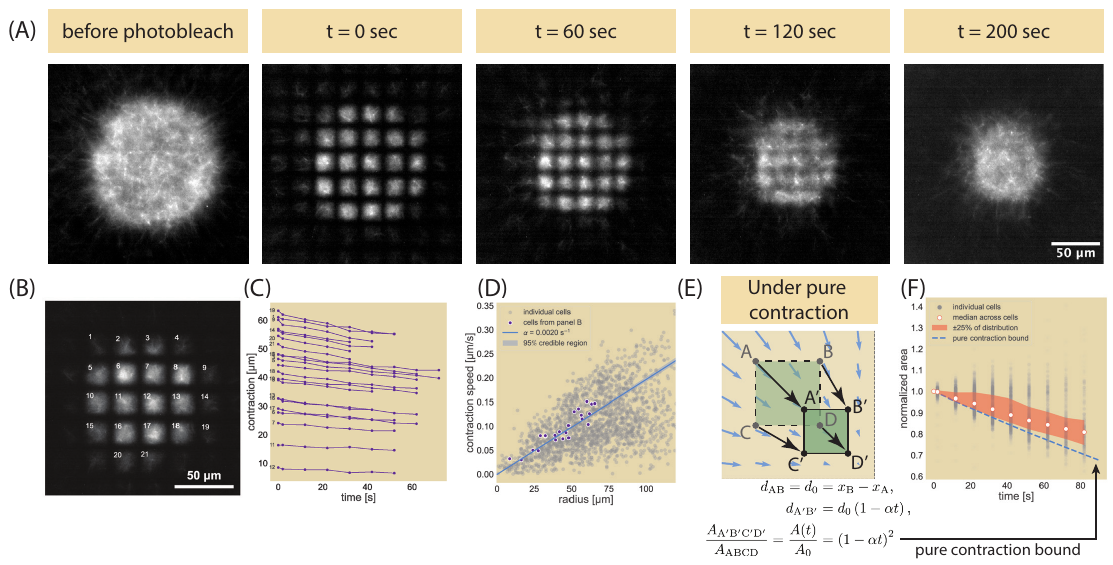}}
\caption{\textbf{Photobleaching a grid pattern onto the contracting microtubule network.} (A) Example dataset, where the microtubule field is photobleached and the deformations of the fluorescent regions observed using Ncd236 and 1.4 mM ATP. (B) Enumeration of individual fluorescent unit cells to (C) compute the distance of their centroids from the center of the network over time. Numbers correspond
to labels from panel (B).
(D) Plot of unit cell contraction speed as a function of their average distances from the center of the network, obtained by fitting the distance vs time data found in (C) to individual lines.  The median contraction rate is
$\alpha = 2.0 \times 10^{-3}$ s$^{-1}$. (E) Schematic of the unit cell
deformation and expected area change under a purely contracting condition. (F) The area of each unit cell is normalized against their initial area as obtained by the unit cell segmentation scheme and plotted as a function of time. The median normalized area is plotted in white among individual unit cells (gray). The red shaded region encompass points between the first and third quartiles of the distribution of all cells.  Dashed blue line corresponds with the normalized area computed in (E) and using the median contraction rate obtained in (D).}
\label{fig:expt_analysis}
\end{figure*}

We find the choice of motor species \cite{miki2001} or
the availability of ATP \cite{yaginuma2014} are key parameters controlling the network dynamics (Fig. \ref{fig:cell_life}(C)). 
For example, by reducing motor speed, whether through decreased ATP concentration or slower motor species, the network globally contracts at a slower rate while the
bulk of the network exhibits a decrease in effective diffusion constant. We further show that contraction rate and effective diffusion constant are linearly proportional measures across all of our conditions and give rise to a tightly bounded
Pecl\'{e}t number slightly greater than one over micron length scales. 
While motors were understood to set the global contraction of the
network, they play a second competing role within the network boundary
that gives rise to a long-time relaxation on the local cytoskeletal structure.
\section*{Results}

\subsection*{Photobleaching a grid pattern} \label{sec:experiment}
To study the local redistribution of microtubules as the network contracts, we designed an augmented optical system that
induces dimerization of kinesin through the iLid-micro system \cite{guntas2015} 
and images the microtubules \cite{ross2019,banks2023}. This modification includes a photobleaching element that allows us to photobleach a grid-like pattern into the microtubule channel (see 
\hyperref[sec:matmeth]{Materials \& Methods} section and SI Sec S1.6). The end result is an array of squares roughly 12 {\textmu}m
in side length and 25 {\textmu}m in center-to-center distance, much longer than the median microtubule length of $\approx 1.5$ {\textmu}m
(see SI Sec S2). Fig. \ref{fig:expt_analysis}(A) shows an example 
of a grid pattern photobleached onto a microtubule network at different time points in its life history and the subsequent
deformations of the bleached lines and fluorescent squares. As the image for the $t=0$ sec timepoint in Fig. \ref{fig:expt_analysis}(A) shows, upon photobleaching the grid pattern, individual fluorescent squares, which we will call unit cells, are produced. 
Over a minute after photobleaching, unit cells contract toward the center of the network while the photobleached lines appear to blur away. By two minutes after photobleaching,
neighboring unit cells appear to blend into each other and at later times any remnants of the photobleached pattern disappear.

\subsection*{Tracking fluorescent squares shows global contraction and local diffusive spread} \label{sec:unitcells}

To better quantify and understand the global network contraction dynamics, 
we segmented individual unit cells and measured their centroids and areas over successive frames (see SI Sec S3 and S4 for analysis). By tracking individual unit cells such as those shown in Fig. \ref{fig:expt_analysis}(B) and computing their distance from the center of the
network over successive frames (Fig. \ref{fig:expt_analysis}(C)), we can determine the local contraction speeds, where we see a rough linear correspondence between distance and time.
We computed the slopes of each unit cell trajectory and compared the resultant speeds as a function of their distance from the network center (Fig. \ref{fig:expt_analysis}(D)) to find that the median contraction 
speed linearly increases with distance from the center, indicating a general linear contraction of the entire microtubule network. We thus fit the velocity against the radius ${\bf{r}}$ with a line passing through the origin (see SI Sec S5.1), giving the expression
\begin{align*}
  {\bf{v}}\!\left({\bf{r}}\right) &= - \alpha {\bf{r}},\numberthis \label{eq:velocity_profile}
\end{align*}
where $\alpha$ is the contraction rate and thus measure a contraction rate and 95\% credible region of 
$\alpha = 2.0 \times 10^{-3}\pm 5 \times 10^{-5}$ s$^{-1}$. Data separated by experimental
replicates are available in SI Sec S6.

Despite the linear global contraction observed for the centroids, a more 
careful examination of the unit cells reveals that the network does not simply undergo purely elastic contraction. Suppose we took two points with
different x-positions but the same y-position $(x_\mathrm{A}, y_\mathrm{A})$ and $(x_\mathrm{B}, y_\mathrm{A})$,
respectively, such as points A and B in Fig. \ref{fig:expt_analysis}(E) that they have a distance $d_0$ of
\begin{equation}
    d_0 = x_\mathrm{B} - x_\mathrm{A}, \label{eq:distance_0}
\end{equation}
where we will take $x_\mathrm{B} > x_\mathrm{A}$. If the two points were strictly subject
to move from the velocity field given by Eq. \ref{eq:velocity_profile}, 
after time $t$ their
positions will have changed such that their distance $d_1$ is now
\begin{align*}
    d_1 &= \sqrt{\left[ \left( x_\mathrm{B} - \alpha x_\mathrm{B} t - x_\mathrm{A} + \alpha x_\mathrm{A} t \right) \right]^2 + \left( y_\mathrm{A} - \alpha y_\mathrm{A} t - y_\mathrm{A} + \alpha y_\mathrm{A} t \right)^2},\\
    &= \left( x_\mathrm{B} - x_\mathrm{A} \right) \left( 1 - \alpha t \right), \\
    &= d_0 \left( 1 - \alpha t \right). \numberthis \label{eq:distance_1}
\end{align*}
So the two points move closer by a factor of $1 - \alpha t$ in that
time. We can make a similar argument for two points vertically separated.
If we imagine this for all four points that make up the corners of a unit
cell (points ABDC transforming to A'B'D'C' in Fig. \ref{fig:expt_analysis}(E)) and look at the change in area, we would expect that under a purely
contractile active system subject to the linear contraction rate measured
from tracking the unit cell centroids, the area $A(t)$ would change
from its initial size $A_0$ by
\begin{equation}
    A(t) = A_0 \left( 1 - \alpha t \right)^2. \label{eq:area_scaling}
\end{equation}

See SI Sec S7 for a more complete derivation.
Fig. \ref{fig:expt_analysis}(F) shows the normalized area of each unit cell as a function of time in gray against this pure contraction scaling given
as a blue dashed line. The median normalized area is shown as a white circle with a red outline.  
As can be seen by comparison with the shaded red region (representing the 50 percent of all cells that fall between the first and third quartiles of the distribution of cell areas), the majority of the experimental observations are above the pure contraction bound. With the area being greater than that for a purely contracting network, we conclude that despite the global contraction of the network, filaments can locally spread and reorganize in the bulk. This observation is further affirmed by the merging of originally distinct fluorescent squares in the 120-second time point of Fig. \ref{fig:expt_analysis}(A), where a purely contractile network would cause fluorescent squares to remain distinct.

\subsection*{The effective diffusion constant is roughly two orders of magnitude lower than free diffusion of a microtubule}

Since a purely contractile description is insufficient to fully capture the observed dynamics, we generalize our treatment of this contractile
effect and accounting for diffusion using an advection-diffusion equation to model
the time evolution of the tubulin concentration $c({\bf{r}},t)$. 
Such a model has a material flux $\bf{J}$ of the form
\begin{align*}
  {\bf{J}} &= - D \nabla c + {\bf{v}}\!\left( {\bf{r}}\right) c, \numberthis \label{eq:flux_term}
\end{align*}
where $D$ is the diffusion constant and ${\bf{v}}\!\left({\bf{r}}\right)$ is the velocity profile of the advective flow as a function of distance
from the center of contraction $r$. As motivated by results shown in Fig. \ref{fig:expt_analysis}, we use the velocity field given by Eq.
\ref{eq:velocity_profile} with $\alpha$ as the computed contraction rate as shown in Fig.
\ref{fig:expt_analysis}(D). When inserted into Eq. \ref{eq:flux_term} and combined with the continuity equation, the
advection-diffusion equation takes the form
\begin{align*}
  \frac{\partial c}{\partial t} &= - \nabla \cdot {\bf{J}},\\
  \frac{\partial c}{\partial t} &= D \nabla^2 c + \nabla \cdot \left( \alpha {\bf{r}} c \right). \numberthis \label{eq:core_pde}
\end{align*}
We perform a series of deep explorations of the model in SI Sec S8-S10 to better understand the time-evolution of the concentration 
profile subject to Eq. \ref{eq:core_pde} and to validate the implementation of our finite element method (FEM) using COMSOL 
Multiphysics$\circledR$. Ultimately, these exercises confirm the importance of using a grid-like photobleaching pattern rather than a circular pattern, the latter of which would have convoluted the contributions of the radially-directed contraction with the more-isotropic diffusive-like spread. After validating our initial simulation, we turn fully to the use of FEM simulations on Eq. \ref{eq:core_pde} to model our experimental results.
\begin{figure}[t!]
\centering{
\includegraphics[scale=0.5, trim={0cm 0cm 0cm 0cm}]{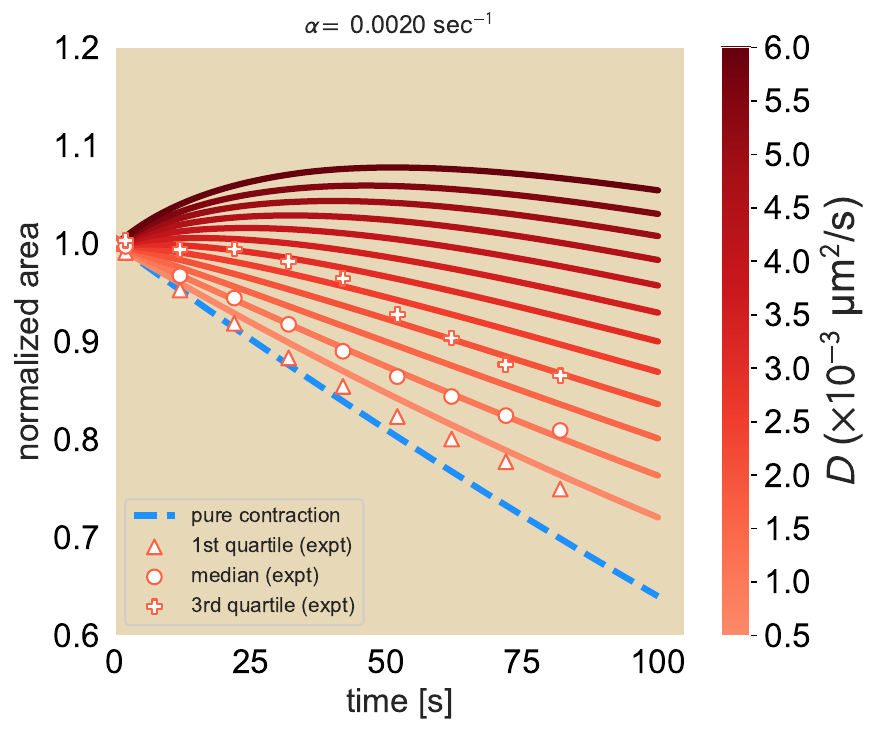}}
\caption{\textbf{Simulated concentration profiles for a linear advection-diffusion equation.} A family of curves for the expected normalized area of fluorescent squares subject to a fixed advection rate $\alpha = 2 \! \times \! 10^{-3}$ s$^{-1}$ and varying diffusion constants. The 25th percentile (triangle), median (circle), and 75th percentile
(plus sign) of the area trajectories are overlayed onto the FEM results for comparison.}
\label{fig:advdiff_grid}
\end{figure}
We perform a FEM simulation of an individual unit cell, where we fix the FEM contraction rate to be the experimentally-measured mean rate and sweep across various diffusion constants (see SI Sec S11 for implementation of a single unit cell in COMSOL).
For example, for the Ncd236 motor at saturated levels of ATP, we show in Fig. \ref{fig:advdiff_grid} the family of normalized area trajectories for a single unit cell subject to the experimentally observed mean contraction rate of
$\alpha = 2\!\times\!10^{-3}$ s$^{-1}$, and for diffusion constants $D$ ranging from $5 \! \times \! 10^{-4}$ to $6 \! \times \! 10^{-3} \, \frac{\text{{\textmu}m}^2}{\text{s}}$.
For comparison to the experimental data, we overlay the median (circle), the 1st quartile (triangle), and 3rd quartile (plus symbol) from the distribution of measured unit cell areas trajectories, where the quartiles give a sense of the trajectory variation. 
Here, we see that by minimizing least squares between the experiments
and the simulation conditions, the median normalized area
trajectory agrees best with the FEM trajectory with an effective diffusion constant of $1.0 \! \times \! 10^{-3}$ $\frac{\text{\textmu m}^2}{\mathrm{s}}$.
The first quartile of area trajectories from measurements lies between the pure contraction limit where $D_\mathrm{eff} = 0$ and an effective diffusion constant $D_\mathrm{eff} = 5.0 \, \times \, 10^{-4}$ $\frac{\text{\textmu m}^2}{\mathrm{s}}$. We interpret the first quartile results to mean the effective diffusion coefficient must be above $5.0 \, \times \, 10^{-4}$ $\frac{\text{\textmu m}^2}{\mathrm{s}}$ in order to model the majority of the data. The third quartile of area trajectories most closely follows the trajectory with an effective diffusion constant of $\approx 2.0 \, \times \, 10^{-3}$ $ \frac{\text{\textmu m}^2}{\mathrm{s}}$ (see SI Sec S4.2 on fitting procedure), serving as a kind of upper bound.   For context, the diffusion coefficients associated with the median and quartiles are all about two orders of magnitude smaller than the diffusion coefficient of a freely diffusing microtubule, which is $\approx 0.1 \, \frac{\text{{\textmu}m}^2}{\text{s}}$  (see SI Sec S2 for microtubule
length; equation for the longitudinal diffusion constant obtained from Ref \cite{ross2019}).

We also explored how well our FEM simluations could capture the qualitative features of the data set shown in Fig. \ref{fig:expt_analysis}(A), such as the merging of unit cells and the time scale of this process (see SI Sec S12).  Our main finding from these efforts is, even for just qualitative comparisons, diffusion must be included in the theoretical description of the dynamics.       
\subsection*{Changing effective motor speed proportionally changes contraction rate and effective diffusion constant}

\begin{figure}[t!]
\centering{
\includegraphics[trim={0cm 0cm 0cm 0cm}, scale=0.38]{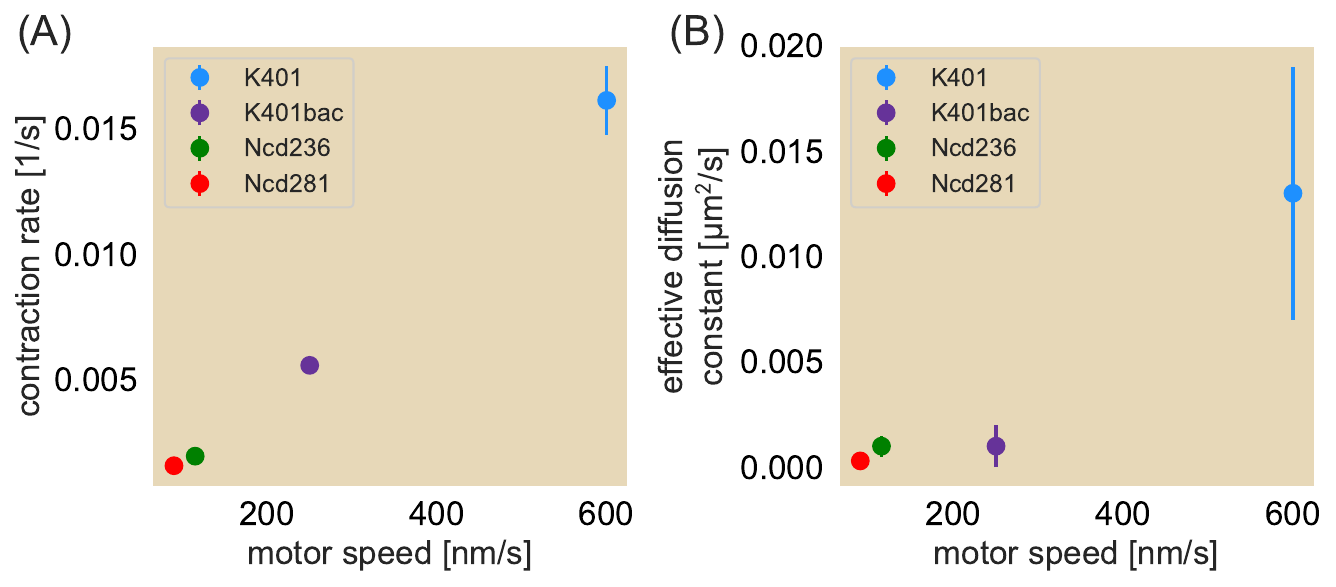}}
\caption{\textbf{Contraction rates and effective diffusion constants for four different motor types.} (A) Contraction rate as a function of motor speed. Here, the motors are (in order of motor speed as found in Table S2 of the SI Text)
Ncd281 (red) \cite{endres2006}, Ncd236 (green), K401 expressed in bacteria (purple), and K401 expressed in Sf9 cells (blue) \cite{banks2023}. (B)
Corresponding effective diffusion constants as a function of motor speed where the circles denote the medians of the experimentally obtained
normalized area trajectories and error bars denote the middle 50\% of the distribution. Error bars for some data points
are smaller than the radius for the size of the dots.}
\label{fig:four_motors}
\end{figure}

To test whether motors play a role in the diffusive-like effect, we next tuned the effective speed of the active elements,
namely, the motors themselves. Two ways in which this can be done is by the choice of motor species or by changing the concentration
of ATP. 
We chose optogenetic versions of previously characterized motor variants that span roughly an order of magnitude in speeds: Ncd236, Ncd281
\cite{endres2006}, K401 expressed in bacteria \cite{ross2019}, and K401 expressed in Sf9 cells \cite{banks2023} (See SI Sec S14 for Ncd281 construct designs and
motor speeds and processivities).
Fig. \ref{fig:four_motors}(A) 
shows the motors speeds for each of these motor types and the associated contraction rate. Interestingly, we observe a roughly linear relationship between contraction rate and  motor speeds. 
The effective diffusion coefficient also demonstrate a roughly linear trend (Fig. \ref{fig:four_motors}(B)). We note that even for the slowest motor Ncd281, 
a non-zero coefficient $D_\mathrm{eff}$ of $3 \times 10^{-4}$ 
$ \frac{\text{\textmu m}^2}{\mathrm{s}}$ is needed to accurately describe the data. 
This general tendency to increase the effective diffusion constant suggests that the motor speed may be responsible for the local microtubule effective diffusion. We note both trends can be explained by just the speeds of the motor species, requiring no consideration of motor processivity.

We next examine how this local diffusive-like effect changes when we decrease the motor speed of a given species by reducing ATP concentrations.
In order to traverse along microtubules, motors must hydrolyze ATP with each step they take. At saturated
concentrations of ATP, motors can hydrolyze ATP at their maximal rate and therefore move at their maximum
speed. At reduced ATP concentrations, the limited availability of ATP causes motors to hydrolyze ATP at
a reduced rate, leading to an effective reduction in motor speed \cite{howard1989,visscher1999,schief2004}. In the context of a contracting
microtubule network, we hypothesize that this decrease in motor speed translates to a reduction in contraction
rate, similar to the effect observed when using a slower motor species. We further hypothesize that for a constant 
motor concentration, reducing the concentration of ATP will increase the fraction of motors that do not move along 
the microtubules and instead behave as passive crosslinkers, causing the areas of the fluorescent unit cells to fall 
closer to the pure contraction bound. 
\begin{figure}[t!]
\centering{
\includegraphics[trim={0cm 0cm 0cm 0cm}, scale=0.18]{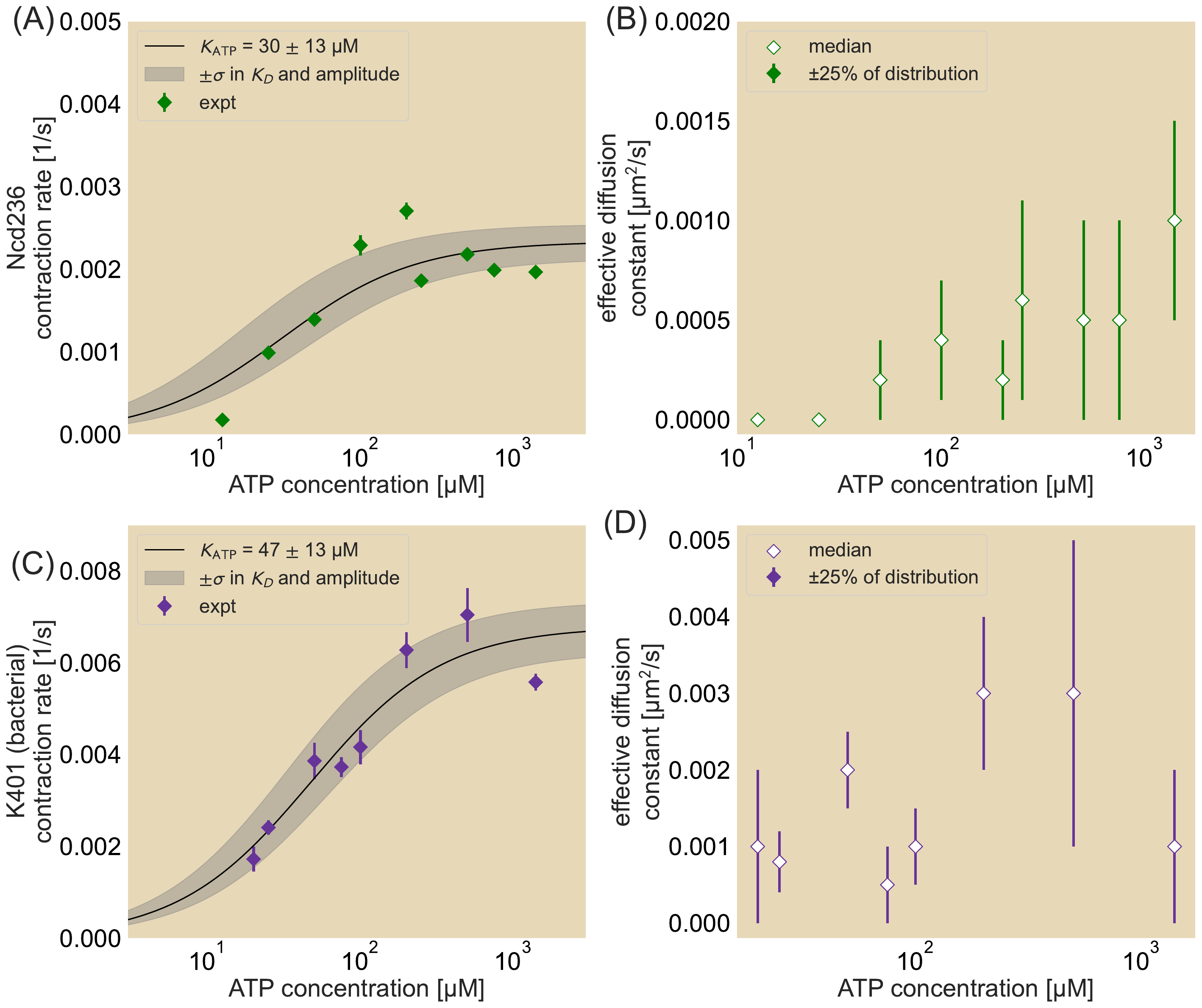}}
\caption{\textbf{Contraction rates and effective diffusion constants
over a range of ATP concentrations.} Contraction rates (A and C) and effective diffusion constants (B and D) as a function of ATP
concentration in the system. Motors used are Ncd236 (A and B) and K401 expressed in bacteria (C and D). Black line represents best fit to a Michaelis-Menten equation. Edges of the gray shaded region bounded to the left (right) using the Michaelis-Menten 
equation where the amplitude is the best fit plus (minus)
one standard deviation and the dissociation constant is the best fit minus (plus) one standard deviation. 
Effective diffusion constants fitted
to the median area trajectories with error bars corresponding to fits spanning the middle 50\% of the distribution motor types are presented.}
\label{fig:atp}
\end{figure}
To test this, we perform our photobleach experiment for Ncd236 and bacteria-expressed K401 at ATP concentrations spanning
two orders of magnitude. We continue to use an ATP regeneration system 
so that the ADP concentration is negligible and therefore does
not compete with ATP for the hydrolysis site \cite{schief2004} (see
Materials and Methods). Fig. \ref{fig:atp}(A) and (C) show that for both Ncd236 (A) and
K401 (C), as the concentration
of ATP is decreased, the contraction rate of the network similarly decreases. At 25 {\textmu}M ATP, the
contraction rate with Ncd is roughly half of that at saturated levels. This concentration is also roughly the measured dissociation constant of
ATP hydrolysis by the motor \cite{foster1998}. However, at ATP concentrations below this dissociation constant, we see that the network contraction,
while still occurring for an ATP concentration of half the dissociation constant for Ncd236, dramatically slows down. 
We fit the contraction rate against ATP concentration to the best
fit of a Michelis-Menten equation to find that the expected dissociation constant for ATP hydrolysis in the contracting network
context (Ncd236: $30 \pm 13$ {\textmu}M; bacterial-expressed
K401: $47 \pm 13$ {\textmu}M) is roughly the same as for measured motor speeds (Ncd236: $\approx \! 23$ {\textmu}M 
\cite{foster1998}; bacterial-expressed K401: $28.1 \pm 0.9$ {\textmu}M \cite{schief2004}). 

Decreasing the ATP concentration overall reduces the effective diffusion
constant for Ncd236 (Fig. \ref{fig:atp}(B)). As the network dynamics scale with
ATP concentration in a similar way to single-motor kinetics, our results show that 
motors are drivers of 
not only the contraction rate but also the local diffusion-like relaxation of the network.
On the other hand, such a trend is less clear for bacteria-expressed K401
at some ATP concentrations (Fig. \ref{fig:atp}(D)), suggesting a need to further explore the underlying
cause for this behavior.

\subsection*{Contraction rate and effective diffusion constant are unified in the P\'{e}clet number}

\begin{figure}[b!]
\centering{
\includegraphics[scale=0.4, trim={0cm 0cm 0cm 0cm}]{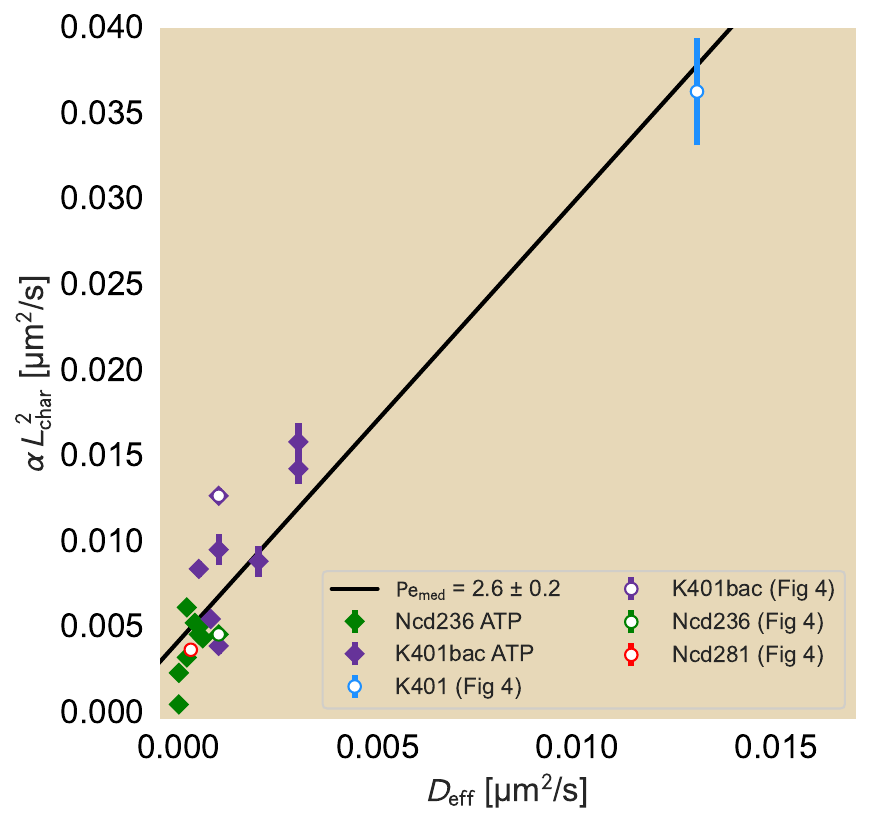}}
\caption{\textbf{Relation of contraction rate and effective diffusion constant}. Comparisons of contraction rate to effective diffusion constant
are made for effective diffusion constants fitted to the median normalized area
trajectories and obtained across all ATP concentration (diamonds) and motor species (hollow circles with colors matching those in Fig. \ref{fig:four_motors})
conditions. Contraction rates are multiplied by the square of a characteristic length scale, in this case roughly the median length of a microtubule
in experiments (1.5 {\textmu}m), to match the units of the effective diffusion constant.
Slopes of lines are best fits of $\mathrm{Pe}$, which are reported with their respective standard deviations
in the legend.}
\label{fig:peclet}
\end{figure}

From tuning the motor type and ATP concentration in our \emph{in vitro} kinesin-microtubule system and measuring
the resultant contraction and diffusion rates,
 we see when parameters increase the contraction rate of the network they also similarly increase the effective diffusion constant. 
This suggests a relationship between the advective and diffusive properties. 

To characterize this, we make use of the Pecl\'{e}t number, a non-dimensional ratio between the rates of advection and diffusion in the system, given by
\begin{align*}
\mathrm{Pe} &= \frac{\text{diffusive timescale}}{\text{advective timescale}},\\
&= \frac{L_\mathrm{char}^2 / D}{L_\mathrm{char} / \mathrm{v}_\mathrm{char}},\\
&= \frac{\mathrm{v}_\mathrm{char} \times L_\mathrm{char}}{D}, \numberthis \label{eq:peclet_def}
\end{align*}
where $L_\mathrm{char}$ and $\mathrm{v}_\mathrm{char}$ are the characteristic length scale
and characteristic advective velocity, respectively. $L_\mathrm{char}$ determines the length scale in the system over which
the advective and diffusive timescales are compared. 
In the system, candidates for $L_\mathrm{char}$ may be as small as a typical microtubule in the assay
and as large as the size of the contracting system. As the microtubule
length is within an order of magnitude of the photobleached line and gives a sense 
of the local competition between diffusion and advection, we choose $L_\mathrm{char} = 1.5$ {\textmu}m. We relate $\mathrm{v}_\mathrm{char}$ to
$L_\mathrm{char}$ through the global contraction rate of the network $\alpha$. 
In other words, $\mathrm{v}_\mathrm{char} = \alpha \times L_\mathrm{char}$. So we have
\begin{equation}
    \mathrm{Pe} = \frac{\alpha \, L_\mathrm{char}^2}{D}. \label{eq:peclet}
\end{equation}

To estimate the Pecl\'{e}t number, we plot the median contraction rates multiplied by the square of a characteristic length scale against the corresponding effective diffusion rates for all experimental conditions (ATP concentration as diamonds and
motor species as hollow circles) in Fig. \ref{fig:peclet}, 
which demonstrates a roughly linear relationship.
The slope of the line through a set of constants gives $\mathrm{Pe}$,
specifically $\mathrm{Pe} = 2.6 \pm 0.2$.
To get a sense for how much $\mathrm{Pe}$ varies due to variability
within conditions,
we find $\mathrm{Pe}_{25} = 4.5 \pm 0.5$ 
and $\mathrm{Pe}_{75} = 2.4 \pm 0.1$ for the first and third quartile datasets, respectively
(see SI Sec S14), suggesting
that $\mathrm{Pe}$ is tightly constrained. In all cases, $\mathrm{Pe} \gtrsim 1$, suggesting that the 
effect of diffusion is smaller than that of advection, but comparable to within an order of magnitude. This makes sense
as the net effect of the unit cells, despite exhibiting
a local diffusive-like effect, shrinks in area. Further,
the rather narrow range in the P\'{e}clet number despite the spread in quartiles
further suggests that the speed of the active elements sets both the
global contraction of the network and the local spread of individual filaments.
We recall that our choice of $L_\mathrm{char}$ allows us to examine the
local competition between advection and diffusion. Had we chosen
the side length of a unit cell or the size of the system as our size scale, 
we would see the increase in $L_\mathrm{char}$ results in 
$\mathrm{Pe} \gg 1$, demonstrating the greater dominance
of the advective component over larger length scales, consistent with the net contraction
in the network. Our results indicate that while advection generally dominates over diffusion, most notably over
longer length scales, the close linear relation between the two rates suggest that 
they are both set by the speed of the motors.
\section*{Discussion}

The dynamic cytoskeleton is critical to carrying out key processes within cells, such as the formation
and maintenance of the mitotic spindle \cite{hueschen2019}, cell division by cytokinesis \cite{miller2012}, and
as centers of morphogenetic information \cite{maroudassacks2021}. Such motor-filament structures are 
vital to a cell or organism, but how the constituent cytoskeletal elements reorganize to reach the same
end configuration due to changes in biochemical conditions has been unclear. 
In order to understand this response by the kinesin-microtubule network, we developed
an experimental framework for probing the bulk redistribution of the filament network 
using a grid photobleaching pattern. By photobleaching the network, we
observe that
microtubules will undergo a diffusive-like spread that locally opposes the global, linear contraction of the system. 
The diffusive-like behavior and contraction rate are jointly tuned by changes to the effective motor speed either from
using different motor species or altering the ATP concentration. These effects also appear to occur from tuning crowding agents, 
a topic that we discuss in SI Sec S15-S16. In short, not only is the contraction an
actively driven phenomenon, but so too is the diffusive-like behavior.

As we observed a general increase in effective diffusion constant with the increase in contraction rate, we further
probed this relationship to find a roughly linear relationship between the two measurements. This strongly suggested that the
P\'{e}clet number, which is the ratio of the diffusive timescale to the advective one, remains roughly constant within the bulk of the
contracting network regardless of the biochemical conditions we used here and further suggests that motor velocity not only
sets the rate that the system contracts but also the effective diffusivity in the bulk.

Active diffusion has been observed in other active systems both \emph{in vitro} \cite{furthauer2020,tayar2021,lemma2021,colin2023}
and \emph{in vivo} \cite{brangwynne2007,weber2012,guo2014}. This phenomenon is particularly exciting in
those cases where the diffusive-like motion exhibits a dependency on the availability of ATP. \emph{In vitro}, a Michaelis-Menten like
relationship has been observed between ATP concentration and the spatial rate of deformation of the system 
\cite{lemma2021}. This relationship between the rate at which deformations occur 
and ATP concentration is consistent with our observations that both the macroscopic deformation
rate and local diffusion is determined by an effective velocity of the motor. 
Furthermore, the fact that this ATP dependence also occurs in \emph{in vivo} systems \cite{weber2012, guo2014}
suggests interesting implications in the ability of cells to carry out key enzymatic reactions.
One key parameter that has not yet been explored is the competition of ADP in the system.
We recall that our work utilized an ATP recycling system that allows us to ignore such competing effects. 
However, cells also have a supply of ADP that may compete with ATP for the ATP hydrolysis
site of motors \cite{schief2004}. A natural extension of this study would involve systematically tuning the ratio of ATP to ADP
to observe the effects of reorganization due to this competition and may provide key insights on the role of
metabolic activity on altering the rates of ATP-dependent processes.

Various quantitative models have been made to recapitulate the experimentally
observed motor-filament dynamics \cite{belmonte2017,schuppler2016,foster2015,furthauer2020,furthauer2021,qu2021}.
These models incorporate contractile stresses in the system, either through motor-driven
activity or Stokeslets in the flow field, resulting in
a net elastic behavior \cite{schuppler2016,foster2015,qu2021}. As we report here, a
long-time relaxation term is required to create the diffusive-like filament redistribution in
the network bulk.
Recent treatments to introduce a long-time relaxation, as found 
in Ref \cite{yang2022}, propose force-balance approaches that incorporate a 
viscoelastic stress to locally oppose the active stress underlying the elastic contraction.
As our findings indicate that the relaxation is motor driven and transports filaments
based on the direction of their orientation, we propose that an active viscoelastic-like stress term would be necessary
and would need to depend
on the motor speed and local polarity of the network.
Indeed, other works account for the orientation $\vec{p}$ of the filaments not only for driving the movement of motors but as time-dependent variables through crosslink-generated
torques \cite{furthauer2020,furthauer2021}, offering encouraging pathways to recapitulate active
cytoskeletal reorganization.

Our work has provided deeper insights into the extent of filament redistribution during
network reorganization, where the active element not only drives the global contractile behavior
but also generates a local redistribution that can be tuned by their effective speeds.
Our findings leave many exciting questions about these self-organizing systems. 
Much is still not known about the origins of the network formation from the initially random
orientation and uniform distribution of filaments prior to contraction. Specifically, the 
key criteria of the formed network, whether in the form of a density or order dependence, to 
drive the contraction process remain unclear.
Photobleaching as applied in our work here provides a helpful macroscopic view of filament 
reorganization that can serve as a complement to other methods that are likely required to probe
the dynamics of the filaments
in the network, such as their orientation when they become coupled by the multimerized motors.
Furthermore, the work presented here offers motivation
for examining redistribution of molecules in other actively driven contexts, such as in systems of opposing 
motors or subject to more complex iLid-micro activation geometries \cite{qu2021} and with the
introduction of ADP to compete against ATP-dependent reactions.

\matmethods{\subsection*{Microscopy set-up} \label{sec:matmeth}
The microscopy elements used to activate the iLid-micro dimerization and image the different fluorescence
channels are similar to those found in Ross \emph{et al.} \cite{ross2019}. Briefly, a digital light processing projector
from Texas Instruments was used to activate the motor dimerization and image the microtubule channels. An excitation filter
wheel was placed in front of the projector to filter out the different channels. Photobleaching was performed using a
diode laser with a center wavelength of 642 nm. A piezoelectric mirror gimbal mount from Thorlabs was placed downstream
of the laser to deflect the beam path over a small range before the laser light passes through a cylindrical lens array
inserted into a direct-drive rotation mount. The gimbal mount can then sweep the projected lines laterally to thicken
the photobleaching lines before the rotation mount is rotated 90$^\circ$ and the gimbal mount changes the deflecting
angle of the beampath in the orthogonal direction. 
Imaging is performed using a 20x objective. More details are available in the SI Sec S1.6.

\subsection*{Microtubule network assay}
The microtubule network formation and contraction assay is set up similarly as in Ross \emph{et al.} 
\cite{ross2019}. Micro- and iLid-tagged motors are mixed in equal motor monomer ratios with GMPCPP-stabilized
microtubules labeled with Alexa 647 in a reaction mix containing among other components ATP, ATP recycling reagents including pyruvate kinase and phosphoenolpyruvate (PEP), 
and pluronic as a crowding agent.
While elements of the oxygen scavenging are kept in the reaction, the glucose oxidase is removed
from the reaction to ensure photobleaching. Removal of these oxygen scavengers minimally affects fluorescence
intensity during imaging from using the projector over the time range over which the data is analyzed, as shown in SI Sec S3. 

\subsection*{Image acquisition arrangement}
Control of the light-dimerizing activation, photobleach laser activation, and imaging are performed through the Micro-Manager (MM)
software \cite{edelstein2010, edelstein2014} while photobleaching is synchronized using a series of in-house compiled executable files
that control the movement of the gimbal and rotation mounts. During acquisition, a beanshell script in MM changes the
projection pattern on the DLP to create a circular light pattern for the iLid activation and full field for the
imaging channels. When the desired state of the microtubule network is reached for performing photobleaching, the
script completes the image acquisition cycle before turning on the photobleaching laser and calling to the executables
to create the grid before the next acquisition cycle.

\subsection*{Motor purification}
Kinesin motors are expressed using the pBiex-1 vector transfected in Sf9 suspension cells. Cells are transfected at 5-7 {\textmu}g for 
every 15$\times10^6$ cells at a starting concentration $10^6$ cells per mL of Sf900-III media using a liposome-based
transfection regent (Escort IV Transfection Reagent). Cells are harvested $\backsim$60-72 hours after transfection
and purified using the FLAG affinity tag and anti-FLAG antibody resins. Proteins are stored in 50\% glycerol by
volume with 1.5 mM DTT, 50 {\textmu}M EDTA, 50 {\textmu}M EGTA, and 15 {\textmu}M ATP and stored at -20$^\circ$C. Full
storage buffers and final concentrations of components are available in the SI.

\subsection*{Data Availability}
All data and code are publicly available. Raw image files and COMSOL simulation file can be downloaded from the 
CaltechDATA research data repository under the DOI:10.22002/f23ds-f2v87. Analyzed data files and code generated
by Python (for analyses) and BeanShell Scripts and C\# (for hardware communication) for the work presented here
are available on the dedicated GitHub repository under the DOI:10.5281/zenodo.12806576.}
\showmatmethods{}

\acknow{We thank members of the Rob Phillips lab for useful discussions. We would also like to thank the David Van Valen 
and Rebecca Voorhees labs for providing resources for performing the protein expression and purification. We also thank
Justin Bois, Griffin Chure, Peter Foster, Sebastian F\"{u}rthauer, Victor Gomez, Stephan Grill, Catherine Ji, Frank J\"{u}licher, Matthias Merkel, Daniel Needleman,
Le\"{i}la Peri\'{e}, Henk Postma, Madan Rao, Shahriar Shadkhoo, and Fan Yang. This work was supported by 1R35 GM118043 and 
2R35 GM118043 Maximizing Investigators’ Research Awards (MIRA) (to R.P.). S.H. was also supported by the Foundational
Questions Institute (FQXI) (R.P.).}
\showacknow{}
\bibliography{advdiff_bib}
\end{document}


\title{Supplementary Information: Motor-driven microtubule diffusion in a photobleached dynamical coordinate system}
\author[a,$\wedge$]{Soichi Hirokawa}
\author[a]{Heun Jin Lee}
\author[b]{Rachel A Banks}
\author[c]{Ana I Duarte}
\author[c]{Bibi Najma}
\author[b]{Matt Thomson}
\author[b,c,1]{Rob Phillips}

\affil[a]{Department of Applied Physics, California Institute of Technology, Pasadena, CA, USA}
\affil[b]{Division of Biology and Biological Engineering, California Institute of Technology, Pasadena, CA, USA}
\affil[c]{Department of Physics, California Institute of Technology, Pasadena, CA, USA}
\affil[$\wedge$]{Present address: Institut de Biologie du D\'{e}veloppement de Marseille, Aix-Marseille Universit\'{e}, Marseille, France}

\date{}
\maketitle
\tableofcontents
\newpage

\section{Materials and Methods} \label{sec:si_am_matmeth}

\subsection{Motor purification} \label{subsec:si_am_motor_pur}

Plasmids containing the gene encoding the motor-fluorescent protein-light-activated dimerization-FLAG tag construct with the
pBiex-1 vector are transfected in Sf9 suspension cells for 60-72 hours at 27$^\circ$C on shakers rotating at 120 rpm.
Cells are then lightly centrifuged at 500 rpm for 12 minutes to remove the supernatant before resuspending in lysis buffer (100
mM NaCl, 2 mM MgCl$_2$, 0.25 mM EDTA, 0.5 mM EGTA, 0.25 \% Igepal, 3.5\% sucrose by weight, 10 mM imidazole
pH 7.5, 10 {\textmu}g/mL aprotinin, 10 {\textmu}g/mL leupeptin, 1 mM ATP, 2.5 mM DTT, and 0.5 mM PMSF) and leaving
on ice for 20 minutes. Cells are then spun down for 30 minutes at 50k rpm after which the lysate is transferred 
to tubes containing mouse monoclonal anti-FLAG resin (Sigma A2220) and slowly rotated at 4$^\circ$C for 1.5$\backsim$3 
hrs to allow protein binding to the resin via the FLAG tag. Resin-bound protein are washed three times by spinning down at 
2000$\times\,g$, clearing the supernatant, then resuspending by tube inversion in wash buffer 
containing 15 mM KCl, 0.5 mM, 0.1 mM EGTA, 0.1 mM EDTA, 2 mM imidazole pH 7.5, 10 {\textmu}g/mL aprotinin, 
10 {\textmu}g/mL leupeptin, 0.3 mM DTT, and ATP in 3 mM, 0.3 mM, and 0.03 mM concentrations for the first, second, 
and third washes, respectively. After the third wash, the protein are spun down again at 2000$\times\,g$ and most of the
supernatant is removed, leaving the resin bed and roughly an equivalent amount of supernatant by volume in the tube. 
The resin bed is resuspended and FLAG peptide (Sigma F4799 or Thermo Scientific A36805) is added at a final concentration
of 0.5 mg/mL before rotating for 3 hrs at 4$^\circ$C. After incubating to allow the peptide to outcompete the protein for resin 
binding, the protein are spun down again at 2000$\times\,g$ with the supernatant extracted and further spun down using 
centrifuge columns with $\backsim$30 {\textmu}m pore sizes to further separate proteins from any collected resin beads.
Flow-through of clarified protein are spin concentrated using a 50 kDa filter tube to a final concentration of 2-2.5 mg/mL before
diluting in 100\% glycerol of the same volume for storage.

\subsection{Stabilized microtubule polymerization} \label{subsec:mt_polymerization}

Fluorescently labeled stabilized microtubules are prepared as in \cite{ross2019,mitchison2012}. After flash thawing at 37$^\circ$C
and kept on ice, a combination of $\approx 1.5$ mg unlabeled and 100 {\textmu}g labeled tubulin are diluted to 7.5 mg/mL and 0.5 
mg/mL, respectively, in M2B 6.8 containing DTT and GMP-CPP at final concentrations of 1 mM and
6mM, respectively. The tubulin mixture is then incubated on ice for 5 minutes in an ultracentrifuge tube before ultracentrifugation at
90,000 rpm at 4$^\circ$C for 8 minutes. Avoiding the pellet at the the bottom, the supernatant containing tubulin monomers are then
placed in a new Eppendorf tube and incubated at 37$^\circ$C for 1 hour, typically in a water bath, during which the tubulin is polymerizing
and stabilizing with GMPCPP. The microtubule mixture is then aliquoted into individual PCR tubes while constantly being suspended in
the mixture by stirring with a pipette tip. PCR tubes are then briefly spun down with a tabletop minicentrifuge before flash-freezing with
liquid nitrogen and placed in a -80$^\circ$C freezer for long-term storage. Microtubules are then prepared for experiments by immersing
the PCR tube in 37$^\circ$C water immediately when taken out of the freezer to quickly thaw.

\subsection{Glass slide treatment} \label{subsec:si_am_glassprep}

Corning glass slides and No. 1.5 Deckgl\"{a}ser coverslips are coated with an acrylamide solution to prevent 
the adhesion of proteins from the light-dimerized activation
assay to the surface. The acrylamide coating is done similarly to that demonstrated in \cite{decamp2016}. Prior to application of 
the solution, slides and coverslips are separated by placement in appropriately sized 
containers and rigorously cleaned through a series of solutions and
sonicating. First, slides are immersed in 1\% Hellmanex to remove dirt particulates, sonicated, repeatedly rinsed with deionized
water (DI H$_2$O), then repeatedly rinsed with ethanol. Slides are then sonicated in 200 proof ethanol before rinsing again with
DI H$_2$O. After rinsing, slides are sonicated in 0.1 M KOH and subsequently rinsed in double-distilled water (ddH$_2$O). Finally,
trace metals are removed by immersing in 5\% HCl for 4 hours. After repeatedly rinsing in ddH$_2$O, slides are stored overnight with MilliQ
ultrapure water.
\\
\\
Upon cleaning and before the acrylamide coating, a silane solution is made
first by mixing 98.5\% 200 proof ethanol and 1\% acetic acid before adding 0.5\% trimethoxysilyl methacrylate and immediately pouring
into the containers holding the slides and coverslips. After roughly 30 minutes, slides are rinsed twice in 200 proof ethanol before drying
with N$_2$ air and baking at 110$^\circ$C for 10-15 minutes to cure silane onto surface with oxygen bonding.
\\
\\
The polyacrylamide solution is made by mixing 950 mL ddH$_2$O with 50 mL 40\% acrylamide and degassing under vacuum for 30
minutes. The solution is then under constant mixing on a stir plate with a stir bar during which time 350 {\textmu}L TEMED and 700 mg
ammonium persulfate (APS) are added to the solution. The acrylamide solution is immediately added to the slides and coverslips
and incubated overnight. Slides are placed in 4$^\circ$C for long-term storage.

\subsection{Flow cell chamber preparation}

Flow cells for all light-dimerized activation assays are prepared by thoroughly rinsing an acrylamide-coated glass slide and
coverslip in ddH$_2$O and air drying with N$_2$ gas. A piece of parafilm with three channels each cut 3 mm wide 
is placed on the glass slide with the long axis of the channels running along the length of the slide. The coverslip is placed on
top of the parafilm with pressure applied to flatten out the film. The flow cell is then briefly placed on a hot plate set at 65$^\circ$C
to warm the parafilm, allowing extra pressure on the contact points between the film and the glass to better seal the chambers.

\subsection{Light-dimerized activation assay preparation}

Photobleaching experiments require an energy mix to maintain stability and function of microtubules and motors while
constantly supplying kinesin motors with ATP to contract the microtubule network. This energy mix is slightly altered from that used by
Ross \emph{et al.} \cite{ross2019} with the major changes being a change in acidity for K-PIPES from pH 6.8 to pH 6.1
and the absence of gluocose oxidase to allow for photobleaching. iLid- and micro-tagged motors with the same fluorescent protein are each 
added to the reaction mixture at final concentrations of 40-100 nM with stabilized microtubules added at a final concentration of
1.5-2.5 {\textmu}M tubulin. Concentrations of motors and tubulin are tuned to ensure that the microtubule network 1) contracts
into an aster, which can fail to occur with too few motors or tubulin, and 
2) without an influx of microtubules from outside of the light-activation region, which can
occur from too much tubulin or too many motors dimerizing in the absense of light.

\subsection{Optical set-up}

The sample is imaged and photobleached using a super planar fluorescence 20x objective from Nikon (numerical aperture 0.45).
Image acquisition is performed using a FLiR Blackfly monochrome camera (BFLY-U3-23S6M-C) with three filters in front
of it: a Semrock Brightline dual-band pass filter centered at 577 nm (28.3 nm FWHM bandwidth) and 690 nm (55.1 nm FWHM 
bandwidth); and a Semrock StopLine single-notch filter at 532 nm (17 nm notch bandwidth) to suppress transmission of the YFP 
laser to the camera.
\\
Activation of motor dimerization and imaging of the microtubules is performed using a digital light projector DLP
Lightcrafter Display 4710 EVM Gen2 from Texas Instruments. The DLP projects white light while a motorized filter wheel sets the
transmissible range of wavelengths onto the sample (beam blocker for no light, 460/50 nm filter for blue light for iLid-micro
dimerization and 630/38 for microtubule imaging). Photobleaching of microtubules is performed using a 645 nm laser. The laser 
path is set to pass through a cylindrical lens
array that transforms the collimated light pattern into a series of lines along one axis. The cylindrical lens array is mounted onto
a rotation mount to allow for photobleaching of vertical and horizontal lines to generate the grid pattern. To ensure that the 
photobleached lines persist for multiple frames of the image, the laser passes through a gimbal-mounted mirror that deflects the
beam over a small range of angles. By deflecting the laser light off of the mirror through two lenses with the same focal length $f$
and a second, stationary mirror placed $4 \times f$ away from the gimbal-mounted mirror before passing the laser through the 
cylindrical lense array, the transformed laser lines can be swept out. We use this beam steering approach to photobleach 
thicker lines.
\\
\\
To perform the activation and imaging patterns, we supply {\textmu}Manager with TIFF image stacks of matching pixel dimensions as
the projector and use a Beanshell script modified from Ross \emph{et al.} \cite{ross2019} to use the correct TIFF image in the stack. The TIFF
stack contains a blank image (all pixel values 0) for when the laser is turned on (which is also used in conjunction with the beam
blocker to prevent light from passing onto the sample outside of the activation and imaging cycles); a maximum pixel intensity
image for the microtubule imaging, and a circular pattern in a blank background for the circular iLid-micro dimerization activation
pattern. The primary modification to the Beanshell script is the incorporation of a timer for when the photobleaching will be
performed. Once the experiment reaches the desired time, the imaging pauses while the Beanshell script turns on the laser and
executes a series of custom written executables that sweep out the laser lines to create thicker parallel photobleached lines, turn
off the laser, rotate the cylindrical lense array, then reactivate the laser and sweep out the laser lines in the orthogonal direction
to generate the grid pattern. Upon finishing this command, the laser is shut off and imaging resumes. The entire photobleaching
is performed within a roughly 10-15 second window.

\section{Microtubule length extraction} \label{sec:am_mt_length}

\begin{figure}[t!]
\centering{
\includegraphics[scale=0.4]{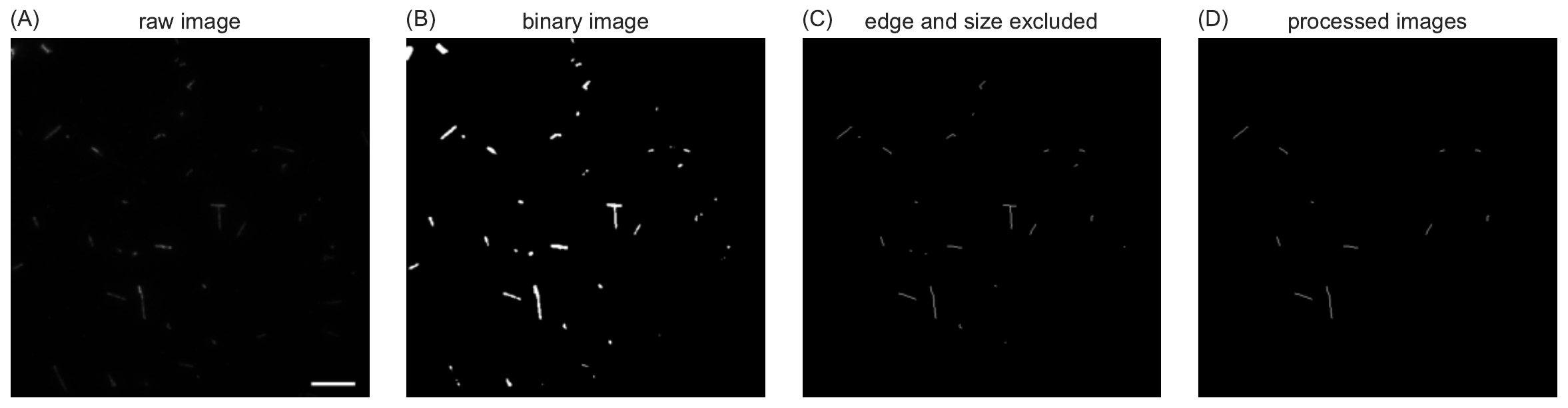}}
\caption{\textbf{Processing steps of microtubule images.} (A) Raw image. Scale bar denotes 10 {\textmu}m. (B) Images processed after
computing a Niblack threshold and using Otsu thresholding on the Niblack threshold array. (C) Putative MTs skeletonized after removing objects
too close to the image border or too small. (D) Removal of any MTs that cross over each other to get the final MTs used for analysis.}
\label{fig:mt_processing}
\end{figure}

Stabilized microtubules imaged under total internal reflection fluorescence (TIRF) microscopy such as the ones shown in Fig.
\ref{fig:mt_processing}(A) were analyzed similar to that discussed
in \cite{ross2019} in order to extract their lengths. Briefly, due to the uneven illumination that can occur in the image, images were
first background corrected using a local thresholding method known as Niblack thresholding \cite{niblack}
with window size of 3 pixels and $k$ value of $0.001$, which determines how many standard deviations below the mean
pixel value that one sets the cut-off in the window. Although the array is a series of pixel values to be weighed against the original image,
we found that this array already improved the image contrast. With this improved contrast but
considering that the result is still a nonbinary image, we used 
Otsu thresholding on the Niblack theshold array to extract the microtubules from the background. The result is shown in Fig. 
\ref{fig:mt_processing}(B).
\\
\\
Using the binary image which contains extracted microtubules, we imposed a morphological closing algorithm to reconnect any microtubules
that were broken during the Niblack thresholding from being picked up as signal. This closing was performed using a 3 pixel x 3 pixel
square array, suggesting that disconnected microtubules needed to be within $3\sqrt{2}$ pixels of each other at their ends to be 
connected again. From here, we removed any microtubules that were too close to the edge of the image as they may extend outside
of the camera field of view and removed any objects that were fewer than 10 pixels in area as we considered them too small to know with enough certainty
whether they were microtubules or small blemishes in the image. Putative microtubules underwent a morphological thinning so that they
were converted to one-pixel wide lines along which we could compute their lengths. The result of the edge and size exclusion and
skeletonizing are shown in Fig. \ref{fig:mt_processing}(C).
\begin{figure}[b!]
\centering{
\includegraphics[scale=0.4]{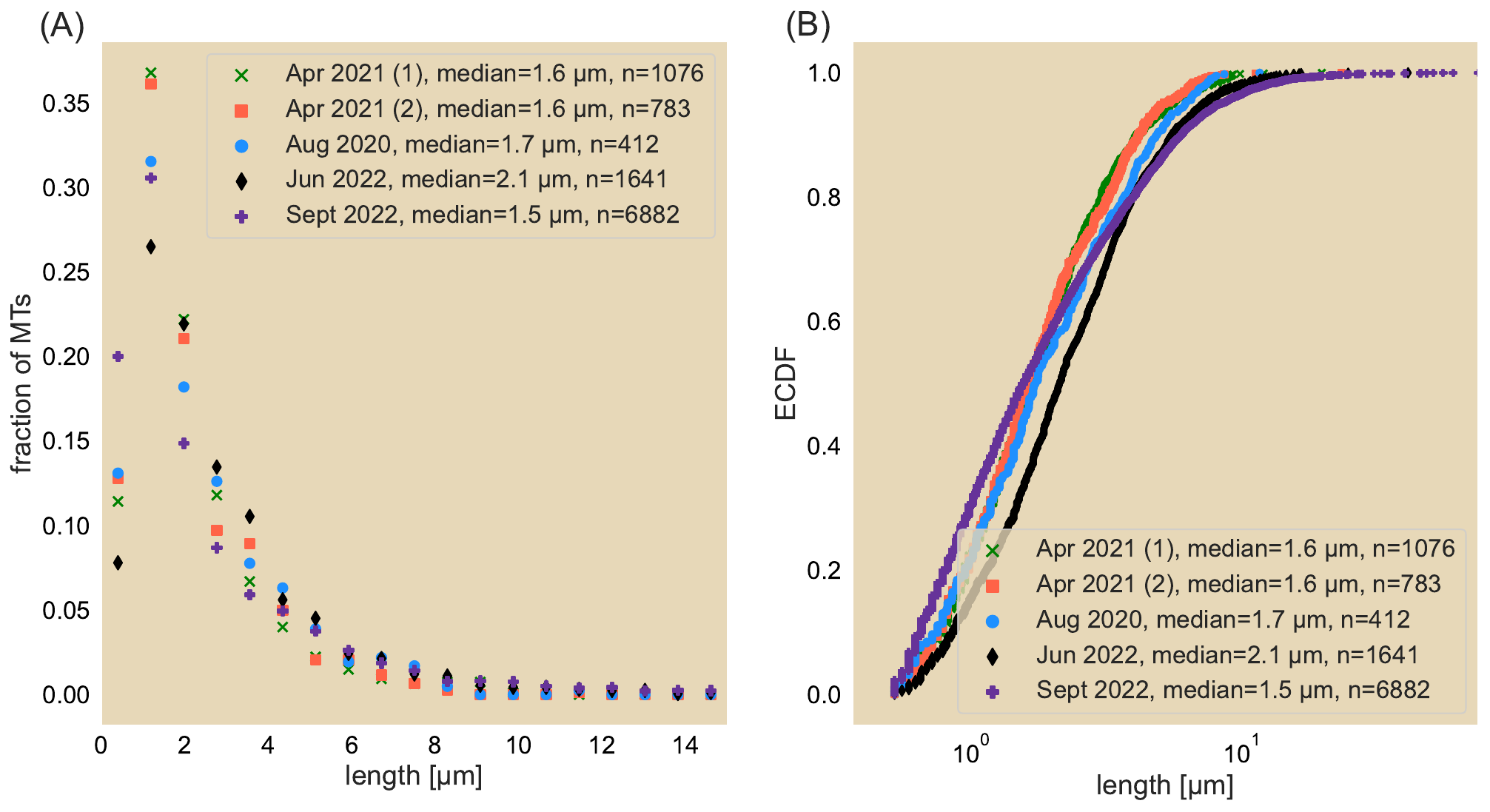}}
\caption{\textbf{Distributions of microtubule length from microtubules stabilized from polymerization preparations 
for experiments used in this manuscript.} Microtubules were prepared five times over the course of the work presented here,
thus shown as five different datasets. Left plot shows the histogrammed length distribution as a linear x-scale of length
while the right plot shows the same data as an empirical cumulative distribution function (ECDF) as a logarithmic x-scale. 
The two polymerization preparations performed in April 2021 were performed separately by
two of the authors of this manuscript on the same day. $n$ denotes the number of microtubules whose lengths were obtained in the 
distributions.}
\label{fig:mt_length}
\end{figure}
\\
\\
As a final step before measuring the lengths, we removed any microtubules that
seem to cross over. This was performed by removing objects where two line segments along the same microtubule strand formed angles
of at least 75$^\circ$, leaving behind a processed image such as Fig. \ref{fig:mt_processing}(D). From here, we used any remaining 
microtubules and measured their lengths and compiled them. Fig. \ref{fig:mt_length}
shows empirical cumulative distribution functions of these microtubules from the five MT polymerization assays performed over the
course of the work presented here. $n$ denotes the number of microtubules that were extracted from the image processing and used in
the distributions for each replicate. Here, we see that for most of the work performed the MTs had lengths between $1-3$ {\textmu}m with median
lengths between $1.5-2.1$ {\textmu}m.

\section{Image processing: global drift correction}
\label{sec:drift_correct}

For computational efficiency in later image processing steps, photobleached images are cropped to contain
only the region where the collective filament network is present. We first find the center of the contracting
network for the image immediately preceding photobleaching. To do so, the image is smoothed with a Gaussian
blur and thresholded with the Yen thresholding method \cite{yen1995}. After
removing objects that are at the image edge or small objects, the largest segmented object is taken. The properties
of this object are then taken, included a pixel-weighted centroid and its major axis length.
\\
\\
For cropping the first photobleached image, we start by cropping a window in the image using the
pixel-weighted centroid as the image center and 130\% of the major axis length as the length of the window. 
This buffer to the window cropping typically ensures
that contracting networks that are drifting can still be easily tracked and cropped. To more efficiently crop the 
image, we then take this cropped window and use a heavy Gaussian blur ($\sigma=30$ pixels) and subtract
this from the cropped photobleached image. We then normalize the image and use Otsu thresholding \cite{otsu1979}
to identify putative fluorescent unit cells. We roughly identify the unit cells by removing those that are close to
the edge of the image as well as objects that are smaller than 36 {\textmu}m$^2$, which would be far smaller than
a unit cell. Unit cells are further cleaned up by filling in any small holes in the unit cell with a morphological closing
before taking the pixel-weighted centroid using all of the unit cells together to get a rough position of the network
center. This process is then repeated on the next photobleached frame using the new centroid for the image center
and the original window length over the desired number of photobleach frames. These cropped images are then
used for further, more careful processing of the unit cells.

\section{Quantifying microtubule unit cell dynamics} \label{sisec:justification}

In this work, we sought to characterize the bulk redistribution of microtubules 
through local deformations and translations within the contracting
network. To develop a processing method that would allow us to quantify
the advective and diffusive components of the network, we first set out to
determine whether the microtubule number is conserved in the system. A part of this
determination, which relies heavily upon the fluorescence signal, depends upon whether the
imaging system also affects the signal over time through passive photobleaching. We seek
to assess these factors in SI Secs. \ref{si:projector} and \ref{si:focus} below.

\subsection{Imaging system negligibly photobleaches microtubules.} \label{si:projector}

One concern in analyzing microtubule 
fluorescence over time is whether the optical system decreases its signal due to secondary photobleaching effects from the projector, which is used to illuminate the field of view for imaging purposes and perform the iLid-micro light stimulation. 
To investigate this, we imaged the microtubule field without activating the iLid-micro dimerization using the same exposure
times (200 ms) and different imaging frequencies depending upon the speed at which contraction takes place, between
3 seconds and 10 seconds per frame. We then examined the mean image intensity and standard deviation
of the pixel intensity as a function of time.
\begin{figure}[t!]
\centering{
\includegraphics[trim={0cm 0cm 0cm 0cm}, scale=0.5]{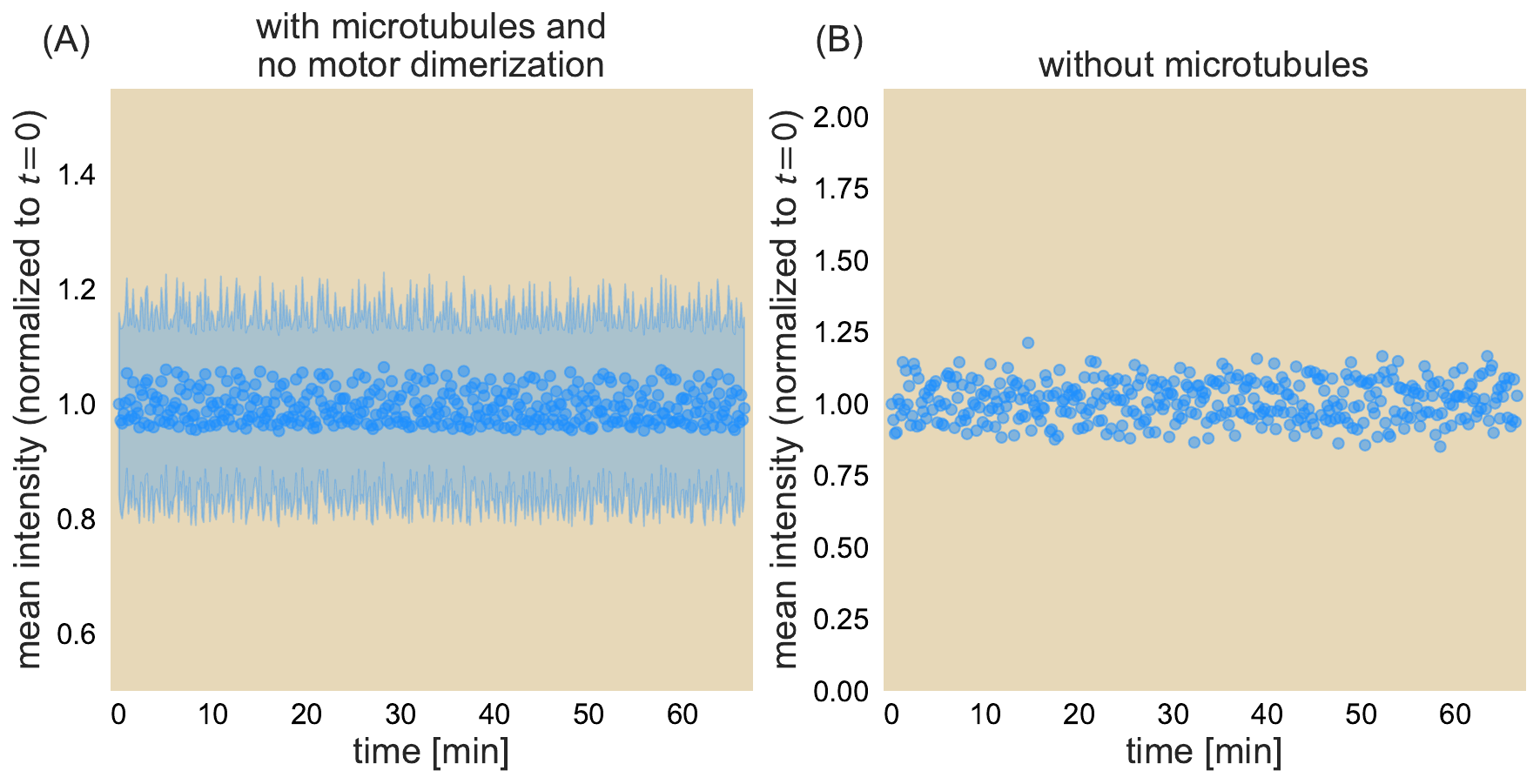}}
\caption{\textbf{Image intensity of the microtubule field as a function of time.} (A) Mean intensity of the microtubule
field normalized against that of the first image. Blue shaded region represents one standard deviation in the mean
intensity (normalized by the same initial mean value). (B) Mean intensity of the same fluorescence channel in the
absence of microtubules.}
\label{sifig:projector}
\end{figure}
\\
\\
Fig. \ref{sifig:projector}(A) illustrates the effects of the projector on the microtubule field. The mean intensity of the
field of view, as normalized against the mean intensity at $t=0$ seconds, indicates that the fluorescence field fluctuates 
within a few percent but does not appear to decrease over an hour. These fluctuations are likely due to fluctuations
from the image acquisition set-up itself, as Fig. \ref{sifig:projector}(B) shows the normalized mean intensity of the 
microtubule fluorescence channel but in the absence of microtubules. Here, we see that that the integrated
intensity fluctuates over the short term but does not appear to exhibit a global decrease, further supporting that 
the small fluctuations in fluorescence intensity in successive imaging
stages comes from the imaging system. Nevertheless, we conclude that the fluorescence intensity is well
preserved over the course of experiments and does not require corrections during image
processing.

\subsection{Net flux of microtubules goes into the imaging plane.} \label{si:focus}

As the projector does not passively photobleach the microtubule
channel (SI Sec. \ref{si:projector}), we 
next ask whether there is a loss of microtubules during the contraction process.
Microtubules may disconnect from the contracting network and diffuse away. 
For example, as we use an epifluorescent
imaging set-up, if microtubules are lost from the network by
moving out of the plane of imaging, they will project a low, more diffuse signal onto the
image. In contrast, microtubules that move into the plane of focus will exhibit a higher signal. Similarly, microtubules lost at the
periphery in the image plane will lead to a reduced integrated
intensity across the entire network.
\\
\\
To determine the effects of flux across the focal plane, we measure the integrated fluorescence signal of the
contracting network after photobleaching. This is done by integrating the fluorescence signal of the activated network as
it contracts away from the unilluminated reservoir. We then examine the normalized integrated signal over
time.
\\
\begin{figure}[t!]
\centering{
\includegraphics[scale=0.7]{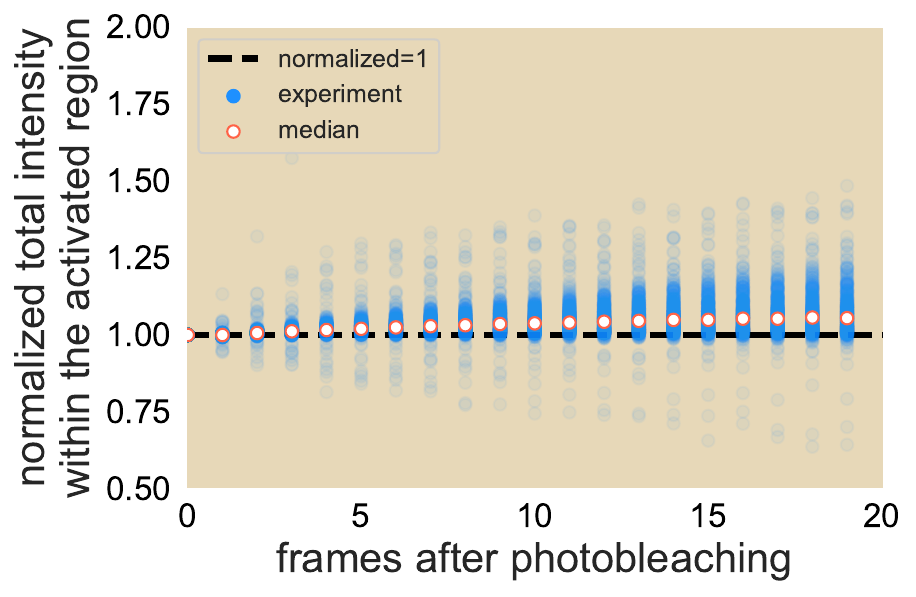}}
\caption{\textbf{Integrated intensity of the photobleached contracting network over image frames.} White dots 
denote the median value across all experiments.}
\label{sifig:focus}
\end{figure}

\noindent
Fig. \ref{sifig:focus} shows that the integrated intensity of the contracting network increases in time. Over 20
imaging frames, the fluorescence increases by about 5\%, suggesting a roughly 0.25\% increase between frames.
While some of the observed increase in intensity can be accounted for by those datasets where the contracting
network is not fully disconnected from the unilluminated reservoir and thus introducing more microtubules at the
periphery, we suspect that the majority of this increase comes from an increase of microtubules that are
entering the imaging plane. This observation makes sense as we expect a growing concentration of microtubules entering 
the imaging plane due to network contraction. Had we accounted for this increase in intensity over successive 
frames, our results would at best have led to a greater area of the unit cells than the ones we computed, which 
would produce greater effective diffusion constants. Even so, we argue that the roughly quarter of a percent
increase between frames is relatively minor and conclude that the
total microtubule count remains roughly constant over the course of
the experiment.

\subsection{Number conservation of unit cells} \label{subsec:unitcell_processing}

Due to the negligible photobleaching effects of the projector on
the network and the small influx of microtubules in the imaging
plane, we make the assumption that the total number of microtubules for the entire network is 
conserved. We further assume
that at the local level, the number of fluorescent microtubules that compose a unit
cell is also conserved. As a result, we choose to identify and track unit cells in time by
conserving their integrated fluorescence intensity. 
\begin{figure}[t!]
\centering{
\includegraphics[scale=0.5]{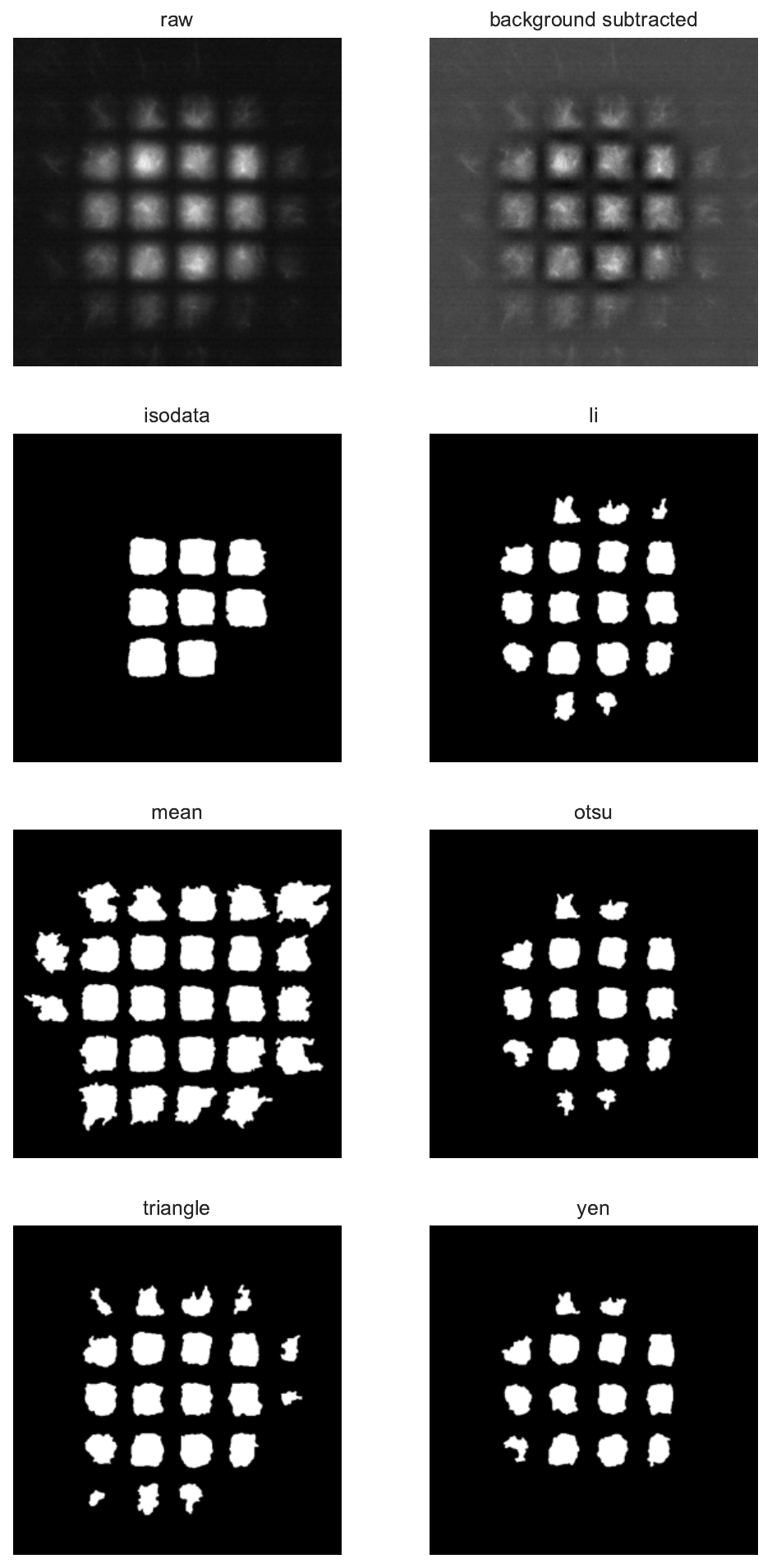}}
\caption{\textbf{Various thresholding schemes of fluorescent squares.} Top two images correspond to the raw (left) and background-subtracted (right) images. 
The thresholding schemes used, in order, were isodata, Li, mean,
Otsu, triangle, and Yen thresholding methods. Due to the under-representation of unit
cell fluorescent signal for all the other methods, we opt for the mean thresholding
scheme to identify unit cells.}
\label{sifig:threshold}
\end{figure}
Fluorescent unit cells of a photobleached microtubule network are thus segmented in the cropped image sets where the microtubules outside
of the activation region are neglected. For each image, we identify the unit cells by first enhancing the contrast between the fluorescence signal of the unit cells and the background through the
subtraction of a heavily Gaussian blurred form of the image ($\sigma=20$
pixels) and subtracting off this blur from the original image. Pixel values are then normalized across the image to fall between 0 and 1. 
\\
\\
In order to
identify each fluorescent square, we tested a variety of thresholding
schemes using the Sci-kit Image package for Python. In summary,
the following thresholding schemes are:
\begin{itemize}
    \item Isodata - identifies those threshold values where, when each pixel is grouped according to whether it lies above or below the threshold value, the threshold value is the average of the two binned groups.
    \item Li - iteratively computes the cross-entropy between the image and a binary image with a different thresholding value. The returned threshold value is that which reduces the cross entropy
    \cite{li1993}.
    \item Mean - computes the mean pixel value across the image \cite{glasbey1993}.
    \item Otsu - finds the threshold that minimizes the sum of the variances of the background and foreground \cite{otsu1979}.
    \item Triangle - computes a line from the peak in the histogram to the last histogram bin (if the peak is shifted to the left of the histogram). A second perpendicular line is drawn from this line toward the first histogram bin it touches. The corresponding x-value gives the threshold \cite{zack1977}.
    \item Yen - computes the minimum cross-entropy between the image and thresholded binary image while accounting for the bit depth of the image \cite{yen1995}.
\end{itemize}
Furthermore, we seek the method that best identifies the unit cells
not only in the bulk that will appear as squares but also those that
lie along the periphery that may not appear as complete squares after the
photobleaching is applied but are nevertheless part of the network.
\\
\\
Fig \ref{sifig:threshold} shows these various thresholding schemes performed on the background-subtracted image (top right) with
comparisons to the original raw image (top left). We see that while the
isodata thresholding approach misses many of the fluorescent squares, other thresholding
schemes reasonably render the threshold of the squares. We notice that the
Li, Otsu, triangle and Yen thresholding schemes miss unit cells on the
periphery of the network, especially if they are not squares as in those found toward
the center of the microtubule network. To keep track of their dynamics, we elect to use
the mean thresholding algorithm, which from visually comparing the threshold to the
raw data better represents the unit cells, including those unit cells on the network periphery. After
the thresholding is applied, the segmented image is cleaned up by removing segmented objects that are too small (less than a third of the
area of a unit cell immediately after photobleaching) and objects that are larger than the area of a unit cell. A morphological closing is performed where any holes smaller than 3 pixels $\times$ 3 pixels within a fluorescent unit cell is closed. 
These small holes may arise from a local minimum in signal that falls in the background regime during thresholding.
With the segmented images from the first frame, the centroid position, area, and total fluorescence of each unit cell 
are computed. For total fluorescence, we compute the pixel intensity by taking the raw image signal
and subtracting the average background 
signal inherent to the camera. Fig \ref{sifig:unitcell_correction}(A)
provides a schematic of the resultant thresholding to initially identify unit cells.
\begin{figure}[b!]
\centering{
\includegraphics[scale=0.6]{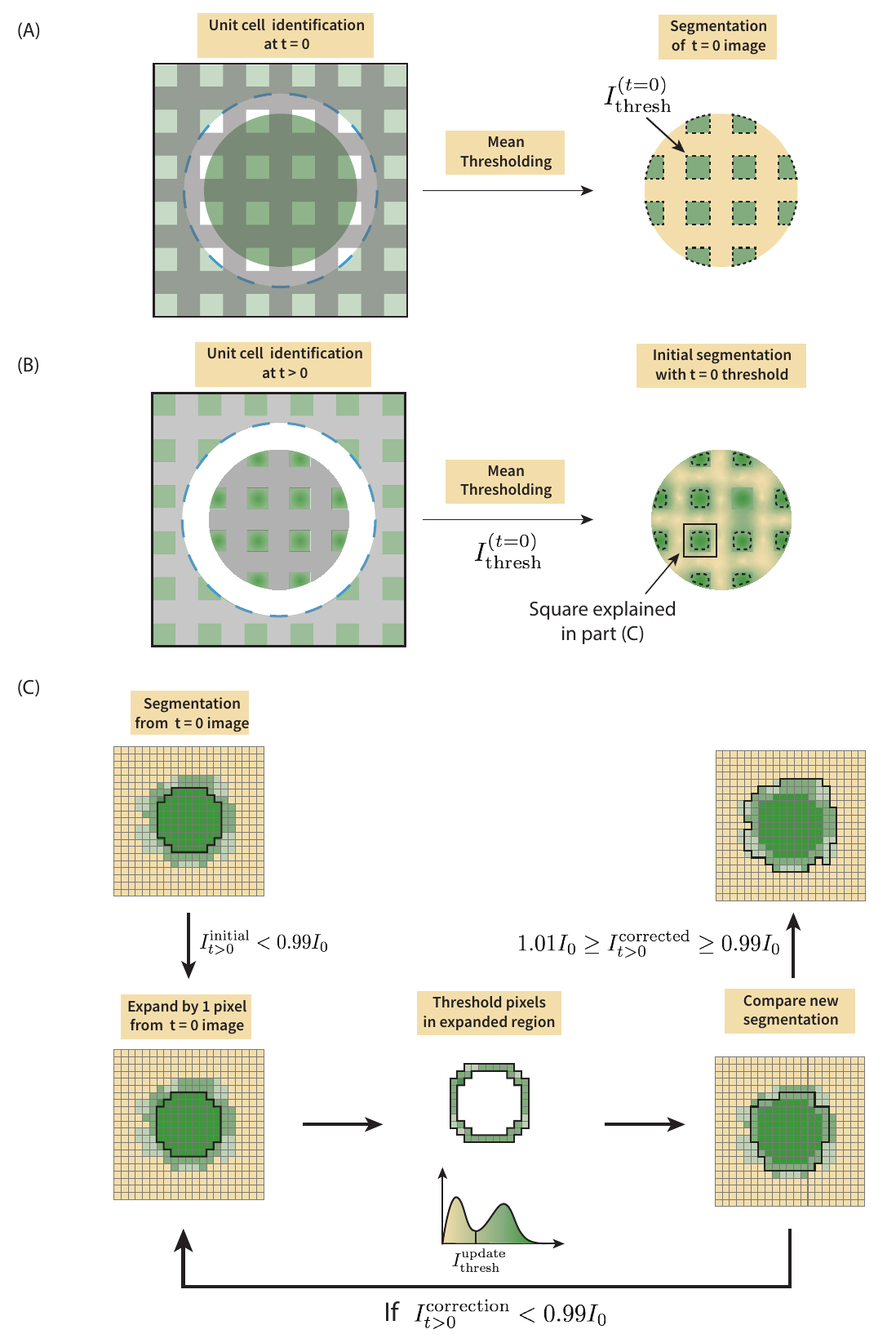}}
\caption{\textbf{Unit cell segmentation correction scheme.} (A) Unit cells in the
first image after photobleaching are segmented using mean thresholding to obtain an initial threshold value $I_\mathrm{thresh}^{(t=0)}$. Dashed blue circle denotes the extent of
the projected light within which motors dimerize, causing the network to couple and
contract (green circle). (B)
Unit cells of later frames are initially segmented using $I_\mathrm{thresh}^{(t=0)}$. (C) The integrated intensity of each unit cell after the initial segmentation
$I_{t > 0}^\mathrm{initial}$ is compared against that for the $t=0$ case, $I_0$. In instances where $I_{t > 0} < 0.99 I_0$, the pixels in a single layer beyond the segmentation
boundary are histogrammed and thresholded to distinguish pixels 
containing microtubules with those regions that make up the
background. These pixels with signal are then added, the integrated
intensity is recomputed and compared again to $I_0$. The process
is repeated until the integrated intensity falls within 1\% of $I_0$.}
\label{sifig:unitcell_correction}
\end{figure}
\\
\\
As schematized in Fig \ref{sifig:unitcell_correction}(B) and (C),
images of subsequent time points are processed with the intention of preserving the integrated
fluorescence of each unit cell, which corresponds with our argument that fluorescent
microtubules are conserved for each unit cell. We first subtract the background signal 
and segment unit cells with the same threshold value as for the $t=0$ timepoint. However, as 
unit cells begin to deform due to their fluorescent microtubules dispersing, 
the integrated fluorescence signal of the newly segmented unit cells
will differ from that of the first time point, which translates
to a different number of fluorescent
microtubules.
As a result, for images after the first frame
immediately succeeding photobleaching, we expand or reduce the segmentation of unit cells by 
adding or subtracting pixels around their boundaries until we obtain 
the same total fluorescence as the $t=0$ timepoint. Fig. \ref{sifig:unitcell_correction}(C) elaborates on the
scheme for correcting to obtain the same integrated intensity as in
the initial time frame to within a specified tolerance. To do this, each unit
cell is then paired with itself from the previous time step by determining nearest centroids. Due to the minimal reduction in
fluorescence intensity from the projector during imaging as discussed in Sec \ref{si:projector}, we compare the total fluorescence
intensity of the segmented unit cell in the frame of interest to that of the same unit
cell from the first frame. 1 layer of pixel beyond (within) the boundary of the unit cell are histogrammed and Otsu thresholded to distinguish microtubule regions
to background regions. The pixels that make up the foreground (background) according
to the thresholding are then
added (subtracted) until the integrated fluorescence falls within 0.01
tolerance of the original
fluorescence intensity. 
\\
\\
To understand how the choice of relative
tolerance in the integrated fluorescence affects that computed effective
diffusion constant $D_\mathrm{eff}$, we performed the unit cell
segmentation and tracking under different tolerance
levels. Fig \ref{sifig:threshold_diff} shows that while a tolerance below
0.015 leads to a constant effective diffusion constant, increasing the
tolerance above this point leads to a monotonic decrease in the $D_\mathrm{eff}$. 
This suggests that the area trajectories of unit cells can be highly sensitive to the tolerance given to the microtubule preservation
count. This control also indicates that the area trajectories of unit
cells are not markedly different below the 0.015 tolerance and yields
robust measurements of the area trajectories and by extension fits of the effective diffusion constants. 
\begin{figure}[b!]
\centering{
\includegraphics[scale=0.7]{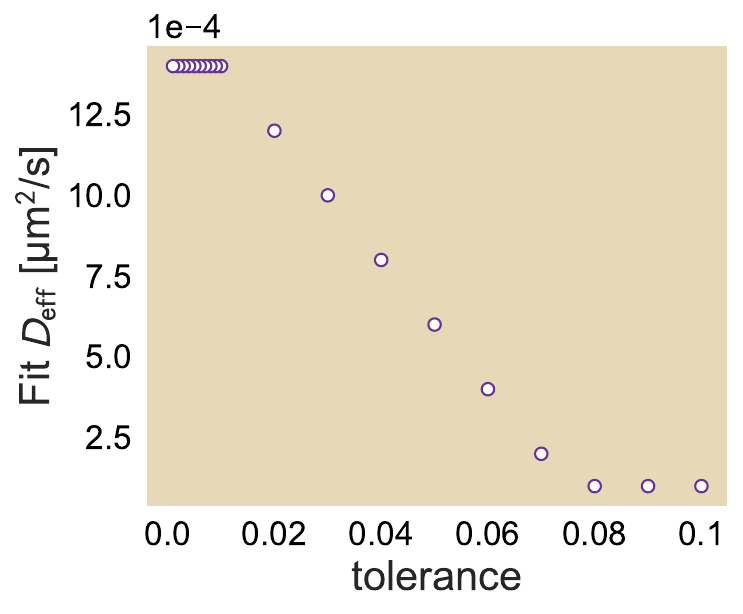}}
\caption{\textbf{Effective diffusion constant fits against various tolerances in the
relative unit cell fluorescence.} The tolerance is the fractional difference in
fluorescence intensity between the unit cell in the first frame and the unit cell
at a later time point. Dataset used on Ncd236 at saturated ATP concentration
(1.4 {\textmu}M).}
\label{sifig:threshold_diff}
\end{figure}
\\
\\
Unit cell centroids, areas, and fluorescence intensities are then computed
in addition to the pixel-weighted center of the entire contracting network after this intensity-adjusted processing for all of the unit cells.
Image processing of a unit cell terminates when it is found to overlap with another unit cell during the fluorescence intensity correction
scheme as this indicates that the unit cells have begun to merge and by the next time point thus microtubules from one unit cell can no longer be distinguished from those of the other.
\section{Data analysis} \label{sec:am_data_analysis}

\subsection{Contraction rate computation} \label{subsec:amsi_contraction}

In the main text, we use the centroids of fluorescent unit cells obtained in SI Sec. \ref{subsec:unitcell_processing} to demonstrate that
contraction speed of the microtubule network scales linearly with distance from the network center. We first obtain the speed that
each unit cell centroid is moving toward the center as a function of time. For each unit cell, we observe a linear relation between the centroid
distance from the network center and time after photobleaching of the form
\begin{equation}
  r = v_c \, t + r_0, \label{sieq:line}
\end{equation}
where $r$ is the unit cell centroid distance from the network center, $v_c$ is the speed of the unit cell (which will take to be
positive here but directed toward the origin), $t$ is the time 
since photobleaching, and $r_0$ is the initial centroid distance from the network center immediately after photobleaching. 
\\
\\
Based on the extracted contraction speed and distances for all of the unit cells for a given motor type, we noted a linear relation between radius $r$ and centroid
speed $v_c$ of the form
\begin{equation}
  v_c = \alpha \, r + v_0, \label{sieq:line2}
\end{equation}
where $\alpha$ is the contraction rate in units of inverse time and $v_0$ is the contraction speed at the network center. To this end, we
aim to compute $\alpha$ and $v_0$. Although we expect the speed 
at the network center to be $0$, we relax this assumption for our analysis.
To more carefully compute the rate of contraction of the network and determine the range of credibility of the computed rate, 
we use a Bayesian approach. Specifically, we compute the probability of $\alpha$ and $v_0$ given the
contraction speed and distance of each unit cell from the network center $\left( r_0, v_c \right)_i$, 
$P \left( \alpha, v_0 | \left\{ \left( r_0, v_c \right)_i \right\} \right)$, where $i$ denotes each unit cell. Here, we use the centroid distance
immediately after photobleaching but found that another criterion such as the median of the centroid distance over the course
of the time window analyzed does not dramatically affect the results due to the relatively small travel ($\frac{\Delta r}{r_0} <10$\% for
$\Delta r$ the distance traveled over the entire time course) the unit cells undergo.
\\
\\
We note from Bayes' Theorem that 
\begin{align*}
  P \left[ \alpha, v_0 | \left\{ \left( r_0, v_c \right)_i \right\} \right] &= \frac{P \left[ \left\{ \left( r_0, v_c \right)_i \right\} | \alpha, v_0 \right] P \left( \alpha, v_0 \right)}{P \left[ \left\{ \left( r_0, v_c \right)_i \right\} \right]},\\
  &= \frac{\prod_i P \left[ \left( r_0, v_c \right)_i | \alpha, v_0 \right] }{\prod_i P \left[ \left( r_0, v_c \right)_i \right]}P \left( \alpha, v_0 \right),\\
  &\propto \prod_i P \left[ \left( r_0, v_c \right)_i | \alpha, v_0 \right] P \left( \alpha, v_0 \right), \numberthis \label{eq:bayes}
\end{align*}
where we drop the denominator on the right-hand side as it does not involve the parameters we want to find, thus making
the two sides proportional to each other. Here, $P \left[ \left( r_0, v_c \right)_i | \alpha, v_0 \right]$ is the likelihood distribution
of getting the $\left( r_0, v_c \right)_i$ that we did given $\alpha$ and $v_0$ while $P \left( \alpha, v_0 \right)$ is the prior distribution of our 
two parameters. 
\\
\\
We expect that our priors on $\alpha$ and $v_0$ are independent of each other, so we can break up the probability function into
a product of two functions
\begin{equation}
  P \left( \alpha, v_0 \right) = p \left( \alpha \right) p \left( v_0 \right). \label{eq:priors_split}
\end{equation}
Meanwhile, we can rearrange each likelihood function as a product of two probabilities. The probability of getting $\left( r_0, v_c \right)_i$
given our parameters is also the probability of getting $v_{c,i}$ given our parameters and $r_{0,i}$ times the probability of getting
$r_{0,i}$, or
\begin{align*}
  P \left( \left( r_0, v_c \right)_i | \alpha, v_0 \right) &= P \left( v_{c,i} | \alpha, v_0, r_{0,i} \right) P \left( r_{0,i} \right),\\
  &\propto P \left( v_{c,i} | \alpha, v_0, r_{0,i} \right), \numberthis \label{eq:multiply_probs}
\end{align*}
where we change to a proportionality again as $P \left( r_{0,i} \right)$ is independent of our parameters. Here, we expect that our
contraction speed for a given unit cell $v_{c,i}$ comes from a Normal distribution where the mean value is $\alpha \, r_{0,i} + v_0$
and standard deviation $\sigma$. This means that we will also need a prior on $\sigma$. This means that our distribution really takes the
form of
\begin{equation}
  P \left( \alpha, v_0, \sigma | \left\{ \left( r_0, v_c \right)_i \right\} \right) \propto P \left( \alpha \right) P \left( v_0 \right) P \left( \sigma \right) \prod_i P \left( v_{c,i} | \alpha, v_0,\sigma, r_{0,i} \right). \label{eq:stats_eq}
\end{equation}
As a result, we say that our likelihood takes the form
\begin{equation}
  v_{c,i} \backsim \text{Normal} \left( \alpha r_{0,i} + v_0, \sigma^2 \right). \label{eq:likelihood}
\end{equation}
We then defined our priors to be that $\alpha$ is drawn from the half-normal distribution where $\alpha > 0$ as we are working
with speeds of contraction, $\sigma$ is also drawn from a half-normal distribution and enforced to be positive, and $v_0$ is 
drawn from a normal distribution about $v=0$. We make the offset a normal rather than a half-normal distribution as there
may be a value of $r > 0$ for which the contraction stops, which for a positive slope would mean a negative speed at $r = 0$. 
Put together, we have the following priors
\begin{align}
  \alpha &\backsim \text{Half-Normal} \left(0, 1 \right), \label{eq:alpha_prior}\\
  \sigma &\backsim \text{Half-Normal} \left( 0, 1 \right), \label{eq:sigma_prior}\\
  v_0 &\backsim \text{Normal} \left( 0, 1 \right). \label{eq:v0_prior}
\end{align}
We sampled the joint distribution of $\left( \alpha, v_0, \sigma \right)$ by Hamiltonian Markov chain Monte Carlo using the Stan
probabilistic program \cite{stan}. From each $\left( \alpha, v_0 \right)$ that is sampled we compute the mean value $\mu = \alpha \, r + v_0$
for $0 \leq r \leq R$ where $R$ is the distance of the farthest centroid from the network center and report the median and
95\% credible region in Fig. 2 and 4-6 of the main text.

\subsection{Computing the best fit effective diffusion constant} \label{subsec:amsi_diff_fit}

In the main manuscript, we use an advection-diffusion model to compute an effective diffusion constant to quantify the difference in area between the
experimental normalized area trajectories and the pure contraction bound (signifying no diffusion). To do so, we used the finite
element method (FEM) on individual unit cells of initially uniform concentration subject to the advection-diffusion equation as described
in Eq. 3 of the main manuscript. We then
processed the simulated concentration field data with a similar integrated particle
count method as described in SI Sec.
\ref{subsec:unitcell_processing} in order to compute the area of the unit cells in time. This analysis gives rise 
to a family of normalized area trajectories for a
fixed contraction rate and variable diffusion constant. In order to compute the effective diffusion constant from, say, the median
normalized area trajectories from a given set of experimental conditions, we take the simulated area trajectory for one of the diffusion constants
and the area trajectory of the experimentally-obtained contraction rate and compute the sum of the square of
the difference between the two trajectories across time. For each of the quartiles, the effective diffusion constant is computed
as the one whose area trajectory minimizes the sum of the differences squared.

\section{Experimental variation of contraction speed and normalized area trajectories} \label{sisec:replicates}

In the main manuscript, we computed the contraction rate using all of the replicates of a given set of experimental conditions. However,
to exhibit experimental variation between replicates, we present in Fig. \ref{fig:replicates} the contraction speed and normalized area
data for all of the replicates involving Ncd236 at 1400 {\textmu}M ATP and 1.5 mg/mL pluronic. Note that the line in the contraction speed
is the same as shown in Fig. 2(D) of the main manuscript where the contraction rate $\alpha = 0.002$ sec$^{-1}$ for comparison of how
each replicate compares to the computed line. This contraction is also used for the pure contraction bound shown on the normalized
area data. The time noted at the top of each contraction speed plot marks the time into the experiment that the photobleaching was
performed, with the plots organized in order of ascending time into the experiment of photobleaching.

\begin{figure}[t!]
  \centering{
  \includegraphics[scale=0.63]{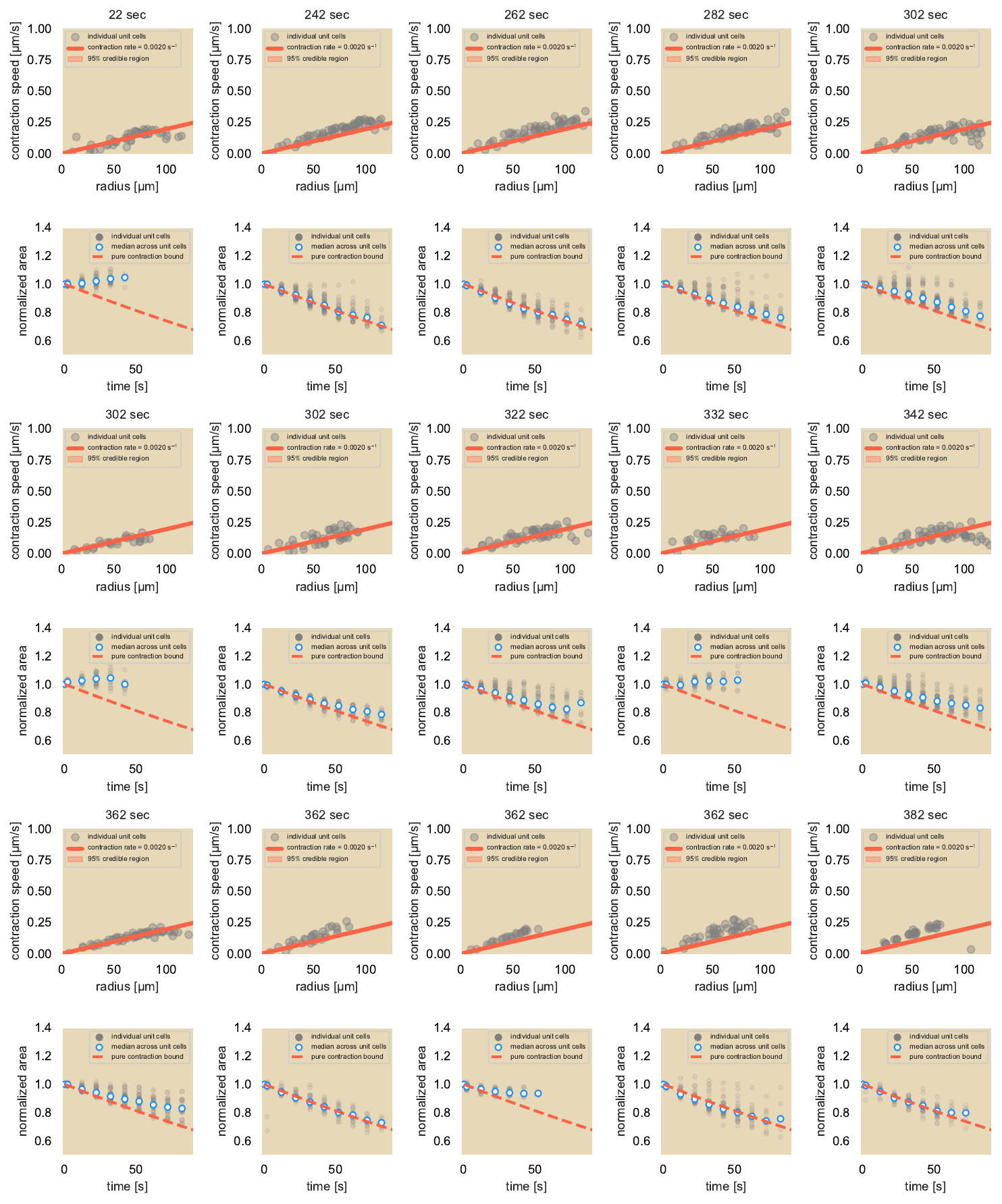}}
\end{figure}
\begin{figure}[t!]
  \centering{
  \includegraphics[scale=0.63]{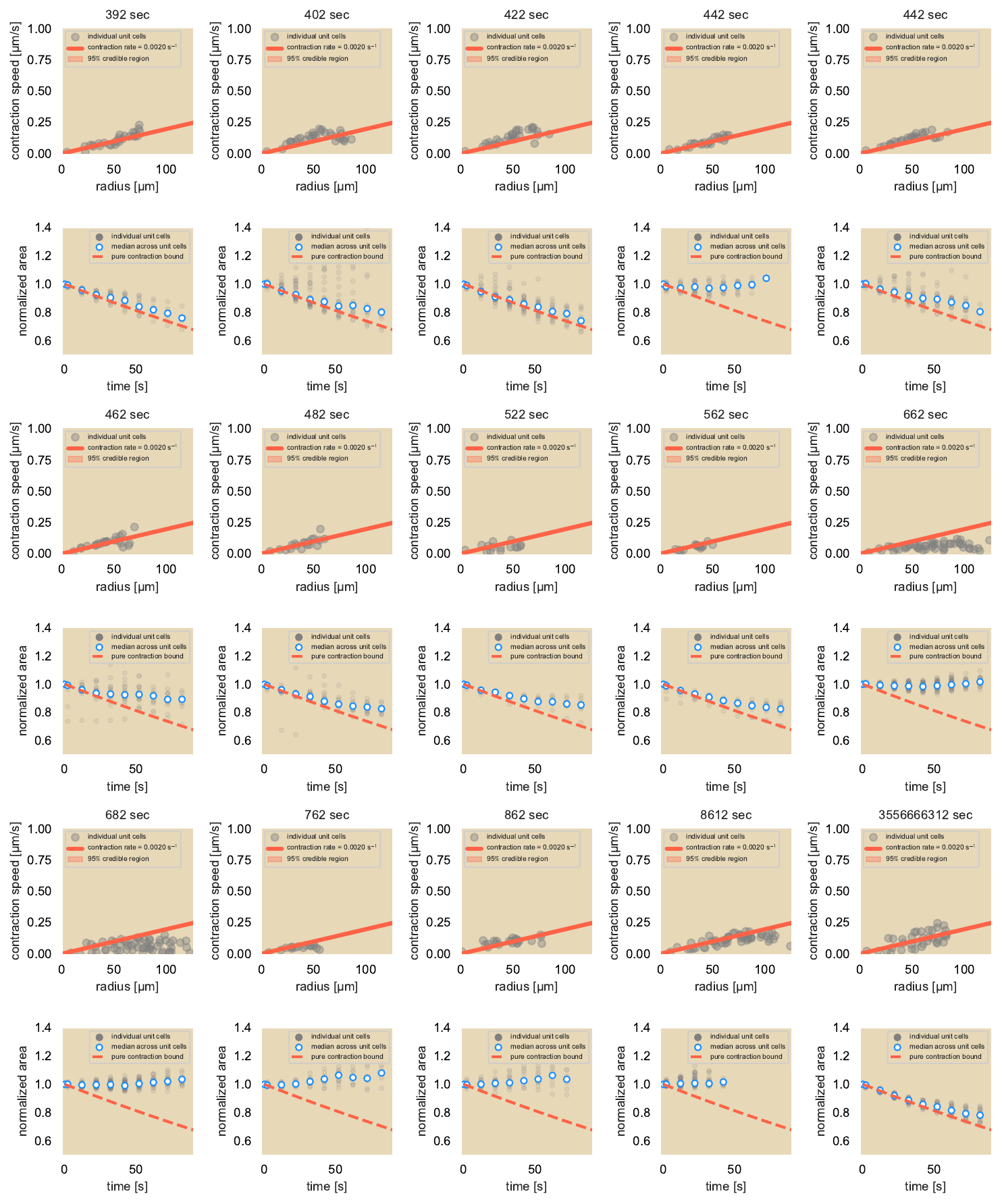}}
  \caption{\textbf{Contraction speed (odd rows) and normalized area trajectory (even rows) of each experimental replicate using 
  1.5 mg/mL pluronic, 1400 {\textmu}M ATP, and Ncd236.} The lines in the plots of contractions speed data and in the plots of the area trajectory are the same as in Fig.
  2(D) and 2(F), respectively, of the main manuscript. The time at the top of each contraction speed plot marks the time into the experiment that the
  photobleaching was performed.}
  \label{fig:replicates}
\end{figure}

\clearpage
\section{Deformation of a square due solely to contraction} \label{sec:am_square_shrink}

\begin{figure}[t!]
\centering{
\includegraphics[scale=0.4]{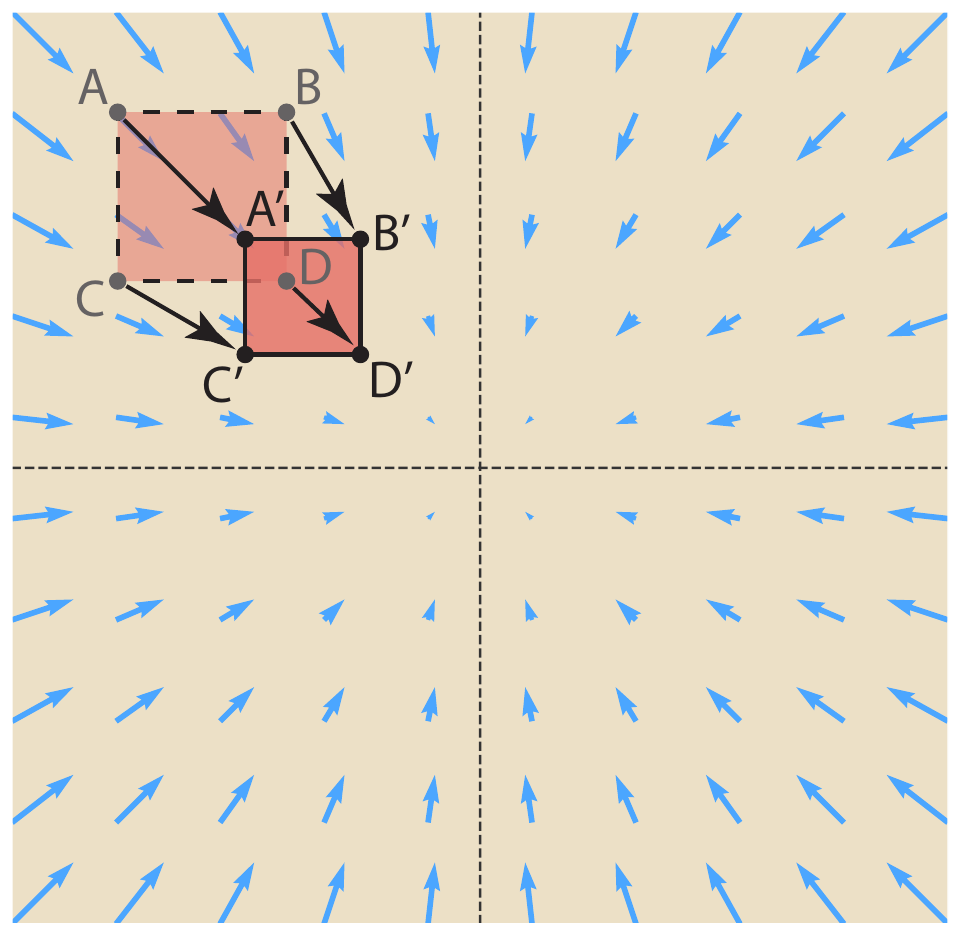}}
\caption{\textbf{Schematic of unit cell contraction due purely to the advective velocity field.} An advective velocity field scales
linearly with distance from the origin while pointing radially inward and is shown in blue. The points at the corners of the
square (A, B, C, D) are mapped after some time $\Delta t$ to the points (A$'$, B$'$, C$'$, D$'$).}
\label{sifig:pure_contraction}
\end{figure}

In the main text, we observed that each fluorescent unit cell on average conserves its area while its center of mass moves
toward the network center with speed that is linearly dependent on the distance from the center. We compute the expected
area of each unit cell had the network elastically contracted due solely to the observed global contraction. We define
the contraction velocity field ${\bf{v}}( x,y )$ as
\begin{equation}
  {\bf{v}}( x,y ) \equiv - \alpha \left( x \hat{x} + y \hat{y} \right), \numberthis \label{eq:velocity_field}
\end{equation}
where $\alpha$ is the contraction rate as computed in SI Sec. \ref{subsec:amsi_contraction} and reported in the main
manuscript. This means that after a time interval $\Delta t$ a point $\left( x,y \right)$ subject to this advective flow will 
be displaced in the x- and y- directions according to
\begin{align*}
  \Delta X &= v_x \Delta t = - \alpha x \Delta t,\\
  \Delta Y &= v_y \Delta t = - \alpha y \Delta t, \numberthis \label{eq:deformation}
\end{align*}
so the point at the later time $(x', y')$ relates to its earlier time point by
\begin{align*}
  x' &= x + \Delta X = x \left(1 - \alpha \Delta t \right) \\
  y' &= y + \Delta Y = y \left(1 - \alpha \Delta t \right). \numberthis \label{eq:point_mapping}
\end{align*}
Suppose we looked at the four corners of a unit cell, labeled as A, B, C, D as depicted in Fig. \ref{sifig:pure_contraction}. If
we assign their coordinates as 
\begin{align*}
  \mathrm{A} &\rightarrow \left( x_\mathrm{A}, y_\mathrm{A} \right),\\
  \mathrm{B} &\rightarrow \left( x_\mathrm{B}, y_\mathrm{B} \right),\\
  \mathrm{C} &\rightarrow \left( x_\mathrm{C}, y_\mathrm{C} \right),\\
  \mathrm{D} &\rightarrow \left( x_\mathrm{D}, y_\mathrm{D} \right).\numberthis \label{eq:coordinates1}
\end{align*}
Under a rectangular geometry, we can choose two vertices diagonally across from each other on the rectangle
and write their x- and y- coordinates with the coordinates of
the other two diagonal vertices, so with a choice of using coordinates from A and D, the coordinates for B and C become
\begin{align*}
  \mathrm{A} &\rightarrow \left( x_\mathrm{A}, y_\mathrm{A} \right),\\
  \mathrm{B} &\rightarrow \left( x_\mathrm{D}, y_\mathrm{A} \right),\\
  \mathrm{C} &\rightarrow \left( x_\mathrm{A}, y_\mathrm{D} \right),\\
  \mathrm{D} &\rightarrow \left( x_\mathrm{D}, y_\mathrm{D} \right).\numberthis \label{eq:coordinates2}
\end{align*}

\noindent Under the deformation mapping, their new coordinates, labeled as A$'$, B$'$, C$'$, and D$'$ get mapped on as
\begin{align*}
  \mathrm{A}' &\rightarrow \left[ x_\mathrm{A} \left(1 - \alpha \Delta t \right), y_\mathrm{A} \left(1 - \alpha \Delta t \right) \right],\\
  \mathrm{B}' &\rightarrow \left[ x_\mathrm{D} \left(1 - \alpha \Delta t \right), y_\mathrm{A} \left(1 - \alpha \Delta t \right) \right],\\
  \mathrm{C}' &\rightarrow \left[ x_\mathrm{A} \left(1 - \alpha \Delta t \right), y_\mathrm{D} \left(1 - \alpha \Delta t \right) \right],\\
  \mathrm{D}' &\rightarrow \left[ x_\mathrm{D} \left(1 - \alpha \Delta t \right), y_\mathrm{D} \left(1 - \alpha \Delta t \right) \right].\numberthis \label{eq:coordinates_mapped}
\end{align*}
Eqs. \ref{eq:coordinates_mapped} tells us that under this particular velocity field, any two points that are horizontally or vertically
aligned will maintain the same horizontal or vertical alignment, respectively, even at later times. Thus, a square will preserve its
shape in time.
\\
\\
We next examine what happens to the area of a unit cell had the only effect been the global contraction. In this case, we can
compare the area of the square before and after the deformation. To compute the area swept out by (A,B,C,D), we multiply the
line segment between B and D, $L_\mathrm{BD}$ with the line segment between C and D, $L_\mathrm{CD}$:
\begin{align*}
  \sigma_{(\mathrm{A},\mathrm{B},\mathrm{C},\mathrm{D})} &= L_\mathrm{BD} \times L_\mathrm{CD},\\
  &= \left[ \sqrt{ \left( x_\mathrm{B} - x_\mathrm{D} \right)^2 + \left( y_\mathrm{B} - y_\mathrm{D} \right)^2} \right] \times \left[ \sqrt{ \left( x_\mathrm{D} - x_\mathrm{C} \right)^2 + \left( y_\mathrm{D} - y_\mathrm{C} \right)^2} \right],\\
  &= \left( y_\mathrm{A} - y_\mathrm{D} \right) \times \left( x_\mathrm{D} - x_\mathrm{A} \right).\numberthis \label{eq:area_undeformed}
\end{align*}
As noted from Eq. \ref{eq:coordinates2}, we used the fact that $x_\mathrm{B} = x_\mathrm{D}$, $y_\mathrm{B} = y_\mathrm{A}$, $x_\mathrm{C} = x_\mathrm{A}$, and $y_\mathrm{C} = y_\mathrm{D}$ to simplify the equation down in terms of two coordinates.
In comparison, the area of the deformed
unit cell swept out by (A$'$, B$'$, C$'$, D$'$) takes the form
\begin{align*}
  \sigma_{(\mathrm{A}',\mathrm{B}',\mathrm{C}',\mathrm{D}')} &= L_{\mathrm{B}'\mathrm{D}} \times L_{\mathrm{C}'\mathrm{D}'},\\
  &= \left[ \sqrt{ \left( x_{\mathrm{B}'} - x_{\mathrm{D}'} \right)^2 + \left( y_{\mathrm{B}'} - y_{\mathrm{D}'} \right)^2} \right] \times \left[ \sqrt{ \left( x_{\mathrm{D}'} - x_{\mathrm{C}'} \right)^2 + \left( y_{\mathrm{D}'} - y_{\mathrm{C}'} \right)^2} \right],\\
  &= \left( y_{\mathrm{A}'} - y_{\mathrm{D}'} \right) \times \left( x_{\mathrm{D}'} - x_{\mathrm{A}'} \right),\\
  &= \left[ y_{\mathrm{A}} \left(1 - \alpha \Delta t \right) - y_{\mathrm{D}} \left(1 - \alpha \Delta t \right) \right] \times \left[ x_{\mathrm{D}} \left(1 - \alpha \Delta t \right) - x_{\mathrm{A}} \left(1 - \alpha \Delta t \right) \right],\\
  &= \left( y_{\mathrm{A}} - y_{\mathrm{D}} \right) \left(1 - \alpha \Delta t \right) \times \left( x_{\mathrm{D}} - x_{\mathrm{A}} \right) \left(1 - \alpha \Delta t \right),\\
  &= \left( y_\mathrm{A} - y_\mathrm{D} \right) \times \left( x_\mathrm{D} - x_\mathrm{A} \right) \left( 1 - \alpha \Delta t \right)^2,\\
  &= \sigma_{(\mathrm{A},\mathrm{B},\mathrm{C},\mathrm{D})} \left( 1 - \alpha \Delta t \right)^2. \numberthis \label{eq:area_deformed}
\end{align*}
Thus we find that the area of the unit cell subject solely to the contraction would decrease by $\left( 1 - \alpha \Delta t \right)^2$
after a time period $\Delta t$. This comes in contrast to the results that we found experimentally where the area of the fluorescent unit
squares remains greater than the pure contraction bound during the contraction process
suggesting a mechanism that disperses microtubules against the global
contraction.
\section{2D Linear Advection-Diffusion Model} \label{sisec:am_2dadvdiff}

In the work presented in the main manuscript, we argue for an
advection-diffusion model to describe the redistribution of 
microtubules in the bulk of the contracting network. In this
section, we explore Eq 3 as shown in the manuscript to examine
whether the model could reasonably recapitulate the experimental
observations as a tool for computing an effective diffusion constant
for the diffusive-like spread of microtubules in the bulk. While
the main manuscript uses the finite element method (FEM) to simulate
the area change of a concentration of particles localized to a square, 
as is the case for microtubules of a unit cell, we first develop
an intuition for this equation for a series of initial conditions and
at steady state. To validate the FEM approach before applying it to
the unit square case, we numerically and analytically solve these
initial conditions and directly compare them.
This theoretical analysis is meant to explore the filament concentration when subject to a
linear contraction velocity profile. We note here that we later
invoke the Sturm-Liouville Theorem, which we elaborate on in 
SI Sec \ref{sec:sturm_liouville}.
\\
\\
In the 2D telescoping case, we assume that we are carrying out an aster assay experiment where we dimerize motors
(and thus couple microtubules) in a circular region of radius $R$. We assume that the distributions of motors and microtubules are
strictly radially dependent and thus have no angular dependence. Finally, we model the velocity profile of the microtubule
movement by assuming radially inward advection of particles where those that lie further away from the origin move
faster than those toward the center as given by
\begin{align*}
  \bf{v} &= - \alpha r \hat{r}. \numberthis \label{eq:advection_2d_profile}
\end{align*}
\noindent
The advection-diffusion equation then takes the form
\begin{align*}
  \frac{\partial c}{\partial t} &= D \nabla^2 c - \nabla \cdot ( {\bf{v}} c ),\\
  &= \frac{D}{r} \frac{\partial}{\partial r} \Big( r \frac{\partial c}{\partial r} \Big) + \alpha \frac{1}{r} \frac{\partial}{\partial r} ( r^2 c ),\\
  &= D \frac{\partial^2 c}{\partial r^2} + \frac{D}{r} \frac{\partial c}{\partial r} + \alpha r \frac{\partial c}{\partial r} + 2 \alpha c,\\
  &= D \frac{\partial^2 c}{\partial r^2} + \Big( \frac{D}{r} + \alpha r \Big) \frac{\partial c}{\partial r} + 2 \alpha c,\\
  \frac{1}{D} \frac{\partial c}{\partial t} &= \frac{\partial^2 c}{\partial r^2} + \Big( \frac{1}{r} + \frac{\alpha r}{D} \Big) \frac{\partial c}{\partial r} + \frac{2 \alpha c}{D}. \numberthis \label{eq:2d_advection_diffusion}
\end{align*}
\noindent
We first follow the procedure of separation of variables $c(r,t) = \Phi(r) T(t)$ and determine that the time-dependent 
component takes on the familiar form of $e^{- D k^2 t}$. This ansatz is then applied to Eq. \ref{eq:2d_advection_diffusion}
and we rewrite the spatial component of the concentration as
\begin{align*}
  -k^2 \Phi &= \frac{\text{d}^2 \Phi}{\text{d} r^2} + \Big( \frac{1}{r} + \frac{\alpha r}{D} \Big) \frac{\text{d} \Phi}{\text{d} r} + \frac{2 \alpha \Phi}{D},\\
  0 &= r \frac{\text{d}^2 \Phi}{\text{d} r^2} + \Big( 1 + \frac{\alpha r^2}{D} \Big) \frac{\text{d} \Phi}{\text{d} r} + \Big( \frac{2 \alpha}{D} + k^2 \Big) r \Phi. \numberthis \label{eq:advdiff_2d_r}
\end{align*}
We will define a new length scale $\lambda^2 \equiv \frac{D}{\alpha}$ as well as a change of variables $\rho \equiv 
\frac{r}{\lambda}$ and $\tilde{k} \equiv \lambda k$. In this case, Eq. \ref{eq:advdiff_2d_r} takes
the altered form
\begin{align*}
  0 &= \rho \frac{\text{d}^2 \Phi}{\text{d} \rho^2} + (1 + \rho^2) \frac{\text{d} \Phi}{\text{d} \rho} + \Big( 2 + \tilde{k}^2 \Big) \rho \Phi. \numberthis \label{eq:advdiff_2d_rho}
\end{align*}
\noindent
By following a prescription on which we elaborate further in SI Sec \ref{sec:sturm_liouville}, we obtain a
weighting function that will help us compute the eigenfunctions using Eq. \ref{eq:weighting_fn}, namely, 
\begin{align*}
  m(\rho) &= e^{\frac{\rho^2}{2}}. \numberthis \label{eq:multiplicative_soln_2d}
\end{align*}
When we multiply Eq. \ref{eq:advdiff_2d_rho} by the multiplicative function, we get
\begin{align*}
  0 &= \rho \, e^{\frac{\rho^2}{2}} \frac{\text{d}^2 \Phi}{\text{d} \rho^2} + (1 + \rho^2) e^{\frac{\rho^2}{2}} \frac{\text{d} \Phi}{\text{d} \rho} + \Big( 2 + \tilde{k}^2 \Big) \rho \, e^{\frac{\rho^2}{2}} \Phi,\\
  \frac{\text{d}}{\text{d} \rho} \Big[ \rho \, e^\frac{\rho^2}{2} \frac{\text{d}\Phi}{\text{d}\rho} \Big] + 2 \rho \, e^\frac{\rho^2}{2} \Phi &= - \tilde{k}^2 \rho \, e^\frac{\rho^2}{2} \Phi. \numberthis \label{eq:sturm_liouville_2d}
\end{align*}
Through the Sturm-Liouville Theorem as described in SI Sec \ref{sec:sturm_liouville}, specifically
Eq. \ref{eq:sturm_liouville_2d}, we find that the weighting function
differs from the multiplicative function due to the inclusion of the prefactor $\rho$. In this case, the weighting function $w(\rho)$
as well as $p(\rho)$ and $q(\rho)$ are given as
\begin{align*}
  w(\rho) = p(\rho) = q(\rho) = \rho \, e^\frac{\rho^2}{2}. \numberthis \label{eq:weighting_soln_2d}
\end{align*}
\noindent
Furthermore, we observe that the eigenvalues take the form $\tilde{k}^2$. Solutions of $\Phi$ from Eq.
\ref{eq:sturm_liouville_2d} are obtained from Wolfram Alpha and take the form
\begin{align*}
  \Phi_\text{ss} (\rho) &= c_\text{ss} \, e^{- \frac{\rho^2}{2}},\\
  \Phi_\text{dyn} (\rho) &= c_1 \, e^{-\frac{\rho^2}{2}}\, {_1 F_1} \Big( - \frac{\tilde{k^2}}{2} ; 1 ; \frac{\rho^2}{2} \Big) + c_2 \, G_{1,2}^{2,0} \Bigg( \frac{\rho^2}{2} \Bigg| {- \frac{\tilde{k}^2}{2} \atop 0,0} \Bigg), \numberthis \label{eq:2d_general_soln}
\end{align*}
\noindent
where $G_{p,q}^{m,n} \Big( z \Big| {a_1, ..., a_p \atop b1,...,b_q} \Big)$ is the Meijer G-function (we split up the eigenfunctions as
dynamic and steady-state terms for now). We note here that the
arguments of the Meijer G-function are such that the function diverges
at the origin. As our system is defined as $0 \leq r \leq R$, we can say that $c_2=0$. Thus, our
eigenfunctions are 
\begin{align*}
  \Phi_\text{ss} (\rho) &= c_\text{ss} \, e^{- \frac{\rho^2}{2}},\\
  \Phi_\text{dyn} (\rho) &= c_1 \, e^{-\frac{\rho^2}{2}} \, {_1 F_1} \Big( - \frac{\tilde{k^2}}{2} ; 1 ; \frac{\rho^2}{2} \Big), \numberthis \label{eq:2d_allowed_soln}
\end{align*}
\noindent
where we note that in the case of $\tilde{k}=0$, we go from the dynamic eigenfunction to the static eigenfunction.

\subsection{No-flux boundary condition} \label{subsec:bc}
In the work presented here, there is no inflow or outflow of material at the boundary. Thus, we impose the boundary
condition ${\bf{J}}\Big|_{r=R}=0$. This means that
\begin{align*}
  J_r \Big|_{r=R} &= D \frac{\text{d} \Phi}{\text{d} r} - v(R) \Phi(R) = D \frac{\text{d} \Phi}{\text{d} r} \Big|_{r=R} + \alpha R \Phi(R) = 0. \numberthis \label{eq:2d_boundary_cond}
\end{align*}
\noindent
We then need to ensure that the boundary condition is satisfied for the dynamic eigenfunction. We start by taking the
derivative of the eigenfunction:
\begin{align*}
  \frac{\text{d} \Phi}{\text{d} r} &= - \frac{c_1 \, \rho}{\lambda} \, e^{- \frac{\rho^2}{2}} \Big[ \frac{\tilde{k}^2}{2} {_1F_1} \Big( 1 - \frac{\tilde{k}^2}{2} ; 2 ; \frac{\rho^2}{2} \Big) + {_1F_1} \Big( - \frac{\tilde{k}^2}{2} ; 1 ; \frac{\rho^2}{2} \Big) \Big],\\
  \frac{\text{d} \Phi}{\text{d} r} \Big|_{r=R} &= - \frac{c_1\, \alpha R}{D} \, e^{- \frac{\alpha R^2}{2D}} \Big[ \Big( \frac{D k^2}{2 \alpha} \Big) {_1F_1} \Big( 1 - \frac{D k^2}{2 \alpha} ; 2 ; \frac{\alpha R^2}{2 D} \Big) +  {_1F_1} \Big( - \frac{D k^2}{2 \alpha} ; 1 ; \frac{\alpha R^2}{2 D} \Big) \Big]. \numberthis \label{eq:2d_eigenfn_deriv}
\end{align*}
\noindent
so when applied to the boundary condition, we get
\begin{align*}
  D \frac{\text{d} \Phi}{\text{d} r} \Big|_{r=R} + \alpha R \Phi(R) = - c_1\, \alpha R &\, e^{- \frac{\alpha R^2}{2D}} \Big( \frac{D k^2}{2 \alpha} \Big) {_1F_1} \Big( 1 - \frac{D k^2}{2 \alpha} ; 2 ; \frac{\alpha R^2}{2 D} \Big) \\
  & - c_1\, \alpha R \, e^{- \frac{\alpha R^2}{2D}}  {_1F_1} \Big( - \frac{D k^2}{2 \alpha} ; 1 ; \frac{\alpha R^2}{2 D} \Big) \\
  &+ c_1 \, \alpha R \, e^{- \frac{\alpha R^2}{2D}} \, {_1 F_1} \Big( - \frac{D k^2}{2 \alpha ; 1 ; \frac{\alpha R^2}{2 D}} \Big). \numberthis \label{eq:solving_BC_2d}
\end{align*}
\begin{figure}[b!]
  \centering{
  \includegraphics[trim={0cm 0cm 0cm 0cm}, scale=1.0]{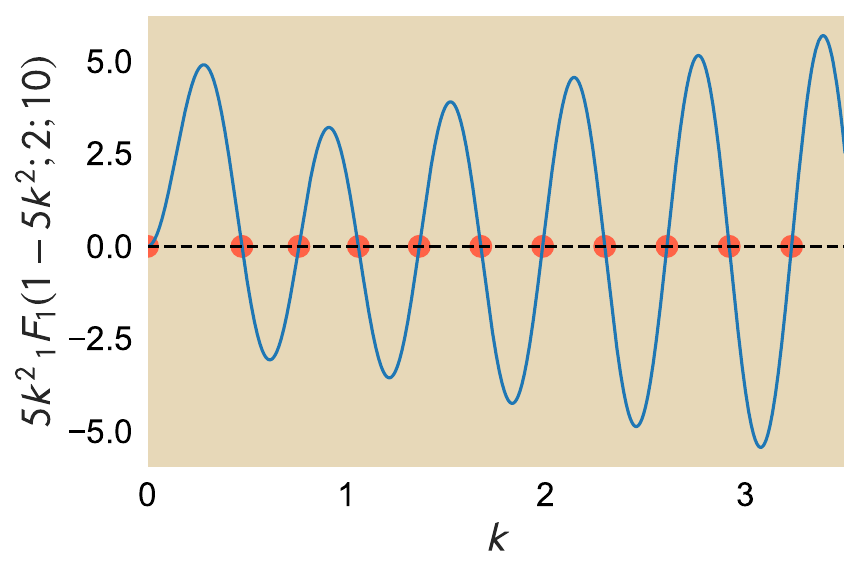}}
  \caption{\textbf{Zeros of $k$ for $\frac{\lambda^2 k^2}{2} \, _1F_1\Big(1 - \frac{\lambda^2 k^2}{2} ; 2 ; \frac{R^2}{2 \lambda^2} \Big) = 0$
  where $\frac{R}{\lambda} = 3.16$.} Red dots are overlayed with the points where the Kummer confluent hypergeometric function
  crosses the $x$-axis.}
  \label{fig:hypergeo_2d_zeros}
\end{figure}
We are then left with the simplified equation,
\begin{align*}
  \Big( \frac{D k^2}{2 \alpha} \Big) {_1F_1} \Big( 1 - \frac{D k^2}{2 \alpha} ; 2 ; \frac{\alpha R^2}{2 D} \Big) &= 0. \numberthis \label{eq:2d_BC}
\end{align*}
\noindent
Here, $k=0$ is satisfied, which yields the steady-state solution. Fig. \ref{fig:hypergeo_2d_zeros} shows
the zeros when we set $\frac{R}{\lambda} = 3.16$. The first few non-zero eigenvalues are then $\tilde{k} = 0.474$, $0.759$,
$1.058$, $1.354$, and $1.672$. Here, we observe a similar oscillator pattern to the zeros of the system.
Once again, we see that there are multiple values of $k$ that satisfy the boundary conditions. This means that the solution
to the advection-diffusion problem once both boundary and initial conditions are satisfied, is a superposition of the different
eigenfunctions:
\begin{align*}
  c(r,t) &= c_\text{ss} \, e^{- \frac{\alpha r^2}{2 D}} + e^{- \frac{\alpha r^2}{2D}} \sum_{i=1}^\infty c_i e^{- D k_i^2 t} \, {_1F_1} \Big( - \frac{ D k_i^2}{2 \alpha} ; 1 ; \frac{\alpha r^2}{2 D} \Big). \numberthis \label{eq:AD_2d_general}
\end{align*}
\noindent 
For simplicity, we will reintroduce the length scale $\lambda \equiv \sqrt{\frac{D}{\alpha}}$ so that the equation is
simplified as
\begin{align*}
  c(r,t) &= c_\text{ss} \, e^{- \frac{r^2}{2 \lambda^2}} + e^{- \frac{r^2}{2\lambda^2}} \sum_{i=1}^\infty c_i e^{- D k_i^2 t} \, {_1F_1} \Big( - \frac{\lambda^2 k_i^2}{2} ; 1 ; \frac{r^2}{2 \lambda^2} \Big). \numberthis \label{eq:AD_2d_general_simple}
\end{align*}
We turn next to the Sturm-Liouville Theory in SI Sec \ref{sec:sturm_liouville} before
applying these equations to three simple initial conditions as validation of the theory
and the finite element methods (FEM) approach.
\section{Sturm-Liouville Theory} \label{sec:sturm_liouville}

The Sturm-Liouville theory says that all well-behaved 
second-order linear ordinary differential equations that can be written in the form
\begin{align*}
 \frac{\text{d}}{\text{d}x} \Big[ p(x) \frac{\text{d}y}{\text{d}x} \Big] + q(x) \, y(x) &= - \lambda \, w(x) \, y(x), \numberthis \label{eq:sturm_liouville}
\end{align*}
\noindent
have real eigenvalues with an orthonormal basis of eigenfunctions.
Curiously, these equations also have a prescription for determining
these eigenfunctions.
Importantly, $w(x)$ is the weighting function, which provides the means for satisfying the orthogonality relations
for finding coefficients of each term in the series solution to the partial differential equation. Specifically, if we were to
write the ODE in the form
\begin{align*}
  P(x) \, y''(x) + Q(x) \, y'(x) + R(x) \, y(x) = f(x), \numberthis \label{eq:sturm_liouville_alt}
\end{align*}
\noindent
for functions $P(x)$, $Q(x)$, $R(x)$, and $f(x)$, then there is a multiplicative function that can be determined by
\begin{align*}
  m(x) = \text{exp} \Big( \int \frac{Q(x) - P'(x)}{P(x)} \text{d}x \Big). \numberthis \label{eq:weighting_fn}
\end{align*}
\noindent
This multiplicative function is then multiplied to Eq. \ref{eq:sturm_liouville_alt} and recast into the form shown in 
Eq.\ref{eq:sturm_liouville}. Thus, with $P({\tilde x})=1$ and $Q({\tilde x})={\tilde x}$,
\begin{align*}
  m({\tilde x}) &= \text{exp} \Big( \int {\tilde x} \, \text{d}{\tilde x} \Big), \\
  &= \text{exp} \Big( \frac{{\tilde x}^2}{2} \Big), \numberthis \label{eq:weighting_soln}
\end{align*}
and the ODE takes the form
\begin{align*}
  0 &= \frac{\text{d}}{\text{d} {\tilde x}} \Big[ e^{\frac{{\tilde x}^2}{2}} \frac{\text{d} \Phi}{\text{d} {\tilde x}} \Big] + \Phi \Big( 1 + {\tilde k}^2 \Big) e^{\frac{{\tilde x}^2}{2}}, \numberthis \label{eq:sturm_liouville_specific}
\end{align*}
\noindent
or in the form of Eq. \ref{eq:sturm_liouville},
\begin{align*}
  \frac{\text{d}}{\text{d} {\tilde x}} \Big[ e^{\frac{{\tilde x}^2}{2}} \frac{\text{d} \Phi}{\text{d} {\tilde x}} \Big] + e^{\frac{{\tilde x}^2}{2}} \Phi &= - {\tilde k}^2 e^{\frac{{\tilde x}^2}{2}} \Phi, \numberthis \label{eq:SL_form}
\end{align*}
\noindent
so that $p(x) = q(x) = w(x) = e^{\frac{{\tilde x}^2}{2}}$ and $\lambda = {\tilde k}^2$. 
\\
\\
Next, we show the orthogonality conditions of the eigenfunctions. Suppose that solving Eq. \ref{eq:sturm_liouville}
creates a series of eigenfunctions $\{ y_j (x) \}$. Suppose that a given eigenfunction $y_i(x)$ has the eigenvalue $\lambda_i$
so that
\begin{align*}
 \frac{\text{d}}{\text{d}x} \Big[ p(x) \frac{\text{d}y_i}{\text{d}x} \Big] + q(x) \, y_i(x) &= - \lambda_i \, w(x) \, y_i(x). \numberthis \label{eq:sturm_liouville_eigen}
\end{align*}
Suppose that each eigenfunction of the system, bounded by $a \leq x \leq b$, obeys the boundary conditions
\begin{align*}
  \alpha_1 y_i(a) + \alpha_2 y_i'(a) &= 0,\\
  \beta_1 y_i(b) + \beta_2 y_i'(b) &= 0. \numberthis \label{eq:SL_BC}
\end{align*}
To test the orthogonality conditions, we multiply both sides by $y_j(x)$, a particular eigenfunction of the differential
equation, and integrate over the entire system,
\begin{align*}
  \int_a^b \frac{\text{d}}{\text{d}x} \Big[ p(x) \frac{\text{d}y_i}{\text{d}x} \Big] y_j(x) + q(x) \, y_i(x) \, y_j(x) \text{d}x &= - \lambda_i \int_a^b w(x) \, y_i(x) \, y_j(x) \text{d}x, \\
  p(x) \frac{\text{d}y_i}{\text{d}x} y_j(x) \Big|_a^b - \int_a^b p(x) \frac{\text{d}y_i}{\text{d}x} \frac{\text{d}y_j}{\text{d}x} \text{d}x + \int_a^b q(x) \, y_i(x) \, y_j(x) \text{d}x &= - \lambda_i \int_a^b w(x) \, y_i(x) \, y_j(x) \text{d}x. \numberthis \label{eq:solving_ortho}
\end{align*}
\noindent
Had Eq. \ref{eq:sturm_liouville_eigen} involved $y_j(x)$ and we multiplied both sides of the equation by $y_i(x)$, then 
Eq. \ref{eq:solving_ortho} would have the subscripts reversed:
\begin{align*}
  p(x) \frac{\text{d}y_j}{\text{d}x} y_i(x) \Big|_a^b - \int_a^b p(x) \frac{\text{d}y_i}{\text{d}x} \frac{\text{d}y_j}{\text{d}x} \text{d}x + \int_a^b q(x) \, y_i(x) \, y_j(x) \text{d}x &= - \lambda_j \int_a^b w(x) \, y_i(x) \, y_j(x) \text{d}x. \numberthis \label{eq:solving_ortho_switch}
\end{align*}
\noindent
Suppose we subtracted Eq. \ref{eq:solving_ortho_switch} from Eq. \ref{eq:solving_ortho} and applied our boundary
conditions:
\begin{align*}
  - (\lambda_i - \lambda_j) \int_a^b w(x) \, y_i(x) \, y_j(x) \text{d}x &= p(x) \frac{\text{d}y_i}{\text{d}x} y_j(x) \Big|_a^b - p(x) \frac{\text{d}y_i}{\text{d}x} y_j(x) \Big|_a^b,\\
  - (\lambda_i - \lambda_j) \int_a^b w(x) \, y_i(x) \, y_j(x) \text{d}x &= p(b) \Big[ \frac{\text{d}y_i}{\text{d}x} \Big|_b y_j(b) -  \frac{\text{d}y_j}{\text{d}x} \Big|_b y_i(b) \Big] - p(a) \Big[ \frac{\text{d}y_i}{\text{d}x} \Big|_a y_j(a) - \frac{\text{d}y_j}{\text{d}x} \Big|_a y_i(a) \Big] ,\\ 
  - (\lambda_i - \lambda_j) \int_a^b w(x) \, y_i(x) \, y_j(x) \text{d}x &= p(b) \Big[ \frac{\beta_1}{\beta_2} y_i(b) \, y_j(b) - \frac{\beta_1}{\beta_2} y_i(b) \, y_j(b) \Big] - p(a) \Big[ \frac{\alpha_1}{\alpha_2} y_i(a) \, y_j(a) - \frac{\alpha_1}{\alpha_2} y_i(a) \, y_j(a) \Big], \\
  - (\lambda_i - \lambda_j) \int_a^b w(x) \, y_i(x) \, y_j(x) \text{d}x &= 0.  \numberthis \label{eq:subtract_ortho}
\end{align*}
\noindent
If $i=j$, then the left-hand side is already zero. 
\begin{align*}
  - \lambda_i \int_a^b w(x) \Big[ y_i(x) \Big]^2 \text{d}x &= p(x) \frac{\text{d}y_i}{\text{d}x} y_i(x) \Big|_a^b - \int_a^b p(x) \Big[ \frac{\text{d}y_i}{\text{d}x} \Big]^2 \text{d}x + \int_a^b q(x) \Big[ y_i(x) \Big]^2 \text{d}x. \numberthis \label{eq:ortho_cond}
\end{align*}
We will return to the case where $i=j$ to find the coefficients of eigenfunction. If $i \neq j$, then the eigenvalues
are different here and the integral is zero:
\begin{align*}
  \int_a^b w(x) \, y_i(x) \, y_j(x) \text{d}x &= 0, \text{for } i \neq j. \numberthis \label{eq:ortho_rule}
\end{align*}
\noindent
Eq. \ref{eq:ortho_cond} serves as a convenient equation for analytically solving
the coefficients for each eigenfunction.
\section{The recovery of a typical FRAP-like disc is time-sensitive in the advection-diffusion model.} \label{sec:analytical}

As we derive in the SI Sec. \ref{sisec:am_2dadvdiff}, the general solution to the PDE
\begin{align*}
  \frac{\partial c}{\partial t} &= D \nabla^2 c + \nabla \cdot \Big[\alpha {\bf{r}} c \Big], \numberthis \label{sieq:core_pde}
\end{align*}
assuming no angular dependence, takes the form
\begin{align*}
  c(r,t) &= c_\text{ss} \, e^{- \frac{r^2}{2 \lambda^2}} + e^{- \frac{r^2}{2 \lambda^2}} \sum_{i=1}^\infty c_i e^{- D k_i^2 t} \, {_1F_1} \Big( - \frac{ \lambda^2 k_i^2}{2} ; 1 ; \frac{r^2}{2 \lambda^2} \Big), \numberthis \label{eq:AD_2d_general}
\end{align*}
\noindent 
where $c_\mathrm{ss}$ is the coefficient for the steady-state concentration term, 
$\lambda \equiv \sqrt{\frac{D}{\alpha}}$, $k_i$ are the eigenvalues specific
to the boundary condition, $c_i$ are the coefficients based on initial conditions, and $_1F_1(a;b;z)$ is the Kummer 
confluent hypergeometric function
\begin{align*}
  _1F_1(a;b;z) = \sum_{l=0}^{\infty} \frac{(a)_l}{(b)_l}\frac{z^l}{l!}, \numberthis \label{eq:1f1}
\end{align*}
\noindent 
where the Pochhammer symbol $(a)_l = \frac{(a + l - 1)!}{(a - 1)!}$. The most well-known example of Eq. \ref{eq:1f1} is
the case where $a=b$, which yields ${_1F_1}(a;a;z)=e^z$.
The eigenvalues $\{k_i\}$ are found by satisfying the boundary conditions and are those terms that satisfy the equation
\begin{align*}
  \Big( \frac{\lambda^2 k_i^2}{2} \Big) {_1F_1} \Big( 1 - \frac{\lambda^2 k_i^2}{2} ; 2 ; \frac{R^2}{2 \lambda^2} \Big) &= 0. \numberthis \label{eq:2d_BC}
\end{align*}
\noindent
Eq. \ref{eq:AD_2d_general} shows that the steady-state profile of the concentration is a Gaussian distribution with standard 
deviation $\lambda$.
\\
\\
We now seek to identify the coefficients of the terms, which are specific to the initial conditions. Here, we will analytically 
examine three
cases for initial conditions: 1) uniform concentration, 2) a uniform concentration except with molecules removed in the region 
$r \leq R_0$ as found in many FRAP assays, and 3) a FRAP-like removal of molecules in the region $r \leq R_0$ after the
system initially reaches a steady-state Gaussian concentration profile.
As our goal is to validate our FEM simulations through agreement with some initial conditions that can be analytically determined, we directly compare analytical and FEM solutions.

\begin{figure}[b!]
\centering{
\includegraphics[scale=0.35, trim={0cm 0cm 0cm 0cm}]{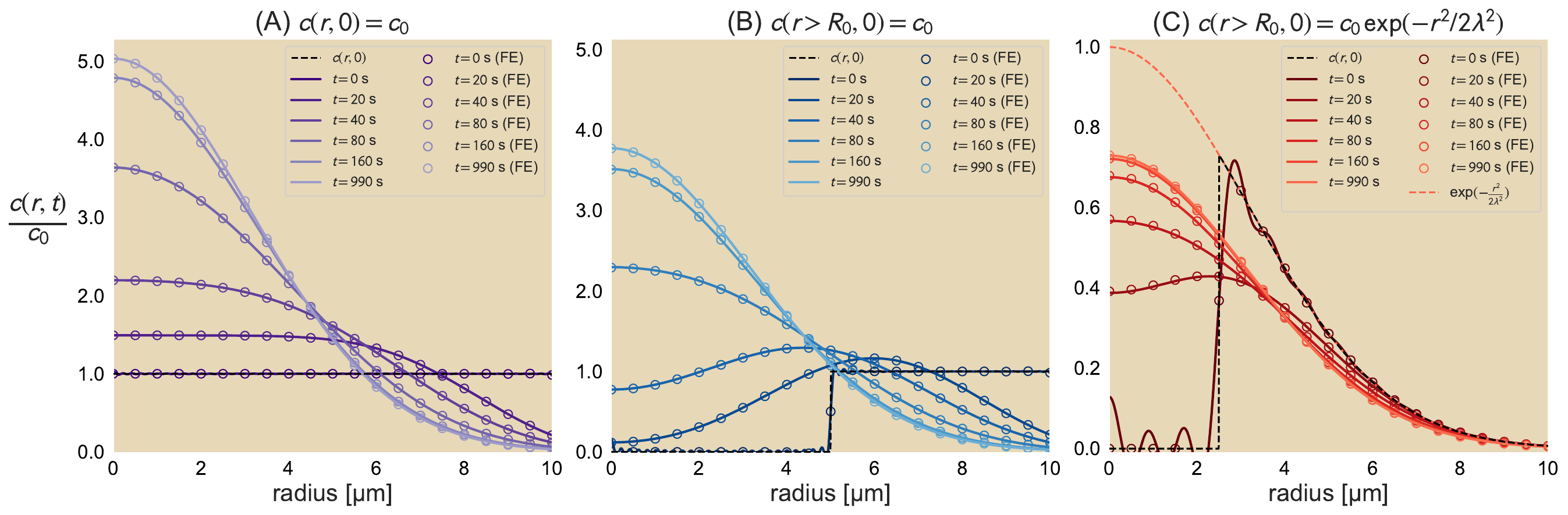}}
\caption{\textbf{Radial advection-diffusion for various initial conditions.} (A) Uniform concentration throughout the system.
(B) Uniform concentration for $r > R_0$ and no molecules for $r \leq R_0$. (C) A Gaussian distribution for $r > R_0$ and no
molecules for $r \leq R_0$. Analytical solutions are presented as solid lines while solutions obtained by finite elements are 
shown as hollow points. The initial condition for each situation is shown as a dashed black line. For all studies, 
$D=0.1 \, \frac{\text{{\textmu}m}^2}{\text{s}}$, $R = 10 \text{ {\textmu}m}$, and 
$v_\mathrm{m} = 0.1 \, \frac{\text{{\textmu}m}}{\text{s}}$. For (B), we set $R_0 = \frac{R}{2}$ while for (C) we set 
$R_0 = \frac{R}{4}$. For (C), the steady-state profile prior to removing molecules for $r \leq R_0$ is shown as a dashed red line.
All analytical solutions use the first 12 eigenvalues that satisfy Eq. \ref{eq:2d_BC}.}
\label{fig:advdiff_simple}
\end{figure}

\subsection{Uniform concentration.} \label{subsec:uniform}
We start with the case where the concentration is uniform everywhere,
\begin{align*}
  c(r,0) = c_0. \numberthis \label{eq:2D_IC_uniform}
\end{align*}
The solution to the PDE with this initial condition takes the form of 
\begin{align*}
  c(r,t) = \frac{c_0}{2} &e^{-\frac{r^2}{2 \lambda^2}} \Bigg\{ \frac{\frac{R^2}{\lambda^2}}{1 - e^{-\frac{R^2}{2 \lambda^2}}} + \sum_{i=1}^\infty \frac{R^2 \, e^{- D\,k_i^2 t} \, {_1F_1} \Big( - \frac{\lambda^2 k_i^2}{2} ; 2 ; \frac{R^2}{2 \lambda^2} \Big)}{\int_0^R r' \, e^{- \frac{r'^2}{2 \lambda^2}} \, \Big[ {_1F_1} \Big( - \frac{\lambda^2 k_i^2}{2} ; 1; \frac{r'^2}{2 \lambda^2} \Big) \Big]^2 \text{d}r'} \times \, {_1F_1} \Big( - \frac{\lambda^2 k_i^2}{2} ; 1 ; \frac{r^2}{2 \lambda^2} \Big) \Bigg\}. \numberthis \label{eq:AD_2d_soln_uniform}
\end{align*}
Fig. \ref{fig:advdiff_simple}(A) shows the concentration profile as a function of radius and for various time points given this
initial condition. Here, we used $D=0.1 \, \frac{\text{{\textmu}m}^2}{\text{s}}$, $R = 10 \text{ {\textmu}m}$, and 
$v_\mathrm{m} = 0.1 \, \frac{\text{{\textmu}m}}{\text{s}}$. Solid lines indicate different time points for the specific analytical
solution given the uniform initial condition. These analytical solutions also show strong agreement with simulations performed by FEM 
which are
denoted by hollow points. Here, we use the first 12 eigenvalues $k_i$ for the analytical solution. Similar to the decomposition
of a square wave into a sum of sinusoidal functions yielding imperfect agreement with the original function, we see here that 
the use of a limited number of eigenvalues that satisfy Eq. \ref{eq:2d_BC} leads to fluctuations about the original
function for $t=0$ (see SI Sec. \ref{sec:am_gibbs} on Gibbs phenomenon). Nevertheless, we see that these fluctuations in the
analytical condition quickly smooth out for $t > 0$.
For the given parameters, the concentration at larger radii decreases quickly due to the higher advection 
overcoming diffusion. As shown at $t=20$ seconds and $t=40$ seconds, the concentration appears roughly uniform at lower
concentrations but the length scale of this uniformity appears to decrease. At $t=990$ seconds, the concentration 
profile reaches the Gaussian steady-state solution where the concentration gradient allows diffusion to counter the advective flow.

\subsection{Uniform concentration for $r > R_0$.} \label{subsec:uniform2}
We apply a similar initial condition as that used in Sec. \ref{subsec:uniform}, but remove any molecules within a distance
$R_0$ from the origin as typically performed in FRAP experiments. This initial condition is mathematically described by
\begin{equation}
  c(r,0) = 
  	\begin{cases}
	0 & \text{if $r \leq R_0$},\\
	c_0 & \text{if $r > R_0$}.
	\end{cases} \label{eq:2D_IC_uniform2}
\end{equation}
The solution for this initial condition is similar to Eq. \ref{eq:AD_2d_soln_uniform} but with different limits of
integration (see SI Sec. \ref{sec:sturm_liouville} on Sturm-Liouville Theory and \ref{sisec:am_2dadvdiff} for application of the theory in 2D),
\begin{align*}
  c(r,t) = \frac{c_0}{2} e^{-\frac{r^2}{2 \lambda^2}} &\Bigg\{ \frac{\frac{R^2}{\lambda^2} - \frac{R_0^2}{\lambda^2}}{1 - e^{-\frac{R^2}{2 \lambda^2}}} + \sum_{i=1}^\infty \alpha_i e^{- D\,k_i^2 t} \, {_1F_1} \Big( - \frac{\lambda^2 k_i^2}{2} ; 1 ; \frac{r^2}{2 \lambda^2} \Big) \Bigg\}, \numberthis \label{eq:AD_2d_soln_uniform2}
\end{align*}
\noindent
where
\begin{align*}
  \alpha_i &= \frac{R^2 \,{_1F_1} \Big( - \frac{\lambda^2 k_i^2}{2} ; 2 ; \frac{R^2}{2 \lambda^2} \Big) - R_0^2 \, {_1F_1} \Big( - \frac{\lambda^2 k_i^2}{2} ; 2 ; \frac{R_0^2}{2 \lambda^2} \Big)}{\int_0^R r' \, e^{- \frac{r'^2}{2 \lambda^2}}  \, \Big[ {_1F_1} \Big( - \frac{\lambda^2 k_i^2}{2} ; 1; \frac{r'^2}{2 \lambda^2} \Big) \Big]^2 \text{d}r'}. \numberthis \label{eq:alpha}
\end{align*}
As $R_0 \rightarrow 0$ in Eq. \ref{eq:AD_2d_soln_uniform2} we recover Eq. \ref{eq:AD_2d_soln_uniform}. Fig. 
\ref{fig:advdiff_simple}(B) shows traces of the concentration profile at the same times as in Fig. \ref{fig:advdiff_simple}(A). Here,
$R_0 = \frac{R}{2}$. Once again, we see that the analytical solution for $t=0$ fluctuates about the defined initial condition but
quickly smooth out and agree well with FEM results (hollow points) for $t>0$.
By removing molecules at $r \leq R_0$, a wave of molecules move toward the origin from a combination of advection toward the
origin and diffusion moving molecules against the concentration gradient while the concentration at 
$r \rightarrow R$ recedes. Once again, we recover a Gaussian profile, but at a lower maximum than that observed in Fig.
\ref{fig:advdiff_simple}(A) due to the lower initial number of molecules.

\subsection{Gaussian profile for $r > R_0$.} \label{subsec:gaussian}
Finally, consider a situation where molecules in this advective-diffusive system are allowed to reach steady-state before 
photobleaching all molecules within a certain radius of the center $r \leq R_0$. The initial conditions would appear as
\begin{equation}
  c(r,0) = 
  	\begin{cases}
	0 & \text{if $r \leq R_0$},\\
	c_0 \, e^{-\frac{r^2}{2 \lambda^2}} & \text{if $r > R_0$}.
	\end{cases} \label{eq:2D_initial_conditions}
\end{equation}
We show analytically that the concentration profile is
\begin{align*}
  c(r,t) &= c_0 e^{-\frac{r^2}{2 \lambda^2}} \Bigg\{ \frac{e^{- \frac{R_0^2}{2 \lambda^2}} - e^{-\frac{R^2}{2 \lambda^2}}}{1 - e^{-\frac{R^2}{2 \lambda^2}}} - \frac{1}{2} \sum_{i=1}^\infty \beta_i e^{- D k_i^2 t}\, {_1F_1} \Big( - \frac{\lambda^2 k_i^2}{2} ; 1 ; \frac{r^2}{2 \lambda^2} \Big) \Bigg\}, \numberthis \label{eq:AD_2d_soln_unity}
\end{align*}
\noindent where
\begin{align*}
\beta_i &= \frac{R_0^2  \, {_1F_1} \Big( 1 + \frac{\lambda^2 k_i^2}{2} ; 2 ; -\frac{R_0^2}{2 \lambda^2} \Big)}{\int_0^R r' \, e^{- \frac{r'^2}{2 \lambda^2}} \Big[ {_1F_1} \Big( - \frac{\lambda^2 k_i^2}{2} ; 1; \frac{r'^2}{2 \lambda^2} \Big) \Big]^2 \text{d}r'}. \numberthis \label{eq:beta}
\end{align*}
\noindent Once again the analytical solution agrees with simulations of the same initial condition shown in Fig. \ref{fig:advdiff_simple}(C)
for $R_0 = \frac{R}{4}$. We note here that as $R_0 \rightarrow 0$ we recover the steady-state solution again as the 
time-dependent terms vanish and the ratio of exponentials in the time-independent term goes to unity. Fig. \ref{fig:advdiff_simple}(C)
shows again the imperfection of the analytical solution for $t=0$ and the initial condition but a strong agreement with FEM
results. In this situation, the concentration toward the outer edge of the system remains largely unchanged as diffusion and 
advection are
balanced toward the boundary. However, at smaller radii of the system, there is a shift in concentration as molecules enter the
$r \leq R_0$ region and for the chosen parameter values, the overall concentration profile returns to a Gaussian distribution 
within 3 minutes.
\\
\\
Across all three initial conditions, we see that the concentration 
builds up toward the contraction center and forms a Gaussian distribution as the steady-state profile. The different time courses in the concentration
profiles for these initial conditions further reveals that in experimental
systems exhibiting such an advective-diffusive behavior the use of FRAP becomes sensitive to the time when photobleaching is
applied. If the concentration profile in the system has already begun to move away from a uniform distribution, such as the initial
contraction of a highly connected filament network, then the molecule redistribution 
until steady state is achieved will show different recovery profiles from that of an experiment 
where photobleaching is applied at a time when the system is already close to reaching the steady-state profile. Such results
provide the two extremes of ``fluorescence recovery'' in potential {\it in vitro} assays that evolve from a uniform concentration
to a Gaussian-shaped distribution subject to this advection-diffusion system.
\\
\\
We show here three cases where analytical solutions to the linear advection-diffusion equation can be determined for direct comparison
to the FEM simulations. As the square unit cell is more complex, we
turn fully to FEM for our measurements and comparisons to the 
analyzed experimental data.
\section{Numerically solving advection-diffusion equations with COMSOL} \label{sec:am_comsol}

Our use of COMSOL Multiphysics$\circledR$ simulations are constructed with consideration of four particular details in mind: design of the geometry; 
set-up of the differential equations, including boundary and initial conditions; choice of mesh size; and sweeping through parameters. 
Elaboration of the mesh size dependence is discussed in Sec. \ref{sec:am_gibbs}.

\subsection{Geometry}
\begin{figure}[b]!
    \centering
    \includegraphics[scale=0.8]{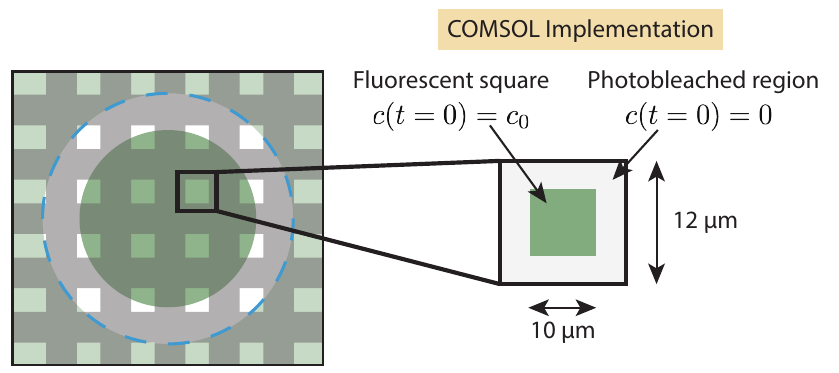}
    \caption{\textbf{Schematic of COMSOL set-up.} To simulate the time evolution of a single unit cell in the advection-diffusion equation, we model a single unit cell as a $10$ {\textmu}m x $10$ {\textmu}m square within a larger $12$ {\textmu}m x $12$ {\textmu}m square.}
    \label{fig:comsolsquare}
\end{figure}
Because we analyzed fluorescent unit cells from our experimental data
until they were no longer distinguishable from neighboring unit cells,
we opted to simplify the FEM numerical simulation by examining the
time course of a single unit cell subject to advection and diffusion.
Even though unit cells in the network may be transported toward the
center of contraction, as we have shown in SI Sec \ref{sec:am_square_shrink}, the unit cell deformation from advection
is not position dependent. This is similarly the case for diffusion,
where its contribution to the flux of molecules is dependent on the gradient of concentration.
As indicated in Fig. \ref{fig:comsolsquare}, the geometry of the system in the COMSOL simulations is a square of side length 20\% longer than the side length of unit cell, which we
take to be 10 {\textmu}m. We then place the smaller square that represents the unit cell inside of the larger square such that it shares the
same center. We then take the union of these two squares before applying the split operation to distinguish the unit cell from the surrounding
region.

\subsection{Setting up the differential equations}

Although there are multiple forms of inputting partial differential equations in COMSOL, for the advection-diffusion equation
studied, that is
\begin{equation}
  \frac{\partial u}{\partial t} = D \nabla^2 u + \alpha \nabla \cdot \left( {\bf{r}} u \right), \label{eq:si_comsol_advdiff}
\end{equation}
we elect to use the coefficient form PDE and define our variable of interest as $u$ with units of mol/m$^3$ and
ensure that each term in the equation carries units of mol/(m$^3\cdot$s). Although our past derivations
use the variable $c$, we use $u$ in the differential equation due to the occurrence of the coefficient $c$ in the coefficient form PDE
in COMSOL. We note that the coefficient form PDE as shown in COMSOL is of the form
\begin{equation}
  e_a \frac{\partial^2 u}{\partial t^2} + d_a \frac{\partial u}{\partial t} + \nabla \cdot \left( - c \nabla u - \eta u + \gamma \right) + \beta \cdot \nabla u + au = f, \label{eq:si_comsol}
\end{equation}
where $e_a$, $d_a$, $c$, $a$, and $f$ are scalar coefficients while $\eta$, $\gamma$, and $\beta$ are vectors. We note here that in COMSOL, the term involving $\eta$ is written as $\alpha$, but to
avoid confusion with the $\alpha$ used throughout our work, we change the
COMSOL notation to $\eta$.
Rewriting Eq. \ref{eq:si_comsol_advdiff} to match the form of Eq. \ref{eq:si_comsol} gives
\begin{equation}
  \frac{\partial u}{\partial t} + \nabla \cdot \left( - D \nabla u - \alpha {\bf{r}} u \right) = 0. \label{eq:si_advdiff_coeffform}
\end{equation}
We can see here that to make Eq. \ref{eq:si_advdiff_coeffform} match Eq. \ref{eq:si_comsol}, then $e_a$, $a$, all of the
elements of $\gamma$, all of the elements of $\beta$, and $f$ are all $0$ while 
\begin{align}
d_a &= 1 \textrm{ s}^{-1}, \label{eq:d_a}\\
c &= D, \label{eq:comsol_D}\\
\eta &= \begin{bmatrix}
		\alpha x\\
		\alpha y
		\end{bmatrix}, \label{eq:comsol_alpha}
\end{align}
where we note that we define $D$ to take on dimensions of length$^2$/time and $\alpha$ to have units of time$^{-1}$
in COMSOL.
\\
\\
In our experiments, we are careful to ensure that there is negligible to no detectable amount of microtubules flowing from outside
of the light-activated region into network. We thus impose a no-flux boundary condition by using the Zero Flux boundary condition
option in COMSOL. 

\subsection{Applying the initial condition}

We opt to make the initial condition of the unit cell of uniform concentration $c_0$ while the concentration in the region outside of the
unit cell is initially set to $0$. However, defining these two initial conditions piecewise with the geometry of the system
outlined above leads to a sharp change in the gradient, which can lead to large errors and negative concentrations at high P\'{e}clet
number, we instead define a rectangle function where the edges of the rectangle function are smoothed over 200 nm and have
well-defined continuous derivatives to second order.

\subsection{Choice of mesh size}

Because we use the total particle number as a conserved quantity for computing the area of the unit cells in time, we wish to minimize
the numerical error in the FEM simulations. Of the various mesh designs, we opt to use the ``Extremely fine'' mesh size with the
boundary between the unit cell and the surrounding system, obtained from the geometry design, to also undergo 6 iterations of
refinement under the ``Control Entities'' tab. This boundary is heavily refined in order to minimize the occurrence and value of
negative concentrations that may arise at high P\'{e}clet number. A more elaborate discussion of mesh size choice is presented
in SI Sec. \ref{sec:am_gibbs} on the Gibbs phenomenon.

\subsection{Parameter Sweep}

To perform the parameter sweep, we include the Parametric Sweep option in the Study section of the simulation and define the 
parameters of interest under Global Definitions $\rightarrow$ Parameters. Within the parameters, we specify the parameters $D$
for our diffusion constant and $\alpha$ for our contraction rate. Under the Parametric Sweep, we can then chose $D$ and $\alpha$
as our parameters to be swept. We select our range of values of $\alpha$ to be the different experimentally-obtained contraction rates 
while $D$ ranged from $0.0001$ {\textmu}m$^2$/s to $0.01$ {\textmu}m$^2$/s in various increments ranging from $0.0001$ {\textmu}m$^2$/s to
$0.0005$ {\textmu}m$^2$/s. All possible combinations of $D$ and $\alpha$ were permitted for the simulations.
\subsection{Gibbs phenomenon in analytical solutions and mesh granularity in FEM} \label{sec:am_gibbs}

\begin{figure}[t!]
\centering{
\includegraphics[scale=0.35, trim={0cm 0cm 0cm 0cm}]{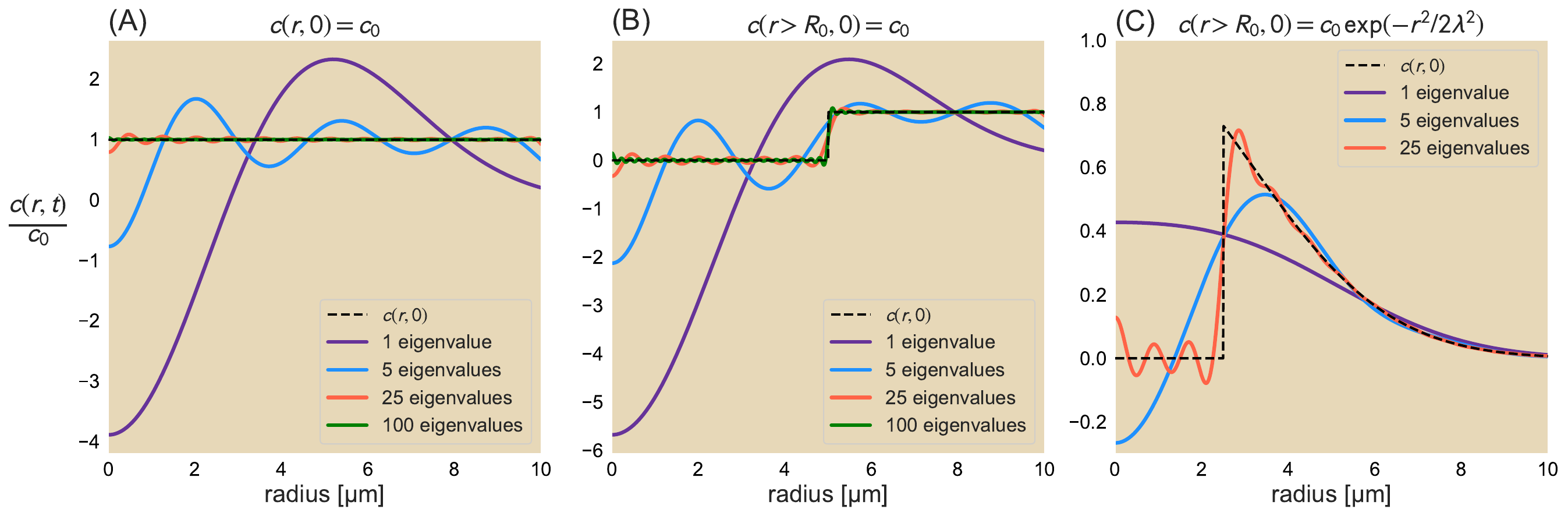}}
\caption{Gibbs phenomenon for analytical solutions. Concentration profiles of the analytical solution for the initial
conditions (A) $c(r,0)=c_0$, (B) $c(r>R_0,0)=c_0$, and (C) $c(r>R_0,0)=c_0 \text{exp}(-r^2/2\lambda^2)$ with the 
steady-state solution and the first nonzero eigenvalue solution (purple line), the first five nonzero eigenvalue solutions 
(blue), the first twenty-five terms (red), and for (A) and (B) the first hundred terms (green). The intended initial conditions 
are represented as dashed black lines.}
\label{sifig:gibbs_analytical}
\end{figure}

As we noted in the SI Sec. \ref{sec:analytical}, upon solving the analytical solutions for
three cases, there was notable discrepancy between the analytically solved concentration
profile at $t=0$ and the defined initial condition. In this section, we address the sensitivity
of the analytical solutions to the number of terms in the infinite series that are kept when
showing the concentration profile over time. We then discuss a similar case of sensitivities
in the finite element method (FEM) which can also affect the accuracy of numerical solutions.
\\
\\
As shown in Fig. \ref{sifig:gibbs_analytical}, 
the analytical solution, which is composed of the first 100 non-zero eigenvalues for the two cases involving a uniform initial 
concentration and the first 25 eigenvalues for the one involving the FRAPed Gaussian profile and the steady-state function,
creates oscillations about the intended initial condition. This disagreement is a demonstration of the Gibbs phenomenon,
as famously revealed by the imperfect decomposition of a square wave into a sum of sinusoidal functions. Fig. 
\ref{sifig:gibbs_analytical} demonstrates the evolution of each of the three analytical solutions examined in the main manuscript
when more eigenvalues are included in the solution. Specifically, for $c(r,0)=c_0$ (Fig. \ref{sifig:gibbs_analytical}(A)),
$c(r>R_0,0)=c_0$ (Fig. \ref{sifig:gibbs_analytical}(B)), and $c(r>R_0,0)=c_0 \text{exp}(-r^2/2 \lambda^2)$ (Fig. \ref{sifig:gibbs_analytical}(C)),
all of which are represented by dashed black lines, more eigenvalues reduce the level of error between the analytical
solution and the initial condition. For the two initial conditions involving a uniform concentration, the use of one eigenvalue
in addition to the steady-state solution (purple line) leads to a large negative concentration at $r=0$ but more closely 
recapitulate the initial conditions after using 100 non-zero eigenvalues. Deviations from the initial condition decrease
dramatically by that point. This is similarly observed for the clipped Gaussian distribution: while the Gaussian tail is
quantitatively captured by the the addition of only a few eigenvalues, the analytical solution begins to better recapitulate
the concentration profile about $r=R_0$ with the addition of more terms in the solution. Curiously, after using
more than 25 eigenvalues, the solution shows large oscillations rather than smaller ones
that are smoothed out rather
quickly after a small amount of time. 
\\
\\
Just as analytical solutions are sensitive to a form of resolution to properly capture the
time evolution of a variable of interest, more concretely shown through the number of 
eigenvalues computed and by extension the number of terms used in the infinite series, so too
are there sensitivities in the FEM solution. These sensitivities must also be addressed during
setup of the FEM solution to ensure that the model equation is being accurately recapitulated. 
In this case, a key consideration is
the choice of granularity in the mesh. As FEM involves solving the governing equation over a particular domain, having
a very fine grained mesh allows for the FEM solution to more accurately reflect the true solution to the problem at the
cost of computational time. On the other hand, a very coarse-grained mesh involves less computing power to solve the
original equations but may coarse grain away details smaller than the element size, requiring a balance between accurately
solving the original PDE(s) and computational efficiency.
\begin{figure}[b!]
\centering{
\includegraphics[scale=0.7, trim={0cm 0cm 0cm 0cm}]{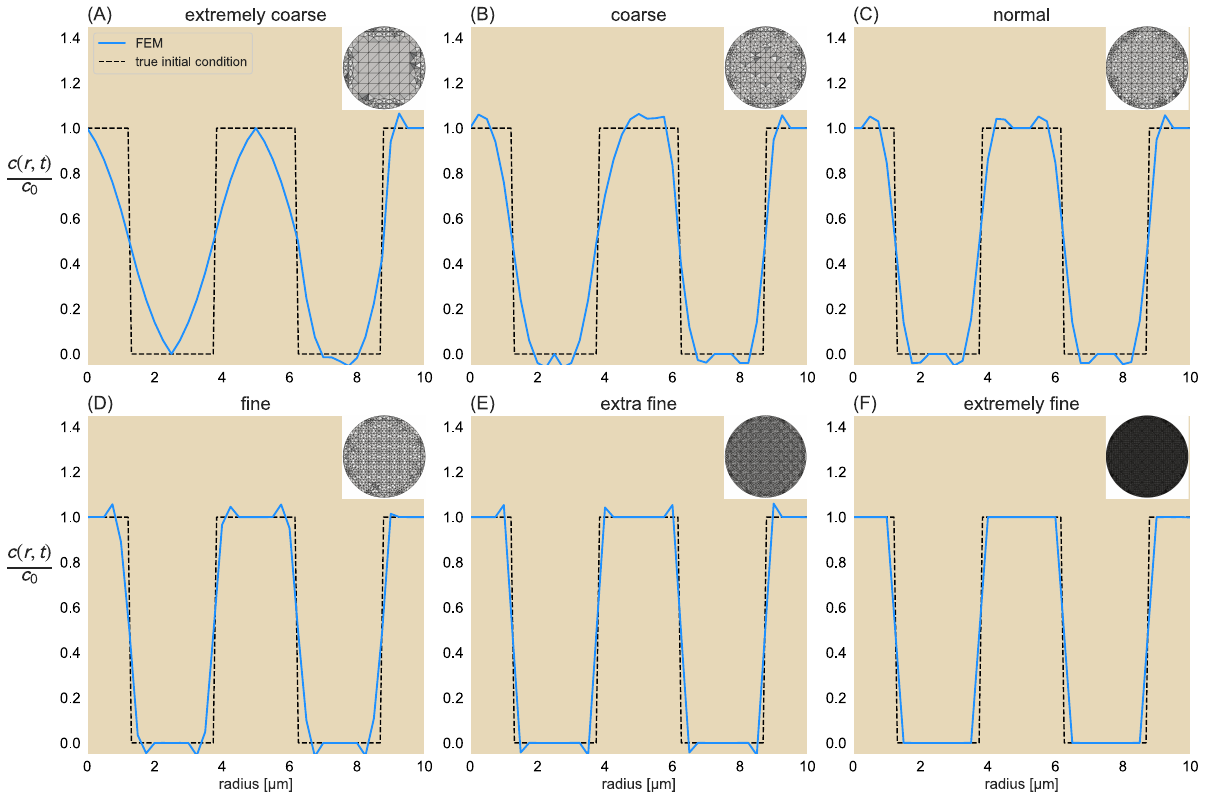}}
\caption{Effects of mesh granularity on FEM solution. Concentration profiles at $t=0$ for six different element sizes as
defined by the COMSOL Multiphysics physics-controlled mesh: (A) extremely coarse, (B) coarse, (C) normal, (D) fine, (E)
extra fine, and (F) extremely fine. Finite elements output is represented by the blues lines while the true initial conditions
are given as the black dashed lines. For visualization purposes, the appearance of the meshes used for the defined
geometry are shown as insets in the upper righthand corner of the respective subfigures. Concentration profile is from
a line trace along the horizontal axis from the origin of the geometry to the boundary.}
\label{sifig:gibbs_fem}
\end{figure}
\\
\\
Fig. \ref{sifig:gibbs_fem} shows how the granularity of the mesh affects the FEM solutions. We compare the concentration
profiles produced by FEM (solid blue lines) against the true initial condition (dashed black lines) for six different element sizes
as found in the physics-controlled mesh feature in COMSOL Multiphysics: (A) extremely coarse, (B) coarse,
(C) normal, (D) fine, (E) extra fine, and (F) extremely fine. We see that while using the most coarse-grained feature poorly matches
the desired initial condition with more of a sine wave than a 
square wave, successively decreasing element size (increase in mesh fineness) allows the FEM solution to more closely reflect the initial
condition. Fig. \ref{sifig:gibbs_fem}(B)-(E) show that increasing the mesh fineness leaves fewer deviations from the true values,
largely located near the discontinuities in the profile. The insets in the upper right of each figure shows the mesh pattern
for the geometry for the study. As Fig. \ref{sifig:gibbs_fem}(F) shows, while the extremely fine mesh does not overshoot above
the $c_0$ values or undershoot the $c(r,0)=0$ regions, the finite size of the elements in the mesh causes the discontinuous
region to take on a value between the two regions instead. As the
FEM simulation is not computationally demanding for the single unit cell case,
we opt to use an Extremely Fine mesh setting.
\section{FEM results of advection-diffusion equation on a simulated unit cell array} \label{sec:am_qualitative}

\begin{figure}[t!]
  \centering{
  \includegraphics[scale=0.7]{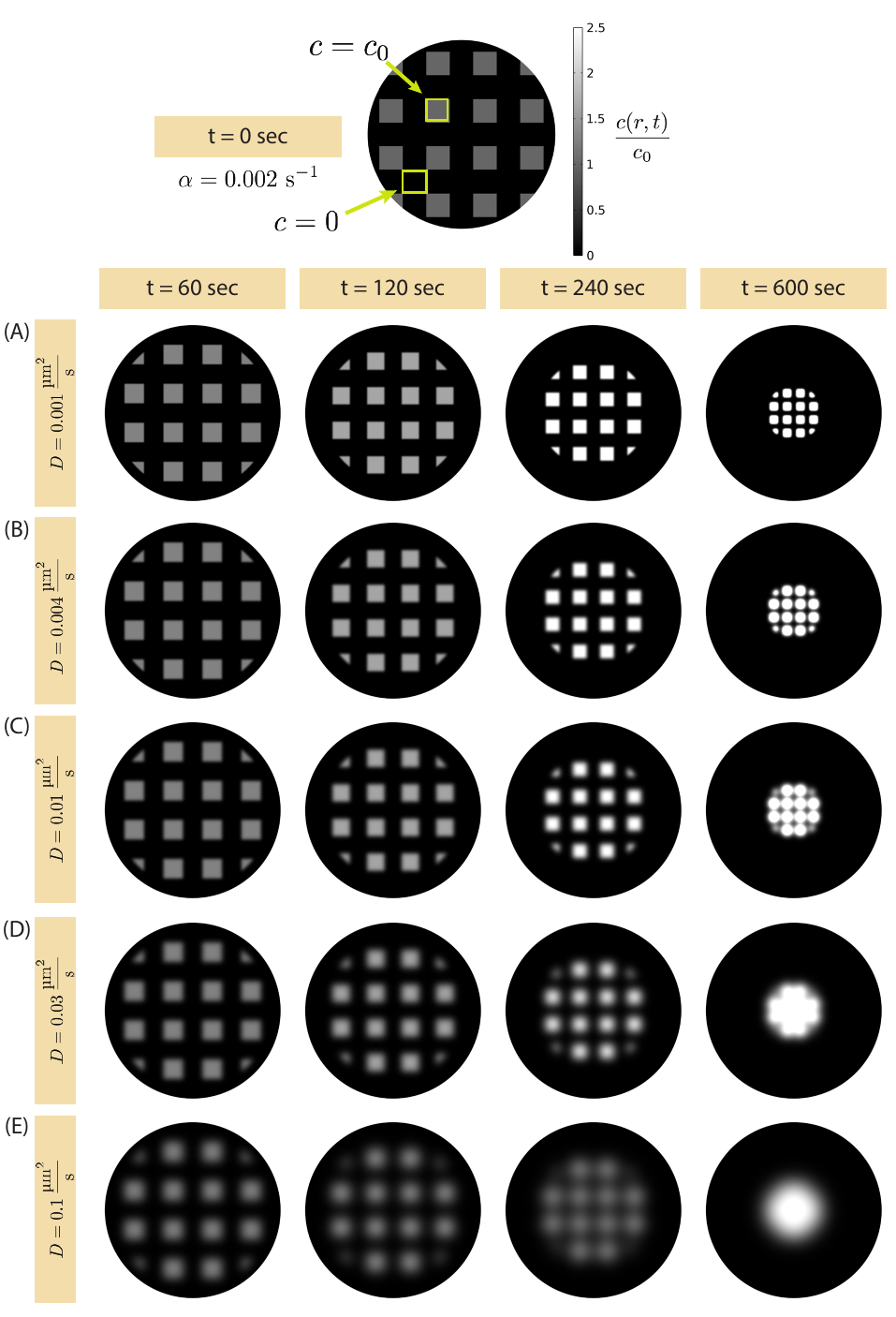}}
  \caption{\textbf{Concentration profiles of an array of unit cells at various time points and diffusion constants.} The FEM simulation
  is the same as that described in SI Sec \ref{sec:am_comsol} but where each square (denoted by initial concentration $c_0$ as
  drawn with the top yellow box in the $t=0$ sec
  schematic) has a side length of 15 {\textmu}m and a center-to-center distance of 30 {\textmu}m, with a concentration of $0$
  in between. In all cases, we use the same advection rate of $0.002$ sec$^{-1}$ and different diffusion coefficients: (A) $0.001 \, 
  \frac{\text{{\textmu}m}^2}{\text{sec}}$, (B) $0.004 \, \frac{\text{{\textmu}m}^2}{\text{sec}}$, (C) $0.01 \, 
  \frac{\text{{\textmu}m}^2}{\text{sec}}$, (D) $0.03 \, \frac{\text{{\textmu}m}^2}{\text{sec}}$, and (E) $0.1 \, 
  \frac{\text{{\textmu}m}^2}{\text{sec}}$.}
  \label{fig:fem_qual}
\end{figure}

In the main manuscript, we measure the area of the fluorescent squares over time and compare the results to numerical simulations
of an advection-diffusion equation through the FEM simulations as described in SI Sec \ref{sec:am_comsol} in order to compute 
effective diffusion constants. While alternative approaches to obtaining the effective diffusion constant exist, we offer this as a direct
comparison to numerical experiments. For a qualitative comparison to the experimentally-observed change in the
photobleached microtubule network, we present in this section the time evolution of the concentration distribution for an array of
unit cells subject to linear advective and diffusive effects. For these simulations, we follow a similar procedure as outlined in SI
Sec. \ref{sec:am_comsol} but on a circle of radius 60 {\textmu}m and squares of side length 15 {\textmu}m with a periodicity of
30 {\textmu}m. Fig. \ref{fig:fem_qual} shows different time points of the concentration profile subject to the same rate of
advection ($0.002$ sec$^{-1}$) but different
diffusion constants, namely, those measured for the median (Fig. \ref{fig:fem_qual}(A)) and 3rd quartile area trajectories (Fig. 
\ref{fig:fem_qual}(B)), an order of magnitude greater
diffusion constants (Fig. \ref{fig:fem_qual}(C)-(D)), and roughly the diffusion coefficient of a free microtubule (Fig. \ref{fig:fem_qual}(E)).
From examining the different concentration profiles in Fig. \ref{fig:fem_qual} in comparison to the experimental results shown in Fig. 2 of the 
manuscript, we see that once again, introducing diffusion to the system is a necessary component to recapitulate the experimentally
obtained results. We further see that by eye the simulated data and experiments look most similar when simulating with an effective diffusion constant of 0.1 $ \frac{\text{{\textmu}m}^2}{\text{sec}}$. However, we note that as discussed in the results section "The effective diffusion constant is roughly two orders of
magnitude lower than free diffusion of a microtubule" in the manuscript and as revealed through Fig. 3, we see that when we careful 
quantify the area trajectories
using the same metric for experiments, the diffusion constants above about $6.0 \times 10^{-3}$ $ \frac{\text{{\textmu}m}^2}{\text{sec}}$
would lead to increasing area trajectories instead. 
\section{Motor Constructs} \label{sec:am_motorconstruct}
\begin{table}[h!]
    \centering
      \begin{tabular}{| c | c |}
        \hline
        \textbf{Motor Construct} & \textbf{Sequence Layout}\\ \hline
        micro variant & pBiex-1:FLAG-GG-mVenus-(GSG)$_2$-micro-(GSG)$_4$-Ncd281\\ \hline
        iLid variant & pBiex-1:FLAG-GG-mVenus-(GSG)$_2$-iLid-(GSG)$_4$-Ncd281\\ \hline
      \end{tabular}
      \caption{\textbf{Ncd281 construct design.} All constructs are designed in the pBiex-1 vector and produced by Twist Biosciences.}
      \label{tab:am_ncd281}
  \end{table}
\begin{table}[h!]
    \centering
      \begin{tabular}{| c | c | c |}
        \hline
        \textbf{Motor Species} & \textbf{Maximum speed} & \textbf{Processivity} \\ \hline
        Ncd281 & 90 nm/s \cite{endres2006} & Nonprocessive \cite{endres2006} \\ \hline
        Ncd236 & 115 nm/s \cite{banks2022} & Nonprocessive \cite{banks2022,hentrich2010} \\ \hline
        bacterial-expressed K401 & 250 nm/s \cite{ross2019} & $\approx 100$ steps \cite{banks2022} \\ \hline
        insect-expressed K401 & 600 nm/s  \cite{banks2022} & \thead{$\approx 100$ steps \cite{banks2022} (Assumed to be the same as \\
        bacterial-expressed K401)} \\ \hline
      \end{tabular}
      \caption{\textbf{Motor variant parameters.}}
      \label{tab:am_motors}
  \end{table}
While several of the motors used here in the analysis are obtained from previous work, including K401 expressed in bacteria \cite{ross2019},
K401 expressed in insects and Ncd236 expressed in insects \cite{banks2022}, we also designed constructs for the study of Ncd281
\cite{endres2006}. Specifically, the sequences are inserted into pBiex-1 vectors and includes a FLAG tag for protein purification, 
mVenus for motor fluorescence visualization, either a micro or iLid domain as described in \cite{guntas2015} and Ncd281 as described in \cite{endres2006}. Between these different domains are multiple repeats of a `GSG' amino acid sequence which offers flexible links
between the regions. Table \ref{tab:am_ncd281} illustrates these sequences. Constructs were produced by Twist Biosciences.

In addition, Table \ref{tab:am_motors} shows the different motors presented in the manuscript, including their processivities and maximum
speeds. Citations of their speeds are added as necessary.

\section{Variability in Pecl\'{e}t number}

\begin{figure}[t!]
  \centering{
  \includegraphics[scale=0.4]{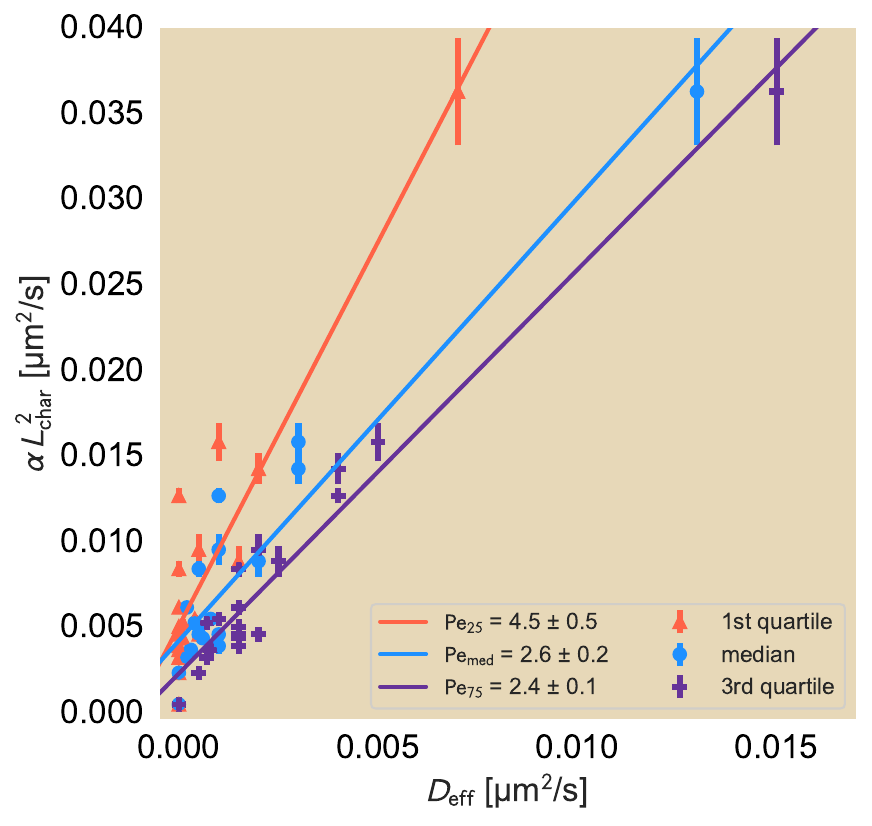}}
  \caption{\textbf{Fits of Pecl\'{e}t number for the first quartile (red),
  median (blue), and third quartile (purple).} We remind the reader that the blue
  datasets are identical to what is presented in Fig. 6 of the main manuscript.}
  \label{fig:pecletquartile}
\end{figure}

In the main manuscript, we argue that a Pecl\'{e}t number $\mathrm{Pe}$ emerges regardless of the
effective motor speed, tuned through ATP concentration or motor species. There, we
presented this using the effective diffusion constants fitted from simuations onto
the median area trajectories for these conditions. To get an idea for how sensitive
$\mathrm{Pe}$ is to the variability found within conditions, e.g. the spread in area
trajectory distribution, we compute $\mathrm{Pe}$ for the first and third quartiles.
Fig. \ref{fig:pecletquartile} shows best linear fits for each of quartiles examined
where the slopes denote the respective values of $\mathrm{Pe}$. Here, we find that
in addition to the median Pecl\'{e}t number $\mathrm{Pe}_\mathrm{med} = 2.6 \pm 0.2$ as noted
in the main manuscript, $\mathrm{Pe}_{25} = 4.5 \pm 0.5$ and $\mathrm{Pe}_{75} = 2.4 \pm 0.1$.
Here, we see that despite the variability in the effective diffusion constant, $\mathrm{Pe}$
is less than a factor of 2 different between the quartiles, suggesting low variability in
this dimensionless number.

\section{Computing depletion forces} \label{sisec:depletion}

One of the most useful effects of crowding agents is their ability to induce entropic
forces upon larger objects when these crowded objects are within the size of the crowding agent from each other.
This may be relevant in \emph{in vitro} active systems where the use of crowding agents help to 
bundle microtubules and promote self-organization. In the case of the work presented here, pluronic ($\backsim12.5$ kDa) 
acts as a crowding agent for microtubules
({\textalpha} and {\textbeta}  tubulin have sizes of 50 kDa each and a one-micron long microtubule consists of
$\backsim1.6\times10^3$ tubulin). 
\begin{figure}[t!]
  \centering{
  \includegraphics[scale=0.5]{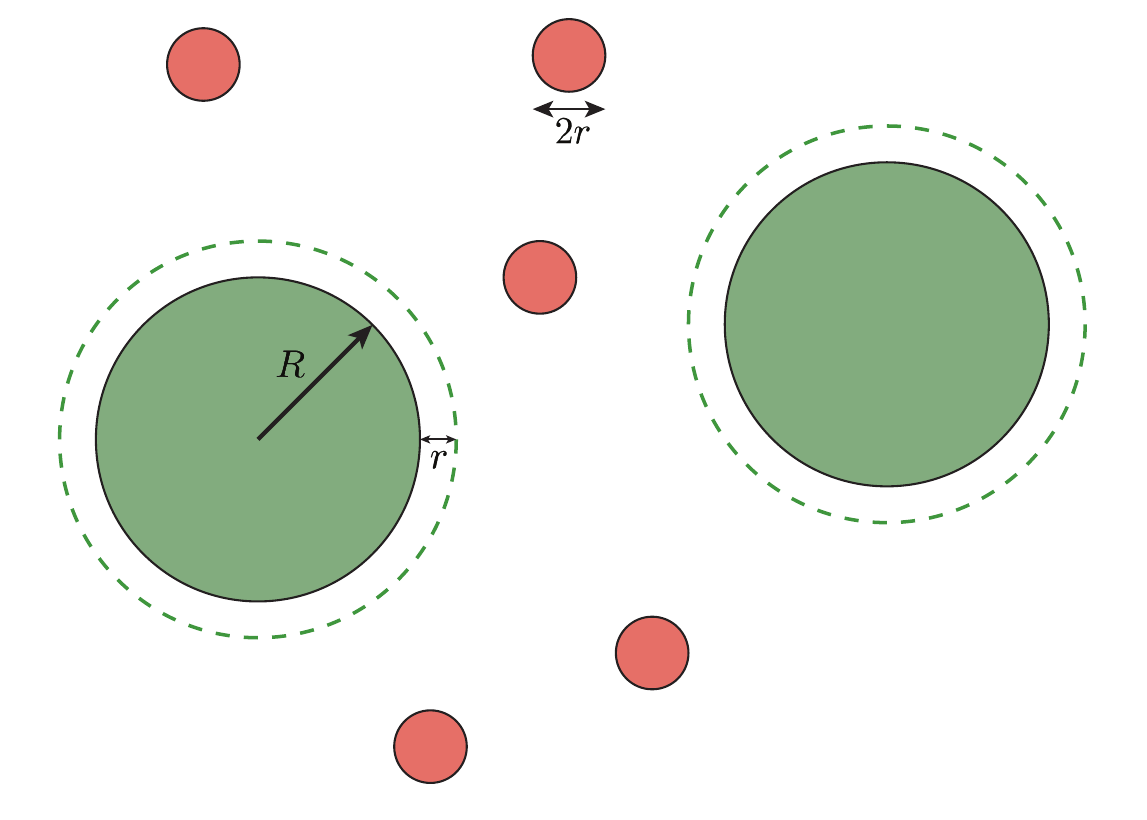}}
  \caption{\textbf{Schematic of crowding action on two larger objects.} The crowders (red) have radius $r$ while the 
  two larger objects (green) have radius $R \gg r$. An additional zone around the large molecules as denoted with
  a dashed outline extends $r$ away from the edge of each molecule and denotes the region within which the centers
  of the crowders cannot enter.}
  \label{fig:crowding}
\end{figure}
Here, to get a sense of the size of these forces, we estimate the entropic forces induced by crowders such as pluronic onto rigid polymers such
as microtubules. To start, we compute the free energy change of the space that crowding agents can occupy when there are 
two larger particles of radius $R$. We will start by solving in two dimensions where we account only for the
cross-sectional area of the microtubules. We will further assume the system does not contain a high density of
crowders, so we will say that there are $N$ crowders that can be distributed across $\Omega \gg N$ lattice sites of 
size $a$. Finally, each crowder will have radius $r$. In the absence of the microtubules, the free energy of the
crowders in a system of size $A_\mathrm{sys}$ is
\begin{align*}
  G_\mathrm{open} &= - N k_B T \,  \mathrm{ln} \left( \frac{A_\mathrm{sys}}{a} \right), \numberthis \label{eq:g_open}
\end{align*}
where $k_B T$ is the thermal energy. Later on we will attempt a
derivation where the number of crowders is dense enough where we need to account for their finite size. With the 
addition of two microtubules, the free energy $G_\mathrm{crowd}$ becomes
\begin{align*}
  G_\mathrm{crowd} &= - N k_B T \, \mathrm{ln} \left( \frac{A_\mathrm{sys} - A_\mathrm{exc}}{a} \right), \numberthis \label{eq:g_crowd}
\end{align*}
where $A_\mathrm{exc}$ is the excluded area unavailable to the crowders. This can be represented as the cross-sectional
areas of the microtubules with an additional radial buffer zone of length $r$ and depends upon the distance the two
cross-sectional areas are from each other. For now, we can compute the free energy change as
\begin{align*}
  \Delta G \equiv G_\mathrm{crowd} - G_\mathrm{open} &= - N k_B T \, \mathrm{ln} \left( \frac{A_\mathrm{sys} - A_\mathrm{exc}}{A_\mathrm{sys}} \right),\\
  \approx N k_B T \, \frac{A_\mathrm{exc}}{A_\mathrm{sys}}, \numberthis \label{eq:delta_g}
\end{align*}
where we assumed that $A_\mathrm{sys} \gg A_\mathrm{exc}$.
\begin{figure}[t!]
  \centering{
  \includegraphics[scale=0.5]{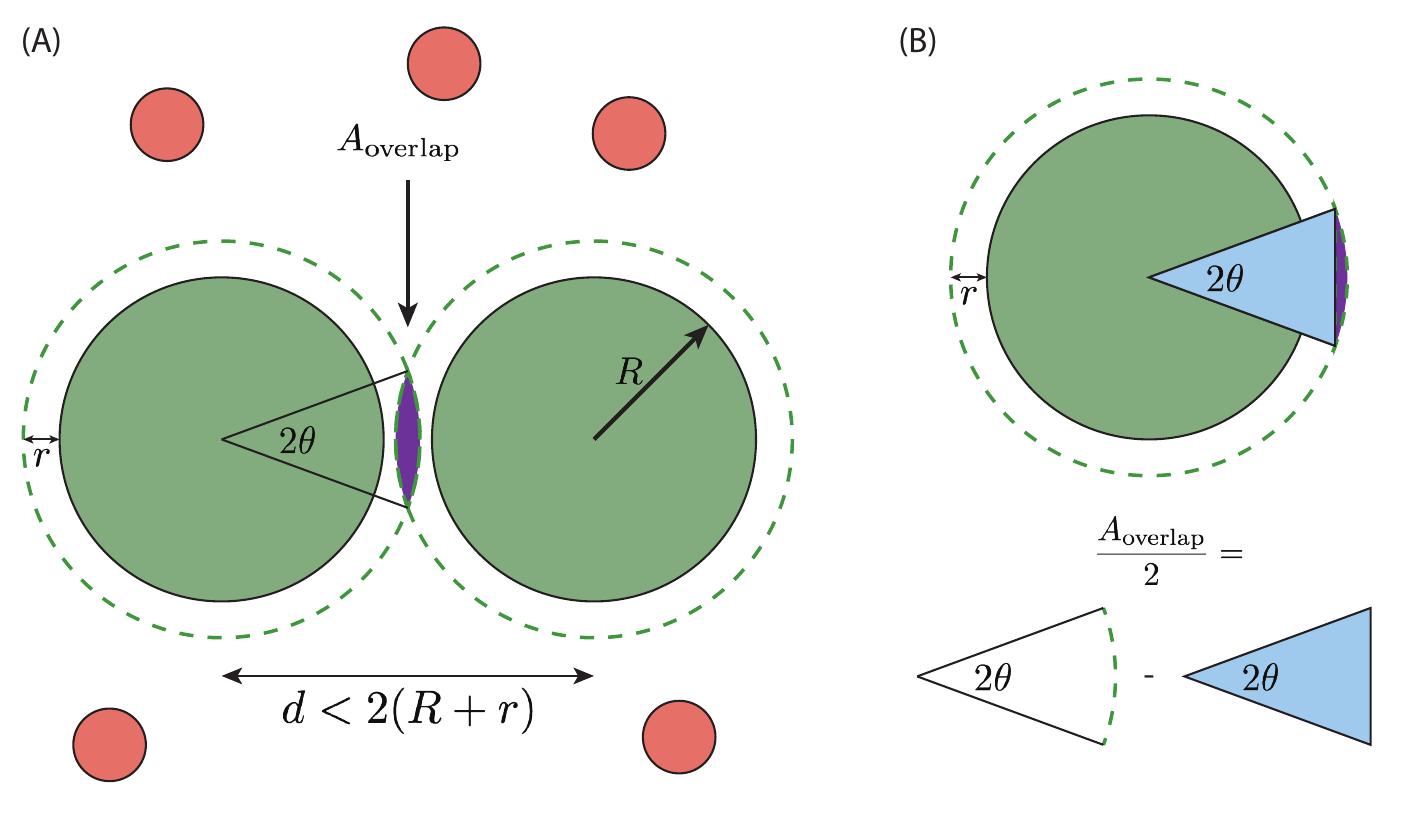}}
  \caption{\textbf{Schematic of the overlap of two molecules.} (A) When the large molecules are separated by a distance
  $d < 2(R + r)$, the exclusion area contains an overlap region that is double-counted in the accounting if the areas of
  the two molecules and their extended zones are added. (B) The overlap area can be computed by subtracting by 
  computing the difference between the slice of the circle whose arclength begins and ends with the two intersection
  points of the overlapping circles (as swept out by the angle
  $2 \theta$) and the triangle whose vertices are the center of the circle and the two points where the overlapping
  circles intersect.}
  \label{fig:overlap}
\end{figure}
\\
\\
\noindent
As noted, the distance between the two microtubules has an effect on the exclusion area. If the microtubules are spaced
such that a crowder can fit between them, then $A_\mathrm{exc}$ is at its maximum, where
\begin{align*}
  A_\mathrm{exc} &= 2 \times \pi (R + r)^2. \numberthis \label{eq:max_Aexc}
\end{align*}
However, if the microtubules are spaced less than a crowder apart, then there is an overlap region that is double-counted
$A_\mathrm{overlap}$ (Fig. \ref{fig:overlap}(A)).
We can compute the area of overlap by recognizing that half of the overlap is the difference between the area swept out
by the portion of the circle whose arclength is marked by the intersections of the two overlapping circles 
and the area of the triangle whose vertices contain these two intersection points and the center of the circle as noted
in Fig. \ref{fig:overlap}(B). We will label the common angle between them as $2 \theta$.
\\
\\
\noindent
We can compute the area swept out by the circular slice as
\begin{align*}
  A_\mathrm{slice} &= \int_0^{2\theta} \mathrm{d}\theta' \int_0^{R+r} r' dr',\\
  &= \theta (R + r)^2, \numberthis \label{eq:A_slice}
\end{align*}
We note that the angle $\theta$ can be obtained with some trigonometry
\begin{align*}
  \mathrm{cos} \theta &= \frac{d/2}{R + r},\\
  &= \frac{d}{2 (R + r)}, \numberthis \label{eq:theta}
\end{align*}
so the area of the slice as a function of the distance $d$ is
\begin{align*}
  A_\mathrm{slice} &= (R + r)^2 \mathrm{cos}^{-1} \left( \frac{d}{2 (R + r)} \right)
\end{align*}
while the area of the triangle is
\begin{align*}
  A_\mathrm{triangle} &= \frac{d}{2} \times \sqrt{\left(R + r\right)^2 - \left( \frac{d}{2} \right)^2}. \numberthis \label{eq:A_triangle}
\end{align*}
\noindent
Then the overlap region is
\begin{align*}
  A_\mathrm{overlap} &= 2 \times \left( A_\mathrm{slice} - A_\mathrm{triangle} \right),\\
  &= 2(R + r)^2 \mathrm{cos}^{-1} \left( \frac{d}{2 (R + r)} \right) - d \times \sqrt{\left(R + r\right)^2 - \left( \frac{d}{2} \right)^2}. \numberthis \label{eq:A_overlap}
\end{align*}
\noindent
Suppose we made a change of variables to $d = 2(R + r) - \epsilon$ where $0 < \epsilon < 2r$. Then we can
modify $A_\mathrm{overlap}$ to be
\begin{align*}
  A_\mathrm{overlap} &= 2(R + r)^2 \, \mathrm{cos}^{-1} \left( \frac{2 (R + r) - \epsilon}{2 (R + r)} \right) - [2 (R + r) - \epsilon] \times \sqrt{\left(R + r\right)^2 - \left( \frac{2 (R + r) - \epsilon}{2} \right)^2},\\
  &= 2(R + r)^2 \, \mathrm{cos}^{-1} \left( 1 - \frac{\epsilon}{2 (R + r)} \right) - \left[ 2 (R + r) - \epsilon \right] \times \left( R + r \right) \sqrt{1 -  \left(\frac{2 (R + r) - \epsilon}{2 (R + r)} \right)^2},\\
  &\approx 2 \left( R + r \right)^2 \sqrt{\frac{\epsilon}{(R + r)}} - \left[ 2 (R + r) - \epsilon \right] \times \left( R + r \right) \sqrt{1 -  \left(1 - \frac{\epsilon}{2 (R + r)} \right)^2},\\
  &\approx 2 \left( R + r \right)^2 \sqrt{\frac{\epsilon}{(R + r)}} - 2 \left( R + r \right)^2 \left(1 - \frac{\epsilon}{2 (R + r)} \right) \sqrt{\frac{\epsilon}{(R + r)} - \left[ \frac{\epsilon}{ 2(R + r) } \right]^2}\\
  &= 2 \left( R + r \right)^2 \sqrt{\frac{\epsilon}{(R + r)}} \left\{ 1 - \left(1 - \frac{\epsilon}{2 (R + r)} \right) \sqrt{1 - \left[ \frac{\epsilon}{ 4(R + r) } \right]} \right\},\\
  &\approx 2 \left( R + r \right)^2 \sqrt{\frac{\epsilon}{(R + r)}} \left[ 1 - \left(1 - \frac{\epsilon}{2 (R + r)} \right) \left( 1 -\frac{\epsilon}{ 8(R + r) }  \right) \right],\\
  &\approx 2 \left( R + r \right)^2 \sqrt{\frac{\epsilon}{(R + r)}} \left[ \frac{5\epsilon}{8 (R + r)}  \right],\\
  &=  \frac{5 \left( R + r \right)^2}{4} \left[ \frac{\epsilon}{(R + r)} \right]^{3/2} \numberthis \label{eq:A_overlap_epsilon}
\end{align*}
where we note that $\epsilon \ll (R + r)$ and expand to enough orders to maintain a dependence on $\epsilon$. 
We also note that for small $x$, $\mathrm{cos}^{-1} (1 - x) \approx \sqrt{2x}$. As a result, the free energy is
\begin{align*}
  \Delta G &\approx N k_B T \frac{A_\mathrm{exc}}{A_\mathrm{sys}},\\
  &= N k_B T \frac{2 \pi (R + r)^2 - A_\mathrm{overlap}}{A_\mathrm{sys}},\\
  &= \frac{N}{A_\mathrm{sys}} k_B T \left\{ 2 \pi (R + r)^2 - \frac{5 \left( R + r \right)^2}{4} \left[ \frac{\epsilon}{(R + r)} \right]^{3/2} \right\},\\
  &= c k_B T \left\{ 2 \pi (R + r)^2 - \frac{5 \left( R + r \right)^2}{4} \left[ \frac{\epsilon}{(R + r)} \right]^{3/2} \right\}, \numberthis \label{eq:G_full}
\end{align*}
where we define the crowder concentration $c = \frac{N}{A_\mathrm{sys}}$.
As expected, we can see that the free energy goes down as the spacing between the microtubules goes down, suggesting
an energetic preference for keeping the microtubules close together.
\\
\\
\noindent
We can compute the entropic force as the negative derivative of the free energy with respect to the distance $d$. We
can then impose the change of variables to see that
\begin{align*}
  F_\mathrm{depletion} = - \frac{\partial \Delta G}{\partial d} &= - \frac{\partial \Delta G}{\partial (2 (R + r) - \epsilon)},\\
  &= \frac{\partial \Delta G}{\partial \epsilon},\\
  &= \frac{15}{8} c k_B T \left[  (R + r) \epsilon \right]^{1/2}. \numberthis \label{eq:force_depletion}
\end{align*}
\noindent
We can then imagine that if we operated in three dimensions, then in the case where two microtubules of length $L$ that
are aligned would have a depletion force that goes as
\begin{align*}
  F = \frac{15}{8} c k_B T L \left[ (R + r) \epsilon \right]^{1/2}. \numberthis \label{eq:force_depletion3d}
\end{align*}

\begin{figure}[t!]
  \centering{
  \includegraphics[scale=0.4]{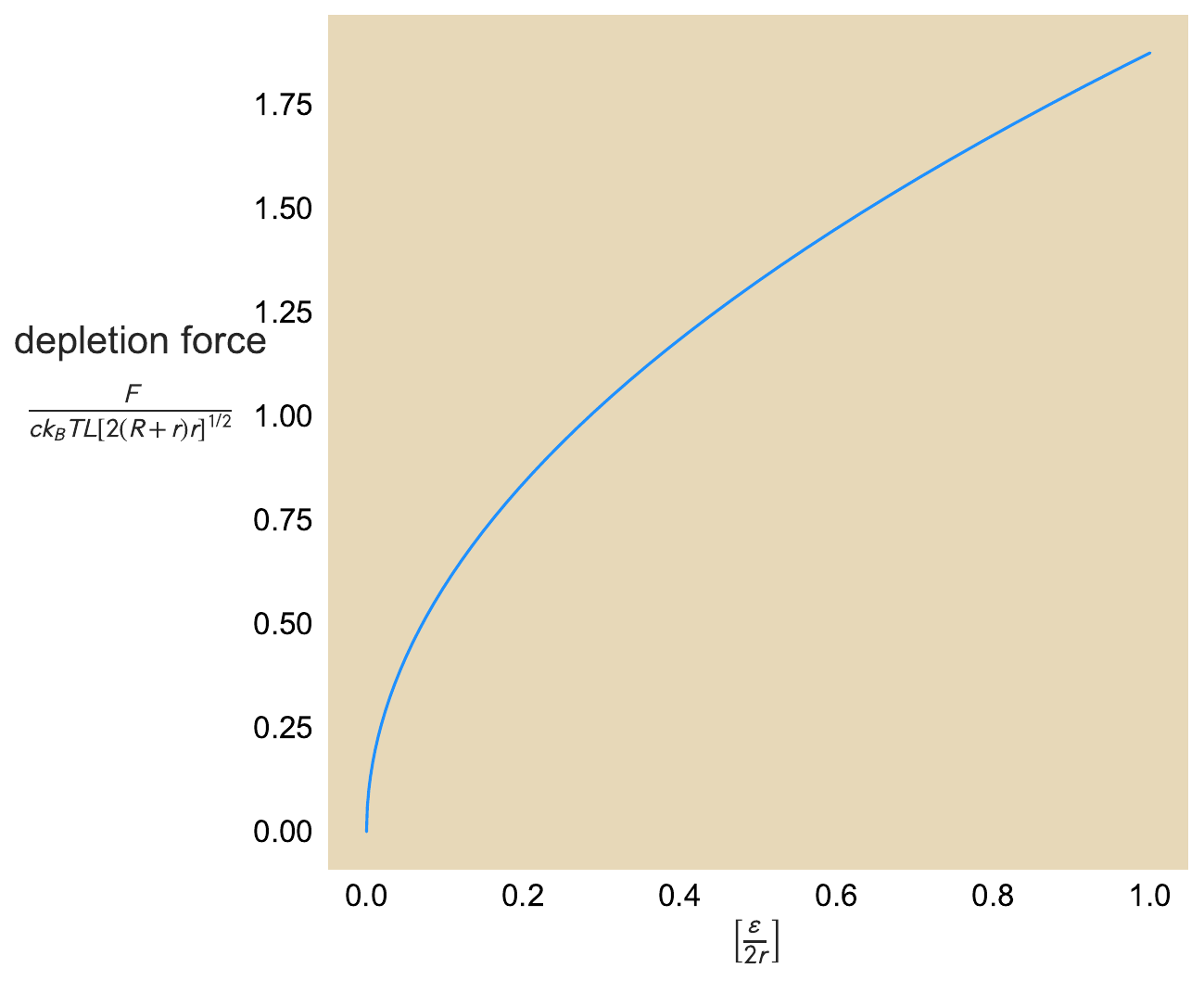}}
  \caption{\textbf{Depletion force as a function of overlap distance $\epsilon$.}}
  \label{fig:depletion}
\end{figure}

\noindent
Fig. \ref{fig:depletion} shows the relation between the depletion force and the overlap length $\epsilon$. Here, we
normalize both sides according to the axes labels. As expected, the depletion force increases as the two microtubules
become closer to each other.
\\
\\
\noindent
If we estimate that a 1 {\textmu}m-long microtubule has an outer radius of $\backsim10$ nm and pluronic, with
a mass of 12.5 kDa and a final concentration of 0.5 mg/mL in the experiments (making it 40 {\textmu}M), has a
radius of $\backsim1$ nm, then the depletion force due to pluronic is
\begin{align*}
  F_\mathrm{depletion}^{\mathrm{pluronic}} &\backsim \frac{15}{8} \times \frac{40 \times 10^3}{\mu\mathrm{m}^3} \times (4 \, \mathrm{ pN} \cdot \mathrm{nm}) \times (10 \, \mathrm{ nm} \times 1 \, \mathrm{ nm})^{1/2} \times 1 {\mu \mathrm{m}},\\
  &\backsim \frac{40 \times 10^4}{\mu \mathrm{m}^2} \times 4 \, \mathrm{ pN} \cdot \mathrm{nm}^2,\\
  &\backsim 1 \, \mathrm{pN}. \numberthis \label{eq:pluronic}
\end{align*}
\noindent
Thus, we can see that microtubules under the standard conditions are subjected to roughly pico-Newton forces, within
the range of forces expected to be exerted by motors. We note, however, that the size of pluronic is
even larger, most likely underestimating the computed entropic force.

\section{Microtubule bundling can affect both contraction speed and filament redistribution} \label{sisec:pluronic_data}
\begin{figure}[b!]
\centering{
\includegraphics[trim={0cm 0cm 0cm 0cm}, scale=0.5]{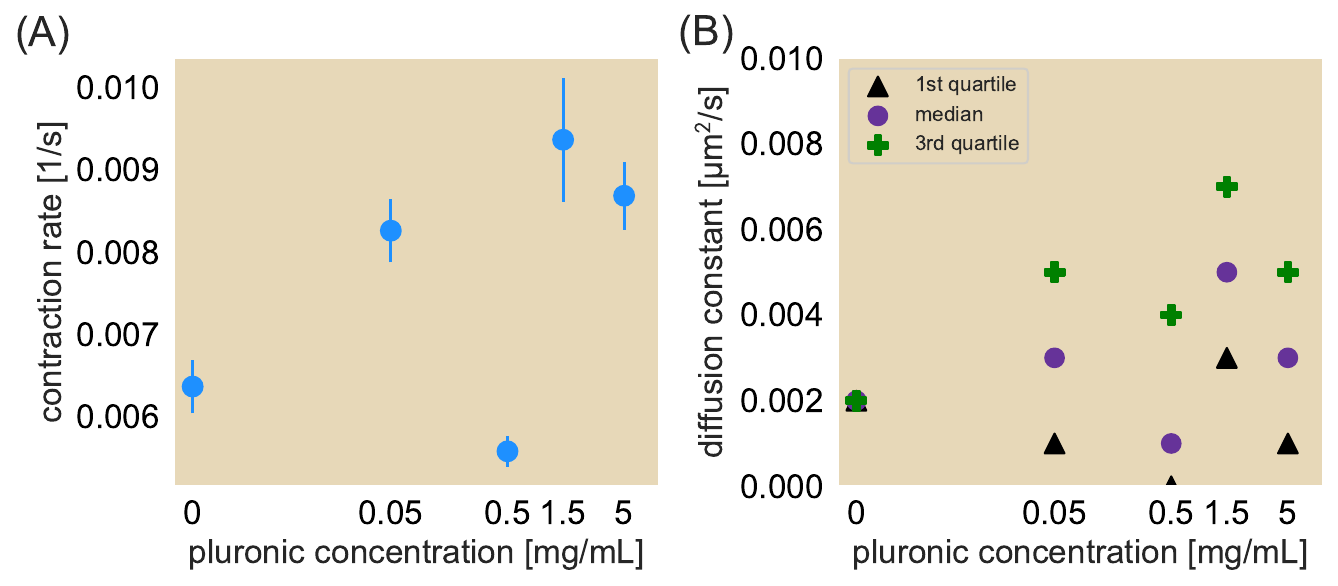}}
\caption{\textbf{The effect of pluronic on the network contraction rate and 
effective diffusion constant.} (A) Contraction rate and (B) diffusion constant
as a function of pluronic concentration as presented here use bacteria-expressed K401
motors. The effective diffusion constants shown here are obtained from best fits to the
1st quartile (triangle), median (circle), and 3rd quartile (plus symbol) of the normalized area trajectories. We note the outlier at 0.5 mg/mL in panel (A) likely
corresponds with different storage conditions of the pluronic than from the
rest of the other pluronic concentrations used in this study, which may have
had a biochemical impact in the assay.}
\label{fig:pluronic}
\end{figure}
Depletion agents such as pluronic or polyethylene glycol (PEG) play central roles in pushing active systems into contractile
or extensile regimes \cite{sanchez2012,najma2022}. These polymers help to induce entropic forces between filaments to form
bundles, which can help allow active or passive crosslinkers to induce filament coupling over larger length scales. 
The motor-microtubule system examined thus far includes 0.5 mg/mL pluronic, a concentration that can induce picoNewton-scale
forces between microtubules (see SI Sec S7). Here, we ask what happens to the contraction and bulk filament redistribution
when these entropic forces are tuned to an alteration to the concentration of pluronic. We thus implement our photobleaching
scheme and track the movement and areas of the fluorescent unit cells when the system is altered over a range of pluronic
concentrations, from a complete removal of the depletion agent to a 10-fold increase in concentration, while keeping all else
fixed, including motor and microtubule concentrations.
Fig. \ref{fig:pluronic} shows the contraction rate and effective diffusion constant for the bacterial-expressed 
K401 across a range of
pluronic concentrations, including its complete absence. We find that increasing the
pluronic concentration leads to a general increase in the contraction rate until 1.5 mg/mL, after which contraction does
not appear to occur any faster. In the absence of pluronic, the network contracts more slowly, with a rate roughly 2/3 the
rate of the 1.5 mg/mL pluronic concentration. We note the dramatic decrease at the standard experimental conditions using
0.5 mg/mL pluronic, which lies below even the complete absence of pluronic. We hypothesize that this inconsistency comes
from the storage of pluronic in the standard set of experiments being different than the storage conditions used for the pluronic
when performing the titration series. Briefly, under standard conditions, the
pluronic is stored in the base reaction buffer used in the experimental assay
involving K-PIPES, MgCl$_2$, EGTA, and KOH. It is possible that under long-term
storage in this media, the pluronic behaves differently and as a result 
exhibits a different effect for the standard reaction.
\\
\\
\noindent
When we computed the effective diffusion constants for the different quartiles as shown in Fig. \ref{fig:pluronic}(B), we found with
the increase in pluronic
a general increase in effective diffusion constant for the 3rd quartile and median data, but a roughly constant effective diffusion
constant for the 1st quartile data. Interestingly, we note that the general increase and decrease of the effective diffusion constants
also appears to follow the contraction rate at the corresponding pluronic concentration, suggesting a close relation between the two.
\\
\\
\noindent
While crowding is commonly implemented
in inducing organization in \emph{in vitro} active matter systems and has become a focus of
attention as a tunable parameter \cite{nasirimarekani2021,najma2024}, only recently has crowding
been systematically studied to understand its effects on bulk reorganization of 
a cytoskeletal network \cite{chew2023}.
Nevertheless, to our knowledge, we show some of the first experimental studies systematically tuning the effects of
crowding on bulk reorganization and observe that entropic forces have more of a binary effect on the
contraction rate: in the absence of pluronic, the network contracts more slowly and by adding even 0.1 
mg/mL of crowding agent the network contracts more quickly without much more increase in contraction
dynamics at higher concentrations. This suggests that entropic forces on the order of pico-Newton
scales are sufficient to aid in the formation of a contracted filament network. This is roughly in the same 
order of magnitude 
as stall forces for motors, further supporting the role of crowding as generating similar effects to passive 
crosslinkers.
